\documentclass[aps,nofootinbib,showpacs,preprintnumbers,amsmath,amssymb]{revtex4}
\usepackage{epsfig}
\usepackage{epsf}
\usepackage{amssymb}
\usepackage{amsmath}

\newcommand{\A}{\mathcal{A}}
\newcommand{\tA}{{\widetilde {\mathcal{A}}}}
\newcommand{\tlt}{{\widetilde t}}
\newcommand{\tlA}{\mathfrak A}
\newcommand{\be}{\begin{equation}}
\newcommand{\ee}{\end{equation}}
\newcommand{\ba}{\begin{eqnarray}}
\newcommand{\ea}{\end{eqnarray}}
\newcommand{\bMS}{{\overline {\rm MS}}}
\newcommand{\bL}{{\overline \Lambda}}
\newcommand{\bM}{{\overline M}}
\newcommand{\ovc}{{\overline c}}

\begin{document}
\preprint{USM-TH-195, hep-ph/0608256 v3, to appear in PRD}

\title{Various versions of analytic QCD and skeleton-motivated evaluation of observables}

\author{Gorazd Cveti\v{c}}
  \email{gorazd.cvetic@usm.cl}
\author{Cristi\'an Valenzuela}
  \email{cristian.valenzuela@usm.cl}
\affiliation{Dept.~of Physics, Universidad T\'ecnica
Federico Santa Mar\'{\i}a, Valpara\'{\i}so, Chile} 

\date{\today}

\begin{abstract}
We present skeleton-motivated evaluation of QCD observables. The approach
can be applied in analytic versions of QCD in certain classes of 
renormalization schemes. We present two versions of analytic QCD
which can be regarded as low-energy modifications of the 
``minimal'' analytic QCD and which reproduce the measured value of the
semihadronic $\tau$ decay ratio $r_{\tau}$. Further, we describe an
approach of calculating the higher order analytic couplings
$\mathcal{A}_k$ ($k=2,3,\ldots$) on the basis of logarithmic
derivatives of the analytic coupling $\mathcal{A}_1(Q^2)$. This
approach can be applied in any version of analytic QCD.
We adjust the free parameters of the afore-mentioned two analytic
models in such a way that the skeleton-motivated evaluation
reproduces the correct known values of $r_{\tau}$ and
of the Bjorken polarized sum rule (BjPSR) $d_b(Q^2)$ at a given point
(e.g., at $Q^2=2 \ {\rm GeV}^2$).
We then evaluate the low-energy behavior of the Adler function
$d_v(Q^2)$ and the BjPSR $d_b(Q^2)$ in the afore-mentioned evaluation 
approach, in the three analytic versions of QCD. We compare with
the results obtained in the ``minimal'' analytic QCD and with the
evaluation approach of Milton {\em et al.\/} and Shirkov.
\\
{\noindent
Changes in v3: the values of parameters of analytic QCD models
M1 and M2 were refined and the numerical results modified
accordingly; the penultimate paragraph of Sec.~II and 
the ultimate paragraph of Sec.~III are new; discussion of
Figs.~4 was extended; new references were added.}
\end{abstract}

\maketitle

\section{Introduction}
In perturbative QCD (pQCD), the coupling parameter 
$a(Q^2) \equiv \alpha_s(Q^2)/\pi$
[where: $Q^2=-q^2=-(q^0)^2 + {\bf q}^2$] is obtained on
the basis of the perturbative $\beta$-function which is
a (truncated) polynomial of $a$. As a consequence,
$a(Q^2)$ has Landau singularities in an infrared space-like
zone ($Q^2 > 0$), and therefore these singularities are unphysical.
This problem was fully recognized and a solution found
about ten years ago by Shirkov and Solovtsov \cite{ShS}. 
The solution found was minimal in the sense that the analytization 
$a(Q^2) \mapsto \mathcal{A}_1(Q^2)$ was performed
by removing the Landau-cut singularities, while keeping
the singularities on the time-like axis unchanged.
Further, completely analogous minimal analytization
was performed for the higher powers 
$a^k \mapsto \mathcal{A}_k$ ($k\geq 2$) 
and this replacement was performed term-by-term
in the simple truncated perturbation series 
(STPS -- in powers of $a$) of
observables by Milton, Solovtsov, Solovtsova, and Shirkov
\cite{Milton:1997mi,Sh,Milton:2000fi} (``Analytic Pertubation Theory'' --
APT).\footnote{
Analytization of noninteger powers in MA was performed and
used in Refs.~\cite{Bakulev}, representing a generalization 
of results of Ref.~\cite{Broadhurst:2000yc}.} 
The resulting series
have in general better convergence behavior and
much less sensitivity under the variation of the
renormalization scale (RScl) and scheme (RSch).
We will call the analytic QCD model based on the afore-mentioned
analytic coupling the ``minimal analytic'' (MA) model
[$\mapsto \mathcal{A}_1^{\rm (MA)}(Q^2)$], 
and the afore-mentioned evaluation approach (involving the
truncated analytic series) the APT-evaluation approach.

The MA coupling $\mathcal{A}_1^{\rm (MA)}(Q^2)$
contains just one adjustable parameter, the QCD scale $\Lambda$.
Reproduction of the measured values of the higher energy
QCD observables ($|q^2| > 10 \ {\rm GeV}^2$) fixes the
scale parameter to the value $\Lambda_{(n_f=5)} \approx 0.26$ GeV,
corresponding to $\Lambda_{(n_f=3)} \approx 0.4$ GeV.
However, then the well-measured value of 
the massless strangeless semihadronic $\tau$-decay ratio 
$r_{\tau}(\triangle S=0, m_q=0) = 0.204 \pm 0.005$
\cite{ALEPH1,ALEPH2,ALEPH3} (cf.~Appendix \ref{app5})
cannot be reproduced \cite{Milton:2000fi}
unless large values of the
$u$, $d$ and $s$ quark masses are introduced
($m_q \approx 0.25$-$0.45$ GeV) \cite{Milton:2001mq}
and the threshold effects become very important. 
One may want to avoid introduction of such large
quark masses, by modifying the MA
model at low energies while keeping the analyticity
of $\mathcal{A}_1(Q^2)$ in the non-time-like region.
In this work we introduce two somewhat different
modifications $\Delta \mathcal{A}_1(Q^2)$ 
($\mathcal{A}_1 \equiv \mathcal{A}_1^{\rm (MA)} + 
\Delta \mathcal{A}_1$), 
both having power-like behaviors.
We construct in a systematic way
the higher order couplings $\mathcal{A}_k(Q^2)$
based on the logarithmic derivatives of $\mathcal{A}_1(Q^2)$.
Further, we construct a skeleton-expansion-motivated algorithm of
evaluation of QCD observables, which can be applied in
any analytic version of QCD and in a large class of
renormalization schemes. For such an evaluation, we have to
know the first few coefficients of STPS 
and all the leading-$\beta_0$ coefficients of
the full perturbation series. 
We believe that the inclusion in this evaluation of 
the light-by-light contributions, if they contribute, should be avoided.
Such contributions have a different topological structure and
their evaluation should be performed separately in most 
evaluation (resummation) methods -- see, for example, Ref.~\cite{KS}.
Some of the main results of the present work were published by us
in a summarized form in Ref.~\cite{Cvetic:2006mk}.

In Sec.~II, we explain the main features of the
analytic versions of QCD (anQCD), we present the known MA model,
and propose two versions of modified MA -- the models 'M1' and 'M2'
[$\mapsto \mathcal{A}_1^{\rm (M1)}(Q^2)$, $\mathcal{A}_1^{\rm (M2)}(Q^2)$].
In Sec.~III, we introduce the higher order
couplings $\mathcal{A}_k(Q^2)$ ($k \geq 2$) in a way that
can be applied in any version of anQCD, by imposing on them
specific natural behavior under the change of scale
$Q^2$ and of RSch.
In Sec.~IV we then present an algorithm which allows us
to evaluate any QCD observable in any version of anQCD,
an algorithm motivated by the skeleton expansion.
In Sec.~V we fix the free parameters in the M1 and M2 
anQCD couplings $\mathcal{A}_1^{\rm (M1)}(Q^2)$ and
$\mathcal{A}_1^{\rm (M2)}(Q^2)$ in such a way that
the afore-mentioned skeleton-motivated approach gives us
the measured values of $r_{\tau}$ and of the
Bjorken polarized sum rule (BjPSR) $d_b(Q^2)$ at
$Q^2=2 \ {\rm GeV}^2$. 
We then present the resulting
low-energy curves for the $V$-channel Adler function
$d_v(Q^2)$ and of the BjPSR $d_b(Q^2)$ in the
skeleton-motivated approach, in the anQCD versions MA, M1, M2.
We investigate the RScl- and RSch-dependence of the
numerical curves, and in the MA-case we compare the
results of $d_b(Q^2)$ obtained by our skeleton-motivated 
evaluation approach with those of the APT approach
of Refs.~\cite{Milton:1997mi,Sh,Milton:2000fi}.
Numerical calculations were performed using
{\it Mathematica\/} \cite{math}. 
In Sec.~VI we present our conclusions and prospects
for further work in this direction. 
Appendix \ref{app1} contains details of the coefficients appearing in
the evaluation method. In Appendix \ref{app2} we present another
evaluation method that is even more closely related to 
the skeleton expansion.
Appendix \ref{app3} contains a derivation of the 
leading skeleton (LS) characteristic function of the BjPSR,
and relations between the space-like and time-like
formulations for the LS-term.
Appendix \ref{app4} is a compilation of expressions
of some coefficients used in this work, and Appendix \ref{app5}
describes an extraction of the experimental value
of $r_{\tau}(\triangle S=0, m_q=0)$.

\section{Minimal analytic QCD and two extensions of it}
\label{anQCDs}
The perturbative QCD coupling $a(Q^2)
\equiv \alpha_s(Q^2)/\pi$ in the space-like region 
[$Q^2$ not in $(-\infty,0)$] has the scale dependence
governed by the renormalization group equation (RGE)
\ba
\frac{\partial a(\ln Q^2; \beta_2, \ldots)}{\partial \ln Q^2} & = &
- \sum_{j=2}^{j_{\rm max}} \beta_{j-2} \: a^j (\ln Q^2; \beta_2, 
\ldots) \ ,
\label{pRGE}
\ea
where the first two coefficients
$\beta_0= (1/4)(11-2 n_f/3)$ and
$\beta_1=(1/16)(102-38 n_f/3)$
are scheme-independent in mass-independent schemes,
and the other coefficients $\beta_j$ ($j \geq 2$)
characterize the RSch.
In practice, the above sum is truncated at a certain
$j_{\rm max}$ where $j_{\rm max}-1$ is the
loop level. The perturbative RGE (\ref{pRGE})
has a standard iterative solution in the form
\be
a(Q^2)=
\sum_{k=1}^{\infty} \sum_{\ell=0}^{k-1} K_{k \ell} \:
\frac{(\ln L)^{\ell}}{L^k},
\label{apt}
\ee
where $L=\ln(Q^2/\Lambda^2)$
and $K_{k \ell}$ are constants depending on the $\beta_j$
coefficients and on the choice of the scale $\Lambda$.
If the conventional (``$\bMS$'') scale $\Lambda=\bL$ 
\cite{Buras:1977qg,Bardeen:1978yd} is used, the
coefficients $K_{k \ell}$ are
\ba
K_{10}&=&1/\beta_0; \ K_{20}=0; \ K_{21}= - \beta_1/\beta_0^3; \
\nonumber\\
K_{30}&=&- \beta_1^2/\beta_0^5 + \beta_2/\beta_0^4; \
K_{31} = -K_{32} = - \beta_1^2/\beta_0^5; \
\ldots
\label{kijbL}
\ea
Further coefficients $K_{k \ell}$, up to $k=6$, are given
in Appendix \ref{app4}.
The coupling $a(Q^2)$, Eq.~(\ref{apt}), has nonanalytic structure
along the time-like axis $Q^2(\equiv - q^2) < 0$.
In addition, it has singularities in the
space-like region $0 < Q^2 \leq \bL^2$, which 
are formally the consequence of the (truncated) power expansion
structure of the beta-function on the RHS of Eq.~(\ref{pRGE}).
Application of the Cauchy theorem to function $a(Q^2)$
in the $Q^2$-plane gives then
the following dispersion relation for $a$:
\begin{equation}
a(Q^2) = \frac{1}{\pi} \int_{\sigma= - \Lambda^2 - \eta}^{\infty}
\frac{d \sigma \rho^{\rm (pt)}_1(\sigma) }{(\sigma + Q^2)}
\ ,
\label{aptdisp}
\end{equation}
where $\rho^{\rm (pt)}_1(\sigma)$ is the (pQCD) 
discontinuity function of $a$
along the cut axis in the $Q^2$-plane:
$\rho^{\rm (pt)}_1(\sigma)= {\rm Im} a(-\sigma - i \epsilon)$.
In the integration, $\eta$ is positive ($\eta \to +0$ can be taken),
reflecting the fact that the corresponding contour integration
path avoids entirely the singularities of $a(z)$ in the
complex plane, including the singularity at
$z \equiv -\sigma = \Lambda^2$ [cf.~Eq.~(\ref{apt})].
	
By special relativity and causality, observables are
analytic functions of the associated physical
momentum squared $q^2 \equiv - Q^2$ in the $Q^2$-plane with 
the time-like axis ($Q^2 < 0$) excluded.
Since QCD observables are functions of the invariant coupling
$a(Q^2)$, both should have the same analyticity properties.
The singularity sector $0 < Q^2 \leq \Lambda^2$ in
$a(Q^2)$, Eqs.~(\ref{apt}) and (\ref{aptdisp}),
is therefore nonphysical. The most straightforward rectification
of this problem is to eliminate that sector from
the dispersion relation (\ref{aptdisp}) while keeping the
pQCD discontinuity function $\rho^{\rm (pt)}(\sigma; \beta_2, \ldots)$
unchanged on the time-like axis $\sigma > 0$
\cite{ShS}, thus leading to the specific ``minimal analytic''
(MA) coupling
\begin{equation}
\A^{\rm (MA)}_1(Q^2) = \frac{1}{\pi} \int_{\sigma= 0}^{\infty}
\frac{d \sigma \rho^{\rm (pt)}_1(\sigma) }{(\sigma + Q^2)} \ .
\label{MAA1disp}
\end{equation}
In practice, truncated series (\ref{apt}) can be used to
obtain the discontinuity function $\rho^{\rm (pt)}_1(\sigma)$ and
thus the coupling (\ref{MAA1disp}).
Prescription (\ref{MAA1disp}) was investigated from
calculational viewpoints in Refs.~\cite{Magradze,Magradze2,Alekseev:2002zn}.
There exists a practical iterative solution
\cite{Magradze,Magradze2} to RGE (\ref{pRGE}) based on
the Lambert function \cite{Gardi:1998qr}. This solution
is an expansion of a different form than (\ref{apt}).
When the number of terms in the Lambert-based expansion 
and in expansion (\ref{apt}) increases,
the two solutions for $\A_1^{\rm (MA)}$ converge to the
exact numerical solution rapidly for all $Q^2$, but
the Lambert-based expansion converges faster. 
When $k_{\rm max} \geq 4$ in (\ref{apt}),
the corresponding solution $\A_1^{\rm (MA)}(Q^2)$ differs 
in $\bMS$ RSch from the exact numerical solution by less
than one per cent for all $Q^2 > 0$ \cite{Magradze2}.
In the present work, we will use expansion (\ref{apt})
with $k_{\rm max} = 5$ or 6.

Other types of analytization of $a$
can be performed by focussing on the
analyticity properties of the beta function
\cite{Nesterenko,Raczka}, or by subtracting 
certain power correction terms $1/(Q^2)^n$
from the MA coupling $\A_1^{\rm (MA)}$ \cite{Alekseev}.
For a review of various models, see Ref.~\cite{Prosperi:2006hx}.

In general, the discontinuity function can be
different, and the analytic coupling must have the form
\begin{equation}
\A_1(Q^2) = \frac{1}{\pi} \int_{\sigma= 0}^{\infty}
\frac{d \sigma \rho_1(\sigma) }{(\sigma + Q^2)} \ ,
\label{A1disp}
\end{equation}
where $\rho_1(\sigma) = {\rm Im} \A_1(-\sigma - i \epsilon)$ .
Relation (\ref{A1disp}) defines an analytic coupling 
in the $Q^2$-plane excluding the time-like semiaxis
$-s = Q^2 < 0$. On this semi-axis, it is convenient to
define the time-like coupling \cite{time-like1,time-like2,time-like3}
\begin{equation}
\tlA_1(s) = \frac{i}{2 \pi} 
\int_{- s + i \epsilon}^{-s - i \epsilon} 
\frac{d \sigma^{\prime}}{\sigma^{\prime}}
\A_1(\sigma^{\prime}) \ .
\label{tlA1A1}
\end{equation}
The integration here is in the $Q^2 \equiv \sigma^{\prime}$ plane avoiding
the (time-like) cut $\sigma^{\prime} < 0$. The relation between
$\A_1(Q^2)$ and $\tlA_1(s)$ is the same
as the relation between the (vector channel) Adler function $D_V(Q^2)$ 
and its time-like analogue, the $e^+e^-$ hadronic scattering 
cross section ratio $R_V(s)$. 
Therefore, while the leading QCD correction
to $D_V(Q^2)$ in anQCD is $\A_1(Q^2)$ 
[-- the anQCD analogue of $a(Q^2)$],
the leading QCD correction to $R_V(s)$ is $\tlA_1(s)$.
The following additional relations \cite{Sh} hold between $\A_1$, $\tlA_1$ and 
$\rho_1$ in any anQCD:
\begin{eqnarray}
\tlA_1(s) &=& \frac{1}{\pi} 
\int_s^{\infty} \frac{d \sigma}{\sigma}
{\rho}_1(\sigma) \ ,
\label{tlA1}
\\
\A_1(Q^2) & = & Q^2 \int_0^{\infty} 
\frac{ ds {\tlA}_1(s) }{(s + Q^2)^2} \ ,
\label{A1tlA1}
\\
\frac{d}{d \ln \sigma} {\tlA}_1(\sigma) &=&
- \frac{1}{\pi} {\rho}_1(\sigma) \ .
\label{tlA1rho1}
\end{eqnarray} 
The MA coupling (\ref{MAA1disp}) contains only one free
parameter, the value of the ($\bMS$) scale $\bL$,
which is not equal to the value of $\bL$ in pQCD,
but has to be adjusted so that the measured values of 
QCD observables be reproduced. By introducing and using a specific
evaluation method within the MA QCD, the authors of 
Refs.~\cite{Milton:1997mi,Sh,Milton:2000fi}
reproduced the measured values of the higher energy
QCD observables ($|q^2| > 10 \ {\rm GeV}^2$) when the
scale parameter $\bL$ had the value 
$\bL_{(n_f =5)} \approx 0.26$ GeV (where $n_f$ is the number
of active quark flavors). This corresponds
to $\bL_{(n_f =3)} \approx 0.4$ GeV.
However, the measured value of the massless part of the
semihadronic strangeless $\tau$-decay ratio 
$r_{\tau}(\triangle S=0,m_q=0) = 0.204 \pm 0.005$
\cite{ALEPH1,ALEPH2,ALEPH3} 
[cf.~Appendix \ref{app5}, Eq.~(\ref{rtauexp})]
cannot be reproduced with such values of $\bL$ \cite{Milton:2000fi}
unless large masses of 
$u$, $d$ and $s$ quarks are introduced
($m_q \approx 0.25$-$0.45$ GeV) \cite{Milton:2001mq}
and the mass threshold effects become central. 

The above consideration motivates us to introduce
low-energy modifications of the MA coupling. 
Modifications, although simple, introduce additional
parameters which have to be fixed by requiring
reproduction of the measured values of 
low-energy QCD observables, including of $r_{\tau}$.
One possible modification is inspired by the 
well measured \cite{ALEPH1,ALEPH2} isovector hadronic spectral
function $R_V(s)$. At low energies ($s < 1 \ {\rm GeV}^2$)
it is dominated by the $\rho$-resonance ($M_{\rho}=0.776$ GeV),
which, in the narrow width approximation, can be represented
as a delta function $\delta(s - M_{\rho}^2)$ \cite{PPR}.
This is in the spirit of the Vector Meson Dominance (VMD).
If we assume that the $s$-dependence of the time-like
quantity $R_V(s)$ is at least qualitatively described by
the first order time-like coupling $\tlA_1(s)$, Eq.~(\ref{tlA1}),
then the afore-mentioned delta-like structure should appear 
in it. This then leads to the following ansatz (model 'M1'):
\ba
\tlA_1^{\rm (M1)}(s) & = & 
c_f \bM_r^2 \delta(s - \bM_r^2)
+ k_0 \Theta(\bM_0^2 - s) + 
\Theta(s - \bM_0^2) \tlA_1^{\rm (MA)}(s) \ ,
\label{M1tl}
\ea
where $c_f$, $k_0$, $c_r= \bM_r^2/\bL^2$,
$c_0 = \bM_0^2/\bL^2$ are four dimensionless
parameters of the model; $\Theta(x)$ is the Heaviside
step function ($+1$ for $x>0$, zero otherwise).
In this model, the MA behaviour
of $\tlA_1(s)$ at low energies $s < \bM_0^2$ has
been replaced by a constant ($k_0$) plus a delta function
(at $s=\bM_r^2 < \bM_0^2$). The more literal application of
the VMD approach results in $k_0=-1$ \cite{Cvetic:2005my}.
This is so because $R_V(s) = 1 + \tlA_1(s) + {\cal O}(\tlA_1^2)$,
and $R_V(s) \to 0$ when $s \to 0$, implying 
$\tlA_1(s) \to -1$. However, such a model
appears to restrict the low energy behavior of $\tlA_1(s)$
and of $\A_1(Q^2)$ too severely, especially if we
want to impose the condition of merging $\A_1(Q^2)$ of the
model with $\A_1^{\rm (MA)}(Q^2)$ at high $Q^2$. As a
consequence, values of various unrelated low energy observables,
such as Adler function (or: $r_{\tau}$) and Bjorken polarized sum rule,
cannot be reproduced simultaneously in such a model. Therefore, 
unlike the choice $k_0=-1$ in Ref.~\cite{Cvetic:2005my},
we keep here the constant $k_0$ in Eq.~(\ref{M1tl}) free.  
Applying transformation (\ref{A1tlA1}) to expression (\ref{M1tl})
gives the space-like analytic coupling of the model
\ba
\A_1^{\rm (M1)}(Q^2) & = & 
\A_1^{\rm (MA)}(Q^2) + \Delta A_1^{\rm (M1)}(Q^2) \ ,
\label{M1a}
\\
\Delta A_1^{\rm (M1)}(Q^2) & = & 
- \frac{1}{\pi} \int_{\sigma= 0}^{\bM_0^2}
\frac{d \sigma \rho^{\rm (pt)}_1(\sigma) }{(\sigma + Q^2)} +
c_f \frac{ \bM_r^2 Q^2 }{ \left( Q^2 + \bM_r^2 \right)^2 } -
d_f \frac{ \bM_0^2 }{ \left( Q^2 + \bM_0^2 \right) } \ ,
\label{M1b}
\ea
where the constant $d_f$ is
\be
d_f \equiv - k_0 + \frac{1}{\pi} \int_{\bM_0^2}^{\infty} 
\frac{d \sigma}{\sigma} \rho^{\rm (pt)}_1(\sigma) \ .
\label{df1}
\ee

The coupling (\ref{M1a})-(\ref{M1b}) 
can also be rewritten in a somewhat different, but equivalent, form
\begin{eqnarray}
\mathcal{A}_1^{\rm (M1)}(Q^2) =
c_f \frac{\bM_r^2 Q^2}{ (Q^2 + \bM_r^2)^2} + 
k_0 \frac{\bM_0^2}{(Q^2 + \bM_0^2)} + 
\frac{Q^2}{(Q^2 + \bM_0^2)} \frac{1}{\pi} 
\int_{\sigma= \bM_0^2}^{\infty} 
\frac{ d \sigma \rho_1^{\rm (pt)}(\sigma) 
(\sigma - \bM_0^2)}{\sigma (\sigma + Q^2)} \ .
\label{M1c} 
\end{eqnarray}
In general, this coupling differs from the MA coupling (\ref{MAA1disp})
by terms $\Delta \A_1^{\rm (M1)} \sim \bL^2/Q^2$. 
However, we will choose to require $\delta \A_1^{\rm (M1)} \sim \bL^4/Q^4$,
i.e., that M1 effectively merge into MA at higher energies,
as we did in Ref.~\cite{Cvetic:2005my}.
This condition eliminates one of the four new parameters, 
for example $k_0$:
\be
k_0 = - \frac{ c_r c_f }{ c_0 } + 
\frac{1}{\pi} \frac{1}{c_0 \bL^2}
\int_0^{c_0 \bL^2} d \sigma \rho^{\rm (pt)}_1(\sigma) +
\frac{1}{\pi} \int_{c_0 \bL^2}^{\infty}
\frac{d \sigma}{\sigma} \rho^{\rm (pt)}_1(\sigma) \ .
\label{M1MA}
\ee
Since the presented version of M1 merges with MA at higher energies,
the value of the scale parameter $\bL$ remains practically unchanged,
$\bL_{(n_f=3)} = 0.4$ GeV, and the model contains only three
dimensionless parameters $c_f$, $c_r$ and $c_0$.

Another, somewhat simpler, modification of the MA coupling consists
in adding a constant value ($c_v$) in the low-energy
region of the MA time-like coupling (model 'M2'):
\ba
\tlA_1^{\rm (M2)}(s) & = & \tlA_1^{\rm (MA)}(s) + 
c_v \Theta(\bM_p^2 - s) \ ,
\label{M2tl}
\\
\A_1^{\rm (M2)}(Q^2) & = & \A_1^{\rm (MA)}(Q^2) + 
c_v  \frac{ \bM_p^2 }{(Q^2 + \bM_p^2)} \ ,
\label{M2}
\ea
where $c_v$ and $c_p = \bM_p^2/\bL^2$ are two dimensionless
parameters of the model. For simplicity, we will assume that
the scale parameter is unchanged: $\bL_{(n_f=3)} = 0.4$ GeV.
The resulting additional term $\propto 1/(Q^2 + \bM_p^2)$
in $\A_1(Q^2)$ can be interpreted, or motivated, as the leading
power-like modification ($\propto 1/Q^2$)
of the MA coupling such that the condition
$|\A_1(Q^2=0)| < \infty$ is preserved. The latter condition
is regarded as desirable in our approach developed in 
Sec.~\ref{skmotexp}, because the so called leading-skeleton
resummation of observables remains finite in such a model. 

Model M1 was motivated by simulating roughly the $\rho$-resonance
contribution in the {\it one-loop} expression for $R_V(s)$, via
a VMD narrow width approximation ansatz in $\tlA_1(s)$. 
However, this was only a motivation for the construction of
an explicit form of $\tlA_1(s)$ as the starting point of the model, 
and the higher-loop contributions $\tlA_k(s)$
and $\A_k(Q^2)$  ($k \geq 2$) are then constructed 
on the basis of this $\tlA_1(s)$ (see the next Section).
The approximation of the $\rho$-resonance is then expected
to get worse at higher loop level. 
Another possible approach,
which we will not follow here, would be to refine
(retroactively) $\tlA_1(s)$ so that higher-loop
evaluations of $R_V(s)$ give us a given specified approximation of 
the $\rho$-resonance at low energies. A similar approach could
possibly be followed also in M2. In general, reproduction of
the correct low-energy behavior of time-like observables such
as $R_V(s)$ represents a difficult problem. In this work,
we will follow a more modest approach -- in
Sec.~\ref{numres} we will fix the
free parameters of models M1 and M2 by requiring,
at loop-level three or four, the reproduction
of the central experimental values for the Bjorken polarized
sum rule $d_b(Q^2)$ at two (in M1) or one (in M2) values
of scale $Q$ ($\geq 1$ GeV), and the reproduction of the
measured value of $r_{\tau}(\triangle S=0)$.

All the versions of anQCD presented here are
infrared finite, i.e., the zero momentum limits
$\A_1(0) = \tlA_1(0)$ are finite.

\section{Analytization of higher powers of the coupling parameter}
\label{analytiz}

In the previous Section, a few of the possibilities of constructing
the analytic version $\A_1(Q^2)$ of $a(Q^2)$ were
presented. For evaluation of QCD observables, the
analytic versions of higher powers $a^k(Q^2)$
are needed as well. For that, there is no unique way of
constructing the correspondence 
$a^k \leftrightarrow \A_k$. In the MA QCD,
one possibility is to apply the MA procedure (\ref{MAA1disp})
to each power of $a$ \cite{Milton:1997mi}:
\begin{equation}
a^k(Q^2) \: \mapsto \:
\mathcal{A}^{\rm (MA)}_k(Q^2) = 
\frac{1}{\pi}\int_0^\infty \frac{d\sigma}{\sigma+Q^2}\: 
\rho_k^{\rm (pt)}(\sigma)
\qquad (k=1,2,\ldots) \ ,
\label{MAAkdisp}
\end{equation}
where $\rho_k^{\rm (pt)}=\text{Im}[a^k(-\sigma-i\epsilon)]$,
and $a$ is given, e.g., by Eq.~(\ref{apt}).
Other choices would be, e.g. $a^k \mapsto
\A_1^k, \A_1^{k-2} \A_2$, etc.
With construction (\ref{MAAkdisp}), it was shown \cite{Magradze}
that the RGE's governing the evolution of $\A_k$'s are
identical to those governing the evolution of
$a^k$'s in pQCD when the replacements
$a^j \mapsto \A_j^{\rm (MA)}$ are made [cf.~Eq.~(\ref{pRGE})]
\ba
\frac{\partial \A^{\rm (MA)}_k
(\mu^2)}{\partial \ln \mu^2} & = &
- k \sum_{j=2}^{j_{\rm max}} 
\beta_{j-2} \: \A^{\rm (MA)}_{j+k-1} (\mu^2) =
- k \beta_0 \A_{k+1}^{\rm (MA)}(\mu^2) - \cdots
\ ,
\nonumber\\
\frac{\partial^2 \A^{\rm (MA)}_k (\mu^2)}
{\partial (\ln \mu^2)^2} & = &
k \sum_{j,\ell=2}^{j_{\rm max}} \beta_{j-2} \beta_{\ell -2}
( \ell + k - 1) \A^{\rm (MA)}_{j+\ell+k-2}(\mu^2) =
k (k+1) \beta_0^2 \A^{\rm (MA)}_{k+2}(\mu^2) + \cdots \ ,
\quad {\rm etc.}
\label{AkRGE}
\ea
The reason for this lies in the fact that $a^k$, and
consequently $\rho^{\rm (pt)}_k(\sigma)$, fulfill analogous RGE's.
Further, the renormalization scheme (RSch) dependence in pQCD,
i.e., dependence of $a^k$ of $\beta_j$ ($j \geq 2$),
is known \cite{PMS} (cf.~also \cite{Cvetic:2000mz}),
the same dependence holds for the discontinuity
functions $\rho_k^{\rm (pt)}(\sigma,\beta_2,\ldots)$
and thus for the MA couplings (\ref{MAAkdisp}) the
analogous dependence via $a^j \leftrightarrow \A_j^{\rm (MA)}$
is obtained ($k=1,2,\ldots$):
\ba
\frac{\partial \A_k^{\rm (MA)}(\mu^2)}{\partial \beta_2} & = &
\frac{k}{\beta_0} \A_{k+2}^{\rm (MA)}(\mu^2) + 
\frac{k \beta_2}{3 {\beta_0}^2} \A_{k+4}^{\rm (MA)}(\mu^2) 
+ {\cal {O}}(\A_{k+5}^{\rm (MA)}) \ ,
\label{dAkdb2}
\\
\frac{\partial \A_k^{\rm (MA)}(\mu^2)}{\partial \beta_3} & = &
\frac{k}{2 \beta_0} \A_{k+3}^{\rm (MA)}(\mu^2) - 
\frac{k \beta_1}{6 \beta_0^2} \A_{k+4}^{\rm (MA)}(\mu^2) + 
{\cal {O}}(\A_{k+5}^{\rm (MA)}) \ ,
\label{dAkdb3}
\\
\frac{\partial \A_k^{\rm (MA)}(\mu^2)}{\partial \beta_4} & = &
\frac{k}{3 \beta_0} \A_{k+4}^{\rm (MA)}(\mu^2) 
+ {\cal {O}}(\A_{k+5}^{\rm (MA)}) \ .
\label{dAkdb4} 
\ea
The RGE-type relations (\ref{AkRGE})-(\ref{dAkdb4}),
valid in the MA QCD, imply the following important
property: If the evaluation of a space-like QCD observable
quantity ${\cal D}(Q^2)$ is based on the analytization of 
STPS of that quantity according to the rule 
$a^k(\mu^2) \mapsto \A_k^{\rm (MA)}(\mu^2)$ ($k \geq 1$),
then the evaluated value of ${\cal D}(Q^2)$
has a dependence on RScl $\mu$ and on RSch ($\beta_j$, $j \geq 2$)
which is suppressed systematically. The suppression gets
stronger as the number of terms increases, just as in pQCD.
The precision ${\cal O}(\A_n^{\rm (MA)})$ corresponds in pQCD to
the precision ${\cal O}(a^n)$.

Having the STPS with terms up to $\sim a^{n_{\rm max}}$
($n_{\rm max} \equiv n_{\rm m}$), as well as its analytized analog
\ba
{\cal D}_{\rm{STPS}}^{(n_{\rm m})}(Q^2) &=& 
a(\mu^2; \beta_2, \dots) +
\sum_{n=2}^{n_{\rm m}} 
d_{n-1} a^n(\mu^2; \beta_2, \dots) \ ,
\label{STPS}
\\
{\cal D}_{\rm{an.}}^{(n_{\rm m})}(Q^2) &=& 
\A_1(\mu^2; \beta_2, \dots) +
\sum_{n=2}^{n_{\rm m}} 
d_{n-1} \A_n (\mu^2; \beta_2, \dots) \ ,
\label{TAS}
\ea
it is then enough to include in the evolution rules
(\ref{AkRGE}) and 
(\ref{dAkdb2})-(\ref{dAkdb4}) (for $k=1$ only)
terms of up to $\A_{n_{\rm m}}$
on the RHS.
Then the analytized evaluated
values ${\cal D}_{\rm an.}(Q^2)$ will have the
RScl- and RSch-independence precision
$\partial {\cal D}_{\rm an.}^{(n_{\rm m})}(Q^2)/\partial X
\sim A_{n_{\rm m}+1}$ ($X = \ln \mu^2, \beta_j$)
which has its perturbative analog 
$\partial {\cal D}_{\rm STPS}^{(n_{\rm m})}(Q^2)/{\partial X}
\sim a^{n_{\rm m}+1}$.

In view of these considerations, we propose to maintain
evolution relations (\ref{AkRGE}) (for $k=1$) for
any version of anQCD, including models M1 and M2 of the
previous Section, truncating them as just mentioned:
\ba
\frac{\partial \A_1(\mu^2; \beta_2, \ldots)}{\partial \ln \mu^2} & = &
- \beta_0 \A_2 - \cdots - \beta_{n_{\rm m}-2} \A_{n_{\rm m}} \ ,
\nonumber\\
\frac{\partial^2 \A_1(\mu^2; \beta_2, \ldots)}{\partial (\ln \mu^2)^2 } 
& = &
 2 \beta_0^2 \A_3 + 5 \beta_0 \beta_1 \A_4 + \cdots 
+ \kappa^{(2)}_{n_{\rm m}} \A_{n_{\rm m}} \ ,
\quad {\rm etc.} \ ,
\label{AkRGEtr}
\ea
where we have altogether $n_{\rm m}-1$ equations,
and $\kappa^{(\ell)}_n$ are the corresponding coefficients
of the pQCD evolution equations. Eqs.~(\ref{AkRGEtr})
represent {\it definitions\/} of $\A_k$'s ($2 \leq k \leq n_{\rm m}$)
via combinations of derivatives $\partial^n \A_1/\partial (\ln \mu^2)^n$.

On the other hand, evolution equations (\ref{dAkdb2})-(\ref{dAkdb4})
(for $k=1$) for the change of RSch remain of the same form, 
but with aforementioned truncation
\ba
\frac{\partial \A_1(\mu^2; \beta_2, \ldots)}{\partial \beta_2} & \approx &
\frac{1}{\beta_0} \A_3 + \frac{\beta_2}{3 \beta_0^2} \A_5 + \cdots
+ k^{(2)}_{n_{\rm m}} \A_{n_{\rm m}} \ ,
\nonumber\\
\frac{\partial \A_1(\mu^2; \beta_2, \ldots)}{\partial \beta_3} & \approx &
\frac{1}{2 \beta_0} \A_4 - \frac{\beta_1}{6 \beta_0^2} \A_5 + \cdots
+ k^{(3)}_{n_{\rm m}} \A_{n_{\rm m}} \ ,
\quad {\rm etc.} 
\label{dA1dbjtr}
\ea
where we have altogether $n_{\rm m}-2$ equations,
and $k^{(\ell)}_n$ are the corresponding coefficients
of the pQCD evolution equations. Eqs.~(\ref{dA1dbjtr}) are,
in contrast to Eqs.~(\ref{AkRGEtr}), not definitions,
but in general approximations for the evolution under
RSch-changes. The RSch-dependence of $\A_1(\mu^2)$ is treated
in more detail later in this work.

On the basis of Eqs.~(\ref{AkRGEtr})-(\ref{dA1dbjtr}),
expressions for the (truncated) derivatives 
$\partial \A_k/\partial X$, for $k \geq 2$ 
($X= \ln \mu^2, \beta_j$), can be obtained.

In our approach, the basic space-like quantities are
$\A_1(\mu^2)$ of a given anQCD model (e.g., MA, M1, M2)
and its logarithmic derivatives
\be
\tA_{n}(\mu^2) \equiv \frac{ (-1)^{n-1}}{\beta_0^{n-1} (n-1)!} 
\frac{ \partial^{n-1} {\A_1}(\mu^2)}{\partial (\ln \mu^2)^{n-1}} \ ,
\qquad (n=1,2,3,\ldots) \ ,
\label{tAn}
\ee
whose pQCD analogs are
\be
{\widetilde a}_n(\mu^2) \equiv
\frac{ (-1)^{n-1}}{\beta_0^{n-1} (n-1)!} 
\frac{ \partial^{n-1} a(\mu^2)}{\partial (\ln \mu^2)^{n-1}} \ ,
\qquad (n=1,2,3, \ldots) \ .
\label{tan}
\ee
The quantities ($\A_1(\mu^2)$, $\tA_2(\mu^2)$,
$\tA_3(\mu^2)$, ...), all derived from 
$\A_1(\mu^2) \equiv \tA_1(\mu^2)$, 
are known functions of the
space-like momenta $\mu$ in any chosen anQCD version
in a given chosen RSch ($\beta_2, \beta_3, \ldots$).
On the basis of these quantities and the (truncated)
evolution equations (\ref{AkRGEtr}),
any higher order quantity $\A_k(\mu^2)$ ($k \geq 2$)
can be constructed, in the given RSch.
Further, (truncated) equations 
(\ref{AkRGEtr})-(\ref{dA1dbjtr})
then give us the values of $\tA_k(\mu^2)$
and of $\A_k(\mu^2)$ ($k \geq 1$) 
in any other chosen RSch
($\beta_2^{'}, \beta_3^{'}, \ldots$).
We emphasize that in this approach, the higher order
quantities $\A_k(\mu^2)$ ($k \geq 2$) are not as basic,
they are defined via Eqs.~(\ref{AkRGEtr})
for convenience of having
closer notational analogy with pQCD formulas
(and $a^k \leftrightarrow \A_k$). In these definitions
(\ref{AkRGEtr}), as well as in $\beta_j$-running
Eqs.~(\ref{dA1dbjtr}), we could have kept one more
term ($\sim \A_{n_{\rm m}+1}$), in order to
come closer to the exact analogy $\A_k = a^k + {\rm NP}$
for $k \geq 2$, where
NP stands for nonperturbative contributions 
(nonanalytic functions of $a$ at $a=0$).\footnote{
$\A_k = a^k + {\rm NP}$ holds exactly for the construction
Eq.~(\ref{MAAkdisp}), i.e., the construction by Milton {\it et al.\/}
\cite{Milton:1997mi,Sh,Milton:2000fi} in MA.}
However, this is not necessary, as argued below.   

The basic analytization rule we adopt will thus be
\be
{\widetilde a}_n \mapsto \tA_n 
\qquad (n=1,2,\ldots) \ ,
\label{basican}
\ee
where $\tA_n$ and ${\widetilde a}_n$ are defined in Eqs.~(\ref{tAn})
and (\ref{tan}), respectively.

At loop level $n_{\rm max} \equiv n_{\rm m}$, 
and in a chosen 'starting' RSch ($\beta_2, \beta_3, \ldots$),
the truncation ('tr') of the RGE-running of the pQCD coupling
$a(\mu^2)$ is in principle via Eq.~(\ref{pRGE}) with 
$j_{\rm max} = n_{\rm max}+1$  ($a = a_{\rm tr}$,
${\widetilde a}_n = {\widetilde a}_{n, {\rm tr}}$).
The corresponding truncated $\tA_n = \tA_{n, {\rm tr}}$
are then
\ba
\tA_{n}(\mu^2) &=& {\widetilde a}_{n} + {\rm NP}
= {\widetilde a}_n(\mu^2)_{(\infty)} + {\rm NP} + 
{\cal O} \left( \beta_0^{n_{\rm m}-1} a^{n_{\rm m}+n} \right) \ ,
\qquad (n=1,2,\ldots) \ ,
\label{tAntr}
\ea
and we assumed that we are in the
class of the RSch's where $\beta_j \sim \beta_0^j$ in the
large-$\beta_0$ limit. We recall that $\tA_1 \equiv \A_1$
and ${\widetilde a}_1 \equiv a$.
The subscript $(\infty)$ in Eq.~(\ref{tAntr}) means
that this is the quantity obtained by not truncating
RGE beta-function (\ref{pRGE}), i.e., for $j_{\rm max} = \infty$
and keeping the same value of $\Lambda$ in expansion
(\ref{apt}) as in the case of the truncated beta-function
(i.e., $j_{\rm max} =n_{\rm max}+1$). The second
identity in Eq.~(\ref{tAntr}) thus shows, as an
additional reference, the
magnitude of error committed due to the truncation
of the beta-function.
Definitions (\ref{AkRGEtr}) of $\A_n$'s then imply
\ba
\A_n(\mu^2) & = & a^n(\mu^2) + {\rm NP} +
{\cal O} \left( \beta_0^{n_{\rm m}-n} a^{n_{\rm m}+1} \right) 
\qquad (n=2,\ldots,n_{\rm m}) \ .
\label{Antr}
\ea 
Since the RGE-running (\ref{pRGE}) of $a$ is truncated, we have
$a^n = a^n_{(\infty)} + 
{\cal O}(\beta_0^{n_{\rm m}-1} a^{n_{\rm m} + n})$,
and relations (\ref{Antr}) remain unchanged when
$a^n(\mu^2)$ there is replaced by $a^n_{(\infty)}(\mu^2)$.

The $\beta_j$-running Eqs.~(\ref{dA1dbjtr}) are also
truncated, i.e., the RHS's there have errors 
$\sim \A_{n_{\rm m}+1}$, so that the changes of RSch
entail additional errors. It can be verified that
this effect, when going from a chosen
'starting' RSch ($\beta_2, \beta_3, \ldots$) to another
RSch ($\beta_2^{'}, \beta_3^{'}, \ldots$),
 modifies relations (\ref{tAntr}) to
\ba
\tA_1 \left( \equiv A_1(\mu^2) \right) & = & a(\mu^2) + 
{\cal O} \left( \beta_0^{n_{\rm m}-2} a^{n_{\rm m}+1} \right)
+ {\rm NP} \ ,
\nonumber\\
\tA_{n}(\mu^2) &=& {\widetilde a}_{n} + 
{\cal O} \left( \beta_0^{n_{\rm m}-2} a^{n_{\rm m}+n} \right) 
+ {\rm NP} 
\qquad (n=2,\ldots,n_{\rm m}) \ ,
\label{tAntrmod}
\ea 
while relations (\ref{Antr}) do not get modified.
We should keep in mind that there is
yet another truncation involved, namely in the
solution (\ref{apt}) of RGE (\ref{pRGE})
the sum over index $k$ has in the calculational practice finite
number of terms.
In our calculations, we will take there $k_{\rm max}=
n_{\rm max}+2$ ($= j_{\rm max}+1$), which is so high
that it does not affect ``precision estimate''
relations (\ref{tAntrmod}) and (\ref{Antr}).

For example, at loop level three ($n_{\rm max}=3$),
where we include in RGE (\ref{pRGE}) term
with $j_{\rm max}=4$ (thus $\beta_2$), 
relations (\ref{AkRGEtr}) are 
\begin{equation}
{\tA_2}(\mu^2) 
= {\A_2}(\mu^2) +
\frac{\beta_1}{\beta_0} {A_3}(\mu^2) \ , 
\qquad
{\tA_3}(\mu^2) 
= {\A_3}(\mu^2) \ ,
\label{tA2tA3}
\end{equation}
implying
\begin{equation}
{\A_2}(\mu^2) 
= {\tA_2}(\mu^2) -
\frac{\beta_1}{\beta_0} {\tA_3}(\mu^2) \ , 
\qquad 
{\A_3}(\mu^2) 
= {\tA_3}(\mu^2) \ .
\label{A2A3}
\end{equation}
The RSch ($\beta_2$) dependence is obtained from
the truncated Eqs.~(\ref{dA1dbjtr}) and (\ref{AkRGEtr})
\be
\frac{\partial \tA_j(\mu^2;\beta_2)}{\partial \beta_2} = 
\frac{1}{2 \beta_0^3} \frac{\partial^2 \tA_j(\mu^2; \beta_2)}
{\partial (\ln \mu^2)^2} 
\left( \equiv \frac{1}{\beta_0} \tA_3(\mu^2; \beta_2) \right)
\qquad (j=1,2,\ldots) \ ,
\label{dA1dbll3}
\ee
where $\tA_1 \equiv \A_1$.
These are second order approximate partial differential equations
for $\A_1(\mu^2;\beta_2)$, $\tA_2(\mu^2;\beta_2)$, $\tA_3(\mu^2;\beta_2)$.
Higher order terms ($\sim \tA_4$) are neglected on the right-hand side of the
RSch-evolution equation (\ref{dA1dbll3}).

At loop level four ($n_{\rm max}=4$), 
where we include in RGE (\ref{pRGE}) term
with $j_{\rm max}=5$ (thus $\beta_3$), relations analogous to
(\ref{A2A3}) are
\ba
{\A_2}(\mu^2) &=& {\tA_2}(\mu^2) -
\frac{\beta_1}{\beta_0} {\tA_3}(\mu^2) +
\left( \frac{5}{2} \frac{\beta_1^2}{\beta_0^2} - \frac{\beta_2}{\beta_0}
\right) {\tA_4}(\mu^2) 
\ , 
\nonumber\\
{\A_3}(\mu^2) & = & {\tA_3}(\mu^2) - 
\frac{5}{2} \frac{\beta_1}{\beta_0}{\tA_4}(\mu^2) \ ,
\qquad
{\A_4}(\mu^2) = {\tA_4}(\mu^2) \ ,
\label{A2A3A4}
\ea
while the changes of the RSch are governed by (approximate) relations
\ba
\frac{\partial \tA_j(\mu^2)}{\partial \beta_2} & = &
\left[ \frac{1}{2! \beta_0^3} \frac{\partial^2}{\partial (\ln \mu^2)^2}
+ \frac{5}{3!\: 2} \frac{\beta_1}{\beta_0^5}
\frac{\partial^3}{\partial (\ln \mu^2)^3} \right] \tA_j(\mu^2) \ ,
\nonumber\\
\frac{\partial \tA_j(\mu^2)}{\partial \beta_3} & = &
- \frac{1}{3! \: 2 \beta_0^4} 
\frac{\partial^3 \tA_j(\mu^2)}{\partial (\ln \mu^2)^3} 
\qquad (j=1,2,\ldots) \ . 
\label{dA1dbll4}
\ea
Our approach is in a sense maximally truncating. 
Namely, the evolution under the 
changes of the RSch is truncated in such a way
that 
$\partial {\cal D}_{\rm an.}^{(n_{\rm m})}(Q^2)/\partial \beta_j
\sim \A_{n_{\rm m}+1}$. Further, our definition of
$\A_k$'s ($k \geq 2$) via Eqs.~(\ref{AkRGEtr})
[cf.~Eqs.~(\ref{A2A3}) and (\ref{A2A3A4})] 
involve short (``truncated'') series which, however, still
ensure the correct RScl-dependence
$\partial {\cal D}_{\rm an.}^{(n_{\rm m})}(Q^2)/\partial \mu^2
\sim \A_{n_{\rm m}+1}$.
Furthermore, it may seem that, for loop level three ($n_{\rm max}=3$),
the RHS of the first of Eqs.~(\ref{tA2tA3}) 
represents only two perturbative terms $[a^2 +
(\beta_1/\beta_0) a^3]$ plus nonperturbative terms (NP).
However, since taking $j_{\rm max} = n_{\rm max} +1 = 4$
in RGE (\ref{pRGE}) as the basis for
calculation of $\A_1(\mu^2)$, it\footnote{
When the anQCD is not MA, but rather M1 or M2, RGE (\ref{pRGE})
and the (truncated) expansion
(\ref{apt}) still remain the basis for calculation
of the MA-part of $\A_1(\mu^2)$, the difference between
$\A_1(\mu^2)$ and $\A_1^{\rm (MA)}(\mu^2)$ being purely
nonperturbative, cf.~Eqs.~(\ref{M1a}), (\ref{M1b}), (\ref{M2}).}
is straightforward to show that the following holds:
\be
\left( \tA_2(\mu^2) = \right) 
{\A_2}(\mu^2) + \frac{\beta_1}{\beta_0} \A_3(\mu^2) =
a^2(\mu^2)  + \frac{\beta_1}{\beta_0} a^3(\mu^2)
+ \frac{\beta_2}{\beta_0} a^4(\mu^2) +
{\cal O}(\beta_0^2 a^5) + {\rm NP} \ .
\label{RHSpexp}
\ee
Completely analogous result holds at loop level 4
($n_{\rm max}=4$ and $j_{\rm max}=5$).

In the MA QCD, in the approach
of \cite{Milton:1997mi}, here Eq.~(\ref{MAAkdisp}) for $\A_k$, 
a truncation is performed only in expansion (\ref{apt}) for
$a$ ($\rightarrow \rho^{\rm (pt)}_1(\sigma)$, 
apparently with $k_{\rm max} = n_{\rm max}$),
and then powers of this truncated $a$
are used to define $\rho_k^{\rm (pt)}$ and thus
$\A_k$ ($k \geq 2$). This implies that in the MA QCD
our $\A_k$'s ($k=2,\ldots$), on the one hand, 
and those of the approach of 
Milton, Solovtsov, Solovtsova, and Shirkov
(MSSSh) \cite{Milton:1997mi,Sh,Milton:2000fi},
on the other hand,
are not the same, although they must
gradually merge when the loop level is increased.
\begin{figure}[htb]
\begin{minipage}[b]{.49\linewidth}
 \centering\epsfig{file=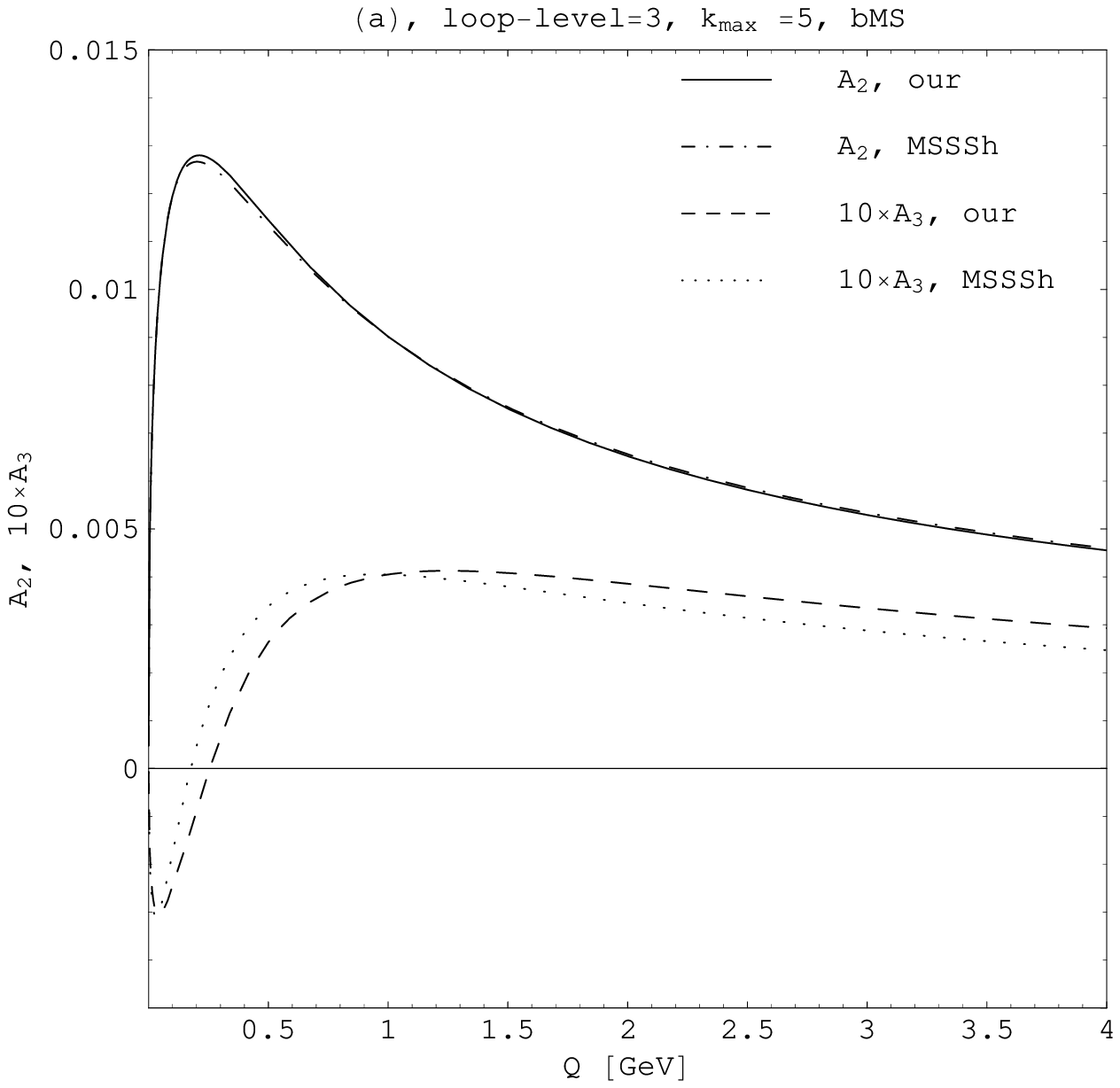,width=\linewidth,height=8.cm}
\end{minipage}
\begin{minipage}[b]{.49\linewidth}
 \centering\epsfig{file=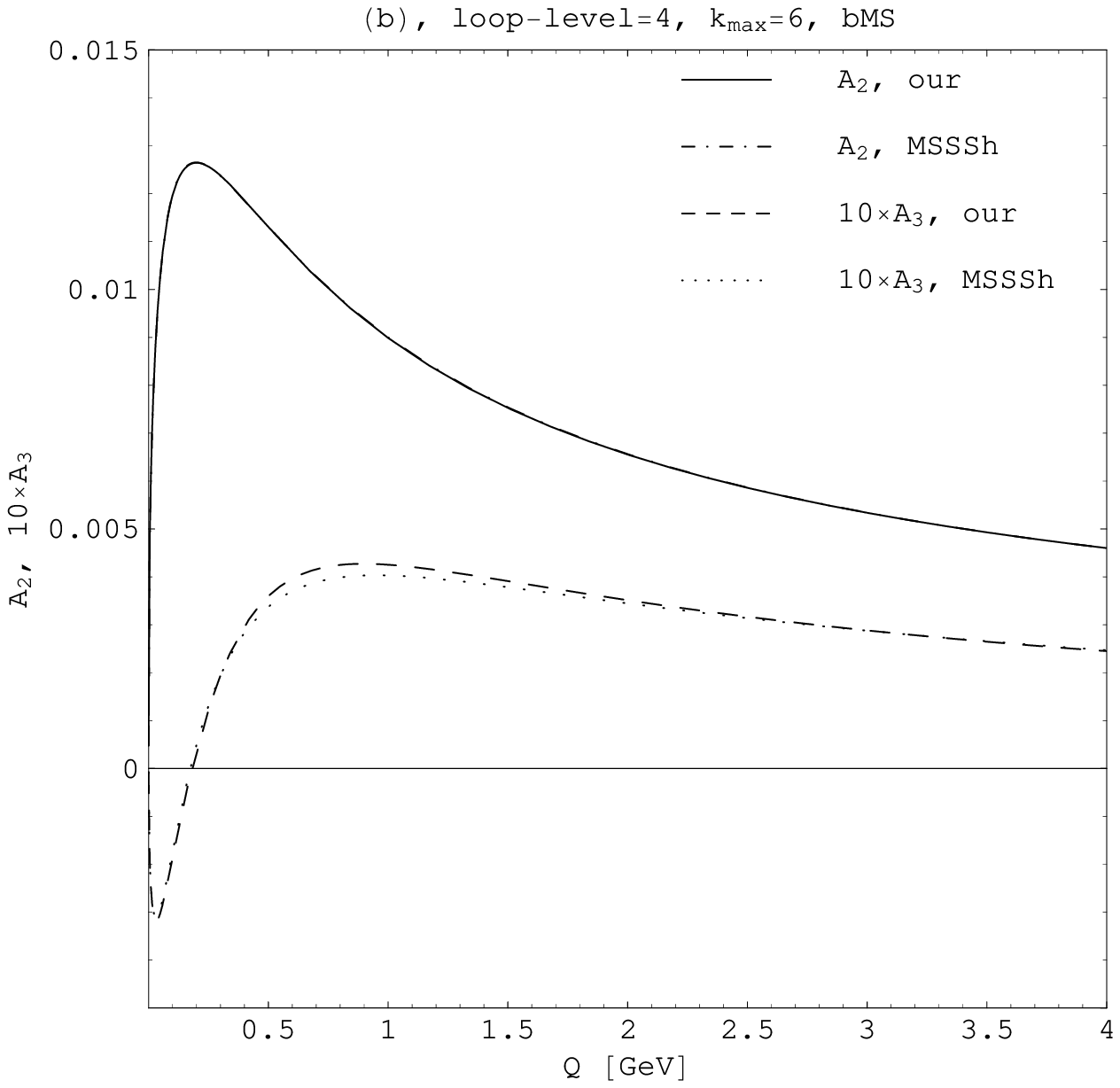,width=\linewidth,height=8.cm}
\end{minipage}
\vspace{-0.6cm}
\caption{\footnotesize The coupling parameters $\A_2(Q^2)$ and
$\A_3(Q^2)$ in MA in $\bMS$ RSch, with $n_f=3$ and
$\bL_{(n_f=3)}=0.4$ GeV, calculated at (a) loop-level=3
(and $k_{\rm max}=5$), and (b) loop-level=4 (and $k_{\rm max}=6$).
Presented are results of construction of Milton {\it et al.\/}
(MSSSh) \cite{Milton:1997mi,Sh,Milton:2000fi}, 
and of our construction.}
\label{FigMSSShbMS}
\end{figure}
\begin{figure}[htb]
\begin{minipage}[b]{.49\linewidth}
 \centering\epsfig{file=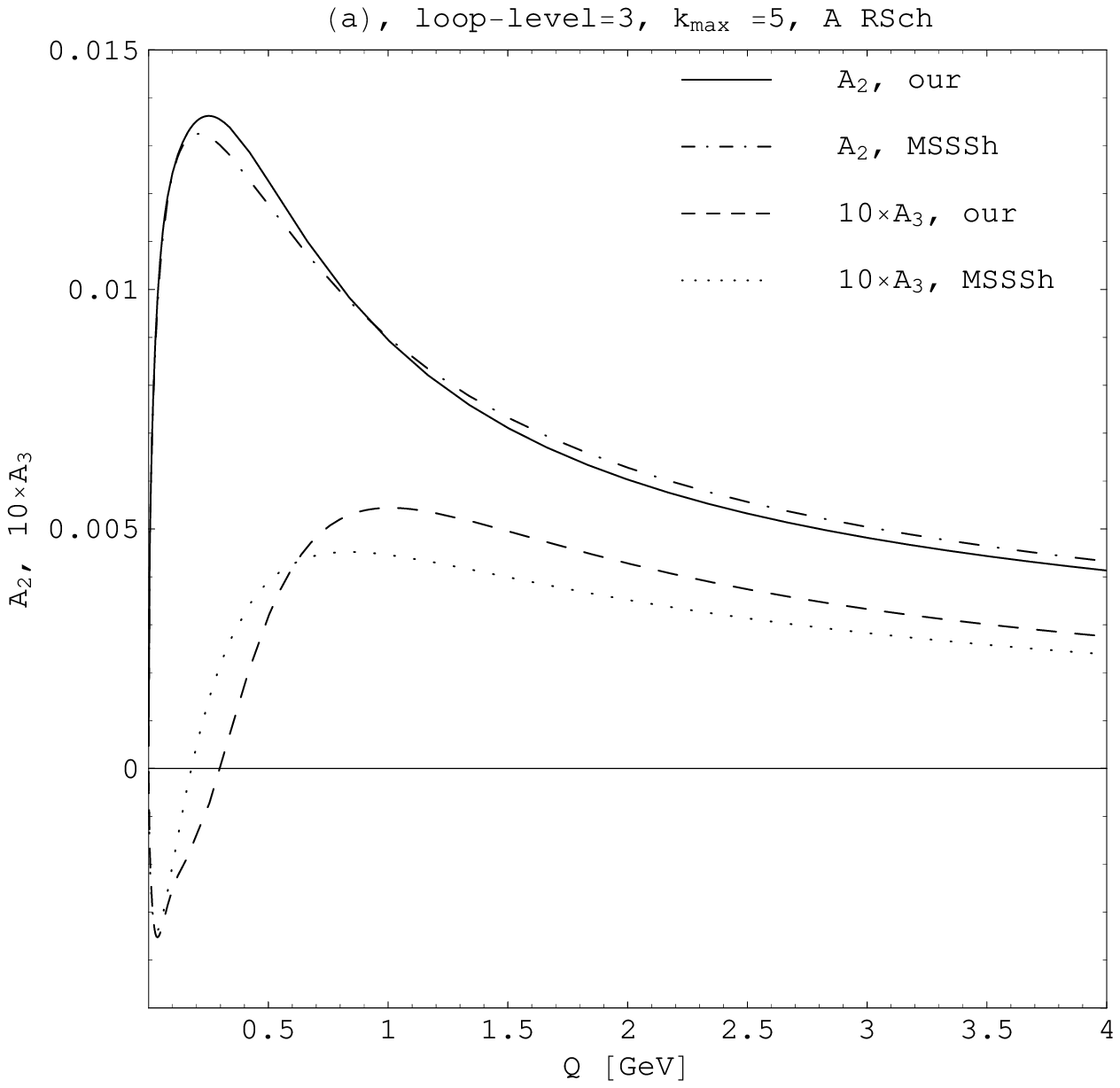,width=\linewidth,height=8.cm}
\end{minipage}
\begin{minipage}[b]{.49\linewidth}
 \centering\epsfig{file=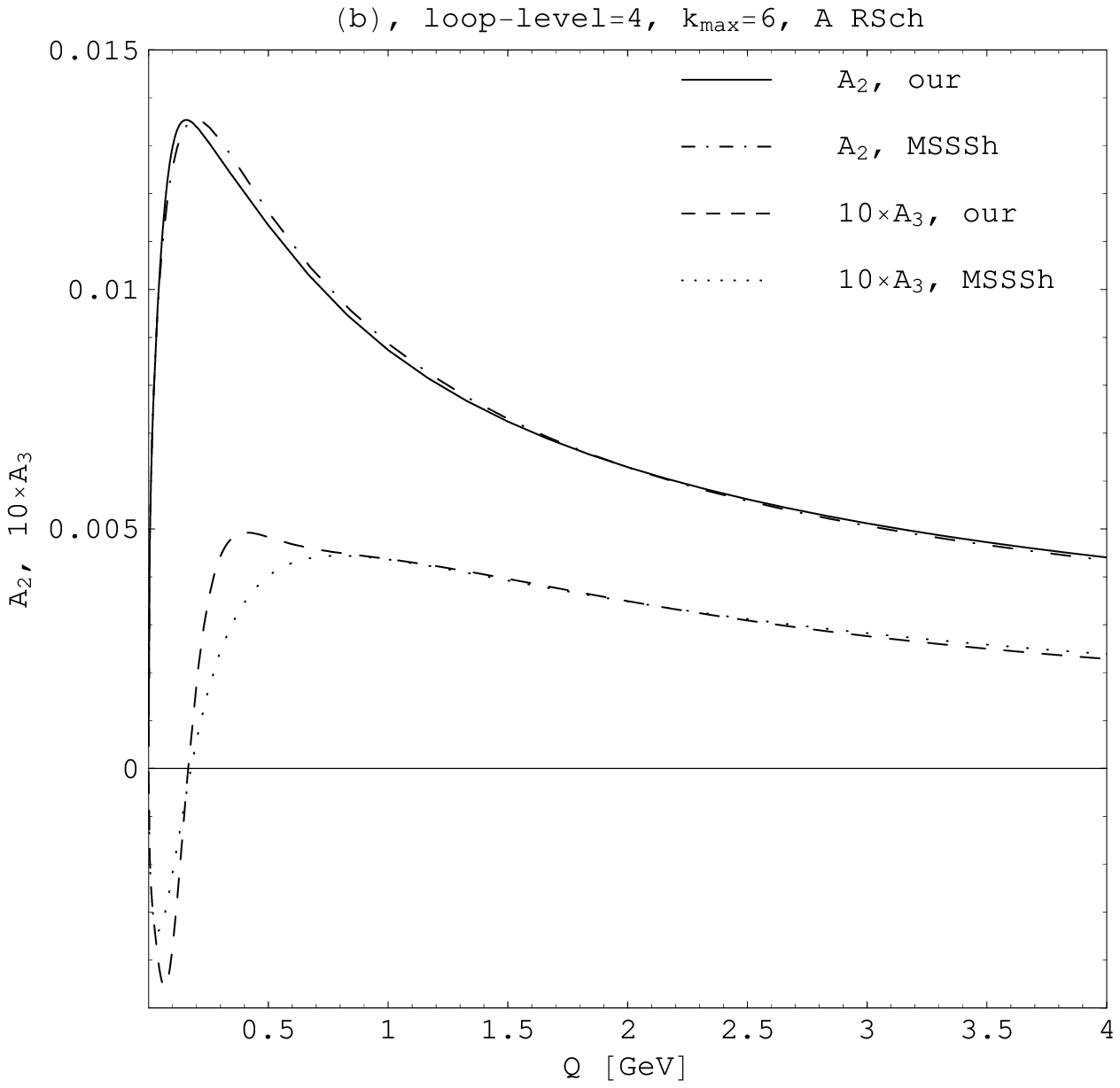,width=\linewidth,height=8.cm}
\end{minipage}
\vspace{-0.6cm}
\caption{\footnotesize Same as in Fig.~\ref{FigMSSShbMS},
but now in RSch A, Eq.~(\ref{ARSch}).}
\label{FigMSSShARSch}
\end{figure}
This is illustrated in Figs.~\ref{FigMSSShbMS} 
and \ref{FigMSSShARSch},
where the MA-coupling parameters $\A_2(Q^2)$ and $\A_3(Q^2)$
of both approaches are compared, for $n_f=3$, 
at loop-level ($=n_{\rm max}$)
three and four, in $\bMS$ and in RSch A, respectively.
The Adler (A) RSch  is defined later in 
Eqs.~(\ref{ARSch}) [cf.~Eq.~(\ref{dvARSch})].
For both $\A_2$ and $\A_3$ one can see a decrease in the 
absolute difference between our and MSSSh methods when 
going from loop-level=3 to 4, Fig.~\ref{FigMSSShbMS} in 
$\bMS$ RSch, and Fig.~\ref{FigMSSShARSch} in RSch A.
The decrease can be understood as coming largely from the
fact that the perturbative part of
this difference is ${\cal O}(a^4)$ when loop-level=3,
and ${\cal O}(a^5)$ when loop-level=4.
Further, inspection of Figs.~\ref{FigMSSShbMS}(a) 
and \ref{FigMSSShARSch}(a) reveals
that the $\A_2$-curves practically merge already at
loop-level=3 if RSch is $\bMS$, but less so if RSch is A.
An indication towards understanding this resides in the fact
that the coefficient at $a^4$ of the difference between the
two curves is proportional to $(2 \beta_0 \beta_2 - 5 \beta_1^2)$,
this being in $\bMS$ about one fifth of the 
corresponding value in RSch A (when $n_f=3$).
In Fig.~\ref{FigM1M2MAARSch}
the coupling parameters
$\A_1(Q^2)$ and $\A_2(Q^2)$
of anQCD models M1, M2 and MA are presented as
functions of the scale $Q$,
for specific chosen fixed parameters of the models M1 and M2
(see Sec.~\ref{numres})
and in the aforementioned specific RSch A. 
Note that we used $k_{\rm max} = n_{\rm max}+2$
in the calculation of $\rho_1^{\rm (pt)}$ via
Eq.~(\ref{apt}) in all cases, i.e., also in the MSSSh cases.
In Fig.~\ref{FigM1M2MAARSch}, loop-level=4 and $k_{\rm max}=6$
was taken (using our described approach). 
In Figs.~\ref{FigMSSShbMS}, \ref{FigMSSShARSch}, \ref{FigM1M2MAARSch},
the basis for calculation was the $k_{\rm max}$-truncated series
(\ref{apt}) in the corresponding RSch.  
\begin{figure}[htb] 
\centering
\epsfig{file=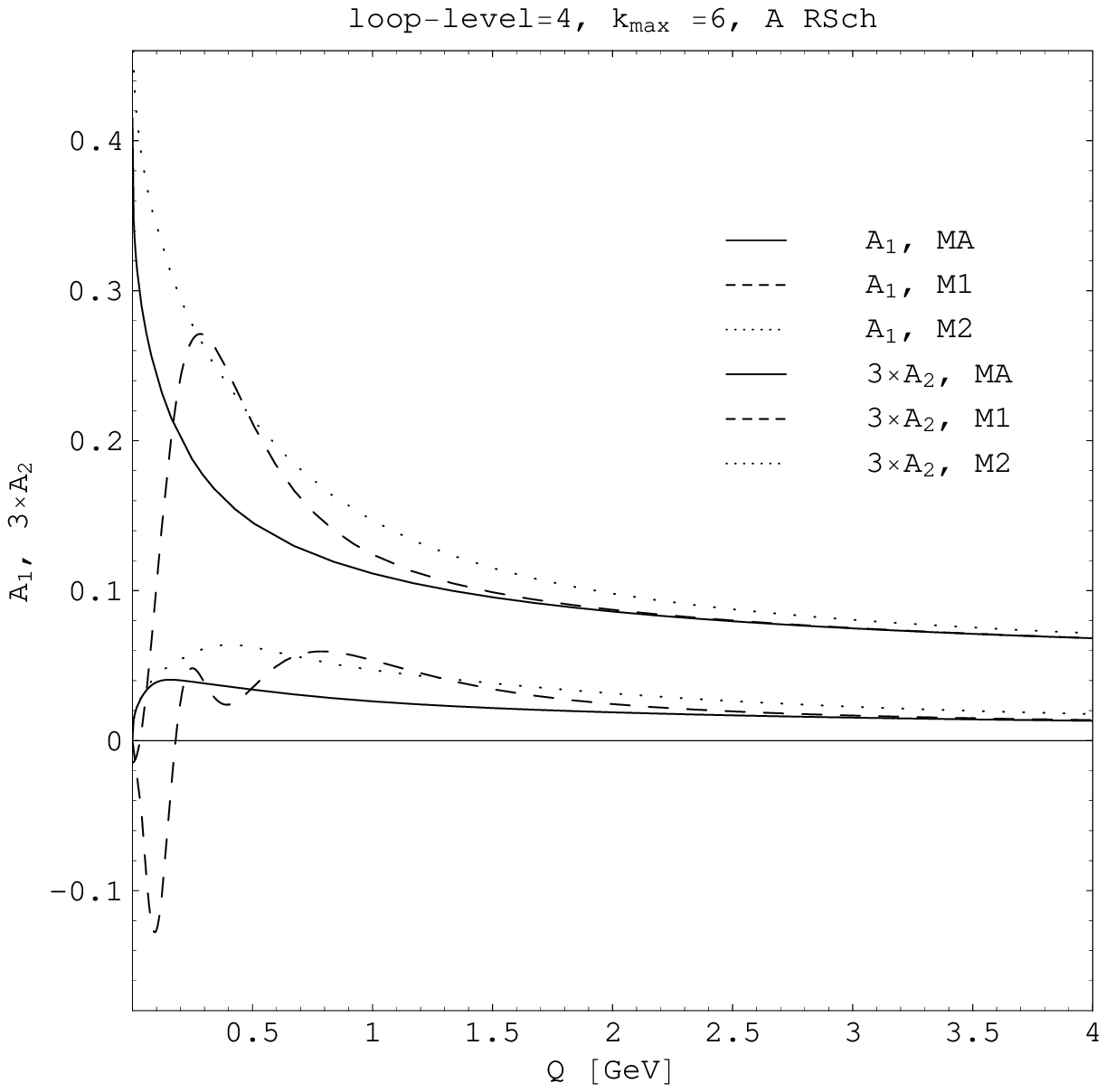,width=13.cm,height=8.cm}
\vspace{-0.8cm}
\caption{\footnotesize Same as in Fig.~\ref{FigMSSShARSch},
but now $\A_1$ and $\A_2$ for various models (M1,M2 and MA) 
with specific model parameters
(see Sec.~\ref{numres}): $c_0=2.94$, $c_r=0.45$, $c_f=1.08$ for M1;
$c_v=0.1$ and $c_p=3.4$ for M2; $n_f=3$ and $\bL_{(n_f=3)}=0.4$ GeV
in all three models. The upper three curves are for $\A_1$,
the lower three curves are for $3 \times \A_2$. All couplings  
are in RSch A, Eq.~(\ref{ARSch}). $\A_2$ is constructed with
our approach.}
\label{FigM1M2MAARSch}
 \end{figure}

Even when already having anQCD coupling $\A_1(Q^2)$,
there is no unique way to merge analyticity requirements with
the perturbative results at higher orders, i.e., 
Eq.~(\ref{Antr}) for $\A_k(Q^2)$ ($k \geq 2$).
The latter relations are ensured by our definitions of 
$\A_k(Q^2)$ for $k \geq 2$ via relations (\ref{AkRGEtr}),
but this is just one of the possibilities of addressing 
the problem. In MA the construction of $\A_1(Q^2)$
is very closely related to the perturbative solution
$a(Q^2)$ via dispersion relation (\ref{MAA1disp}).
Therefore, it is very natural to keep that close analogy at higher
orders, via dispersion relations (\ref{MAAkdisp}).
As a consequence, the RGE-type of relations 
(\ref{AkRGEtr}) are fulfilled in MA \cite{Magradze}.
For a general anQCD model, this approach does not apply.
Deviations of $\A_1(Q^2)$ and $\tlA_1(s)$ from their
MA values imply that the discontinuity function
$\rho_1(\sigma)$ deviates from its MA
analog $\rho_1^{\rm (pt)}(\sigma) = {\rm Im} a(-\sigma - i \epsilon)$
at low values of $\sigma$, 
cf.~Eqs.~(\ref{tlA1rho1}), (\ref{M1tl}) and (\ref{M2tl}).
Therefore, there is no direct natural way of
prescribing the low-$\sigma$ behavior of
the higher order discontinuity functions
$\rho_k(\sigma)$ appearing in the dispersion relations
of the type of Eq.~(\ref{MAAkdisp})
for $\A_k$, i.e., prescribing their deviations
from $\rho_k^{\rm (pt)}(\sigma) = {\rm Im} a^k(-\sigma - i \epsilon)$
for $k \geq 2$. We define $\A_k(Q^2)$ for $k \geq 2$ by
forcing them to obey the truncated RGE-type relations
(\ref{AkRGEtr}). We emphasize that these relations
define, in our approach, the couplings $\A_k(Q^2)$ for $k \geq 2$.
Thus, we indirectly define the
corresponding discontinuity functions $\rho_k$.  
This construction of $\A_k$'s is motivated also by the skeleton
approach as discussed in Ref.~\cite{Cvetic:2006mk}.
Furthermore, as we will see later, this construction of $\A_k$'s
allows us to suppress systematically the RScl- and
RSch-dependence in the evaluated observables with
the increasing order, because an RGE-type of analogy
with pQCD is being preserved.

\section{Skeleton-motivated expansion}
\label{skmotexp}

Consider an observable ${\cal D}(Q^2)$ depending on a single 
space-like physical scale $Q^2 (\equiv - q^2) >0$.
Its perturbation expansion has the form
\be
{\cal D}(Q^2)_{\rm pt} = a + d_1 a^2 + d_2 a^3 + \cdots \ ,
\label{Dpt}
\ee
where $a = a(\mu^2; \beta_2, \ldots)$ is taken at a given
RScl ($\mu$) and RSch ($\beta_2, \ldots$). As mentioned before, we will
take the convention $\Lambda = \bL$, i.e., the $\bMS$ QCD scale as the
reference scale for $\mu$ [cf.~Eq.~(\ref{apt})-(\ref{kijbL})]. 
Further, we will
work in the following classes of RSch: each $\beta_k$ ($k \geq 2$)
is a polynomial in $n_f$ of order $k$; equivalently, it is
a polynomial in $\beta_0$:
\be
\beta_k = \sum_{j=0}^{k} b_{kj} \beta_0^j \ , \qquad k=2,3,\ldots
\label{betak}
\ee
The $\bMS$ clearly belongs to this class of schemes. 
In such schemes, the coefficients $d_n$ of expansion (\ref{Dpt}) 
have the following specific form in terms of $\beta_0$, as
can be deduced from the scheme independence of
observable ${\cal D}(Q^2)$, e.g. by using relations
of Ref.~\cite{PMS}:
\be
d_1 = c^{(1)}_{11} \beta_0 + c^{(1)}_{10},
\qquad d_n = \sum_{k=-1}^{n} c^{(1)}_{nk} \beta_0^k \ ,
\label{dns}
\ee
i.e., each $d_n$ is a polynomial of order $n$ in $\beta_0$
and includes in general, in addition, a term with the
negative power $1/\beta_0$ ($d_1$ does not have it). In
the $\bMS$ scheme, the negative powers do not occur.

We will now construct a separation of the series (\ref{Dpt}) into
a sum of RScl-independent subseries
\be
{\cal D}(Q^2)_{\rm pt} = {\cal D}^{(1)}(Q^2)_{\rm pt} + 
\sum_{n=2}^{\infty} k_n {\cal D}^{(n)}(Q^2)_{\rm pt} \ ,
\label{Dptsubser}
\ee
with the following properties: (a) each dimensionless constant $k_n$ is
RScl-independent; (b) each subseries ${\cal D}^{(n)}_{\rm pt}$
($n \geq 1$) is RScl-independent, and it is normalized
so that ${\cal D}^{(n)}_{\rm pt} = a^n + {\cal O}(a^{n+1})$;
(c) the subseries ${\cal D}^{(n)}(Q^2)_{\rm pt}$ contains
all the leading-$\beta_0$ coefficients of the 
following ``rest'':
\be
\frac{1}{k_n} \left[
{\cal D}(Q^2)_{\rm pt} - {\cal D}^{(1)}(Q^2)_{\rm pt} - \cdots
- k_{n-1} {\cal D}^{(n-1)}(Q^2)_{\rm pt} \right] \ .
\label{rest}
\ee
We will show that these conditions uniquely determine
factors $k_n$ and perturbation expansions of all ${\cal D}^{(n)}(Q^2)$.
Further, we show in Appendix \ref{app2} that the above subseries,
which always exist, would coincide with the expansions of the
corresponding skeleton terms in the skeleton expansion
of the observable if such an expansion existed in the considered RSch.

We consider first the leading-$\beta_0$ part of expansion (\ref{Dpt})
\ba
{\cal D}^{(1)}_0(Q^2)_{\rm pt} &=&
a + \sum_{j=2}^{\infty} a^j \left[ c^{(1)}_{jj} \beta_0^j \right] \ .
\label{D10}
\ea
Under the change of RScl from $\mu^2$ to $\mu_*^2$, using the
notation $L_* \equiv \ln(\mu_*^2/\mu^2)$, we have by RGE (\ref{pRGE})
\ba 
a &=& a_* +  \sum_{n=1}^{\infty} {\widetilde a}_{*n+1} \beta_0^n L_*^n
\nonumber\\
& = &
a_* + a_*^2 \beta_0 L_* + a_*^3 ( \beta_0^2 L_*^2 + \beta_1 L_* ) +
a_*^4 \left( \beta_0^3 L_*^3 + \frac{5}{2} \beta_0 \beta_1 L_*^2 + 
\beta_2 L_* \right)
+ \cdots \ ,
\label{avsast}
\ea
where $a \equiv a(\mu^2)$ and $a_* \equiv a(\mu_*^2)$.
Inserting this into expansion (\ref{Dpt}) we obtain the transformation
rules for the coefficients $c^{(1)}_{ij}$ (\ref{dns}) under the change 
of RScl. Specifically, for the diagonal coefficients
the transformations are
\ba
c^{(1)}_{*kk}& = & \sum_{s=0}^k 
\left(
\begin{array}{c}
k \\
s
\end{array}
\right) 
L_*^s c^{(1)}_{k-s,k-s} \ ,
\label{c1kk}
\ea
where we use the notations $c^{(1)}_{ij} \equiv c^{(1)}_{ij}(\mu^2)$
and $c^{(1)}_{*ij} \equiv c^{(1)}_{ij}(\mu_*^2)$ 
(and $c^{(1)}_{00}=1$ by definition).
Inserting expansion (\ref{avsast}) into expansion (\ref{D10})
we obtain
\ba
{\cal D}^{(1)}_0(Q^2)_{\rm pt} &=&
a_* + a_*^2   \left[ \beta_0 c^{(1)}_{*11} \right] +
a_*^3 \left[ \beta_0^2 c^{(1)}_{*22} + 
\beta_1 ( c^{(1)}_{*11} - c^{(1)}_{11} ) \right] + 
{\cal {O}}(\beta_0^3 a^4) \ .
\label{D10p}
\ea
This implies that the leading-$\beta_0$ series (\ref{D10}) does not maintain
its form under the change of RScl, since a new term
$a_*^3 \beta_1 ( c^{(1)}_{*11} - c^{(1)}_{11} )$ appears at
$\sim a^3$. The RScl-''covariant'' form, up to $\sim a^3$, is then
\ba
{\cal D}^{(1)}_1(Q^2)_{\rm pt} &=&
a + a^2 \left[ \beta_0 c^{(1)}_{11} \right] + 
a^3 \left[ \beta_0^2 c^{(1)}_{22} + \beta_1 c^{(1)}_{11} \right] + 
a^4 \left[ \beta_0^3 c^{(1)}_{33} \right] + {\cal {O}}(\beta_0^4 a^5) \ .
\label{D11}
\ea
We now iteratively repeat the procedure: we insert expansion (\ref{avsast})
into expansion (\ref{D11}) and, after some algebra and using relations
(\ref{c1kk}), obtain
\ba
{\cal D}^{(1)}_1(Q^2)_{\rm pt} &=&
a_* + a_*^2   \left[ \beta_0 c^{(1)}_{*11} \right] +
a_*^3 \left[ \beta_0^2 c^{(1)}_{*22} + 
\beta_1 c^{(1)}_{*11} \right] + 
\nonumber\\
&& a_*^4 \left[ \beta_0^3 c^{(1)}_{*33} +
\frac{5}{2} \beta_0 \beta_1 (c^{(1)}_{*22} - c^{(1)}_{22}) +
\beta_2 (c^{(1)}_{*11} - c^{(1)}_{11}) \right] + {\cal {O}}(\beta_0^4 a^5) \ . 
\label{D11p}
\ea 
The new terms appearing at $\sim a_*^5$ here require the following
restoration of the RScl-''covariance'' up to order $\sim a^5$:
\ba
{\cal D}^{(1)}(Q^2)_{\rm pt} &=&
a + a^2 \left[ \beta_0 c^{(1)}_{11} \right] + 
a^3 \left[ \beta_0^2 c^{(1)}_{22} + \beta_1 c^{(1)}_{11} \right] +
a^4 \left[ \beta_0^3 c^{(1)}_{33} + \frac{5}{2} \beta_0 \beta_1 c^{(1)}_{22} +
\beta_2 c^{(1)}_{11} \right] + {\cal {O}}(\beta_0^4 a^5) \ .
\label{D1}
\ea 
This procedure can be continued to any required order. Expression
(\ref{D1}) is now the {\it RScl-covariant} leading-$\beta_0$ part of
the full perturbation expansion (\ref{Dpt}). This means that
it keeps its form (\ref{D1}) under any change of RScl $\mu^2$.
Variations of $a=a(\mu^2)$ and of $c_{kk}=c_{kk}(\mu^2)$
under the RScl variation are governed by the RScl-invariance of the entire
observable ${\cal D}$ and of its perturbation expansion (\ref{Dpt}),
as reflected by relations (\ref{avsast}) and (\ref{c1kk}).
The additional terms appearing in expansion (\ref{D1}),
in comparison with the original leading-$\beta_0$ series (\ref{D10}),
are subleading in $\beta_0$ and represent effects beyond one loop
involving diagonal coefficients $c^{(1)}_{kk}$.
As shown in Appendix \ref{app2}, 
the covariant leading-$\beta_0$ expansion
(\ref{D1}) is the expansion of the leading skeleton (LS)
term in an assumed skeleton expansion of the observable
${\cal D}$.

Now we subtract the LS expansion (\ref{D1}) from
expansion (\ref{Dpt}), and the difference now involves
only subleading-$\beta_0$ terms
\ba
\left[
{\cal D}(Q^2)_{\rm pt} - {\cal D}^{(1)}(Q^2)_{\rm pt}
\right] & = &
k_2 \left[
a^2 + \sum_{n \geq 1} a^{n+2} d^{(2)}_n
\right]  \qquad (k_2 = c^{(1)}_{10}) \ ,
\label{DD1}
\ea
where the coefficients $d^{(2)}_n$ have a structure
similar to that of $d_n$'s (\ref{dns})
\be
d^{(2)}_n = \sum_{k=-1}^{n} c^{(2)}_{nk} \beta_0^k \qquad (n=1,2,\ldots) \ .
\label{d2ns}
\ee
Coefficients $c^{(2)}_{ij}$ are related to the
original coefficients $c^{(1)}_{ij}$ by relations
\ba
c^{(1)}_{10} c^{(2)}_{1j} & = & c^{(1)}_{2j} - b_{1j} c^{(1)}_{11} 
\quad (j=1,0,-1) \ ,
\nonumber\\
c^{(1)}_{10} c^{(2)}_{2j} & = & c^{(1)}_{3j} - 
\frac{5}{2} b_{1,j-1} c^{(1)}_{22} - b_{2j} c^{(1)}_{11}
\quad (j=2,1,0,-1) \ ,
\label{c2ij}
\ea
and coefficients $b_{kj}$ are those of the
expansion of $\beta_k$ coefficients (\ref{betak}) 
in powers of $\beta_0$ (including the case $k=1$).
Specifically, we have $b_{k,-1}=0$ ($k=1,2,\ldots$).
For $k=1$, we have: $b_{11}=19/4$ and $b_{10}=- 107/16$,
both numbers being RSch-independent.  
Now we repeat the previous construction, but now for
the (canonically normalized) rest 
$(1/k_2)({\cal D} - {\cal D}^{(1)})$ of Eq.~(\ref{DD1})
instead of ${\cal D}$ (\ref{Dpt}). Its RScl-covariant 
leading-$\beta_0$ part ${\cal D}^{(2)}$ 
then turns out to give
\ba
k_2 {\cal D}^{(2)}(Q^2)_{\rm pt} &=&
k_2 {\big \{}
a^2 + a^3 \left[ \beta_0 c^{(2)}_{11} \right] + 
a^4 \left[ \beta_0^2 c^{(2)}_{22} + \beta_1 c^{(2)}_{11} \right] +
{\cal {O}}(\beta_0^3 a^5) {\big \}} \ .
\label{D2}
\ea
Subtracting this from the rest (\ref{DD1}), we obtain
\ba
\left[
{\cal D}(Q^2)_{\rm pt} - {\cal D}^{(1)}(Q^2)_{\rm pt} 
- k_2 {\cal D}^{(2)}(Q^2)_{\rm pt}
\right] 
&=&  
k_3 \left[a^3 + \sum_{n \geq 1} a^{n+3} d^{(3)}_n \right]  \ ,
\label{DD1D2}
\\
k_3 &=& c^{(1)}_{10} \left( c^{(2)}_{10} + \frac{1}{\beta_0} c^{(2)}_{1,-1}
\right) \ ,
\label{k3}
\\
d^{(3)}_1 & = & \beta_0 ( c^{(2)}_{21} - b_{11} c^{(2)}_{11})/c^{(2)}_{10} + 
k_4/k_3 \ ,
\label{d31}
\ea
where $k_4/k_3$ is a number $\sim \beta_0^0$ 
which will be given explicitly below.
The (RScl-covariant) leading-$\beta_0$ part ${\cal D}^{(3)}$
of the canonically normalized expression
$(1/k_3)({\cal D} - {\cal D}^{(1)} - k_2 {\cal D}^{(2)})$
gives
\ba
k_3 {\cal D}^{(3)}(Q^2)_{\rm pt} &=&
k_3 \left\{
a^3 + a^4 \left[ \beta_0 c^{(3)}_{11} \right] + 
{\cal {O}}(\beta_0^2 a^5) \right\}  \ ,
\label{D3}
\\
c^{(2)}_{10} c^{(3)}_{11} & = & ( c^{(2)}_{21} - b_{11} c^{(2)}_{11}) \ .
\label{c311}
\ea 
Defining
\be
{\cal D}^{(4)}(Q^2)_{\rm pt} = a^4 + {\cal {O}}(\beta_0 a^5) \ ,
\label{D4}
\ee
and following the procedure pattern, we subtract expression
(\ref{D3}) from expression (\ref{DD1D2}) and obtain
\ba
{\cal D}(Q^2)_{\rm pt} &=&
{\cal D}^{(1)}(Q^2)_{\rm pt} + k_2 {\cal D}^{(2)}(Q^2)_{\rm pt} +
k_3 {\cal D}^{(3)}(Q^2)_{\rm pt} + k_4 {\cal D}^{(4)}(Q^2)_{\rm pt} +
{\cal {O}}(\beta_0^0 a^5) \ ,
\label{D1-D4}
\ea
where perturbation expansions for ${\cal D}^{(j)}$'s are
given by (\ref{D1}), (\ref{D2}), (\ref{D3}), (\ref{D4});
coefficients $k_2$ and $k_3$ are given by Eqs.~(\ref{DD1}) and (\ref{k3});
coefficients $c^{(1)}_{ij}$, $c^{(2)}_{ij}$, $c^{(3)}_{ij}$
are given by Eqs.~(\ref{dns}), (\ref{c2ij}), (\ref{c311});  
and an explicit expression for the coefficient $k_4$ is
\ba
k_4 & = & c^{(1)}_{10} \left[
c^{(2)}_{20} - b_{10} c^{(2)}_{11} - 
\frac{c^{(2)}_{1,-1}}{c^{(2)}_{10}} ( c^{(2)}_{21} - b_{11} c^{(2)}_{11} )
+ \frac{1}{\beta_0} c^{(2)}_{2,-1} \right] \ .
\label{k4}
\ea
It is straightforward to check that all the coefficients
$k_2$, $k_3$, $k_4$ are RScl-independent [as are the subseries
${\cal D}^{(j)}(Q^2)$]. Thus, identity (\ref{D1-D4}),
obtained by our construction,
represents identity (\ref{Dptsubser}) to order $n=4$.
This construction can be continued to any order.

In practice, we know only all the leading-$\beta_0$ parts of the
coefficients $d_j$ of observable ${\cal D}(Q^2)$ Eq.~(\ref{Dpt}),
i.e., all the coefficients $c^{(1)}_{jj}$; and in addition,
we usually know only one, two or three full coefficients
($d_1$, $d_2$, and possibly $d_3$). This implies that the
first term ${\cal D}^{(1)}$ on the RHS of identity (\ref{D1-D4})
is known to all orders, while the other terms
(${\cal D}^{(2)}$, ${\cal D}^{(3)}$, and possibly
${\cal D}^{(4)}$) are known only in their truncated version.
This means that the rest term in Eq.~(\ref{D1-D4})
is, in such a case, ${\cal {O}}(\beta_0^3 a^5)$, not ${\cal {O}}(\beta_0^0 a^5)$.

The perturbation expansion ${\cal D}^{(1)}_{\rm pt}$
of the ``leading-skeleton'' (LS) term can be written in
a resummed form \cite{Neubert,Neubert2}
\be
{\cal D}^{(1)}(Q^2)_{\rm pt} = 
\int_0^\infty \frac{dt}{t}\: F_{\cal D}^{\cal {E}}(t) \: 
a(t e^{\cal C} Q^2) \ ,
\label{LS1}
\ee
where $F_{\cal D}^{\cal {E}}(t)$ is the LS-characteristic
function\footnote{
The superscript ${\cal {E}}$ means ``Euclidean'', since
the scales involved ($Q^2$, $t e^{\cal C} Q^2$) are space-like.}
which often can be written in a closed explicit form
\cite{Neubert}. In principle, $F_{\cal D}^{\cal {E}}(t)$ can
be obtained for any space-like observable whose
leading-$\beta_0$ parts ($c^{(1)}_{kk}$) of all coefficients are
known. The value of ${\cal C}$ in (\ref{LS1}) depends on the value of
the reference scale $\Lambda$ used in the RGE-running;
in our convention, as mentioned before, we use
$\Lambda = \bL$ which corresponds to 
${\cal C}={\overline {\cal C}} \equiv -5/3$.

At this point, we will turn to the question of
the RSch-dependence of the (truncated) perturbation series (\ref{D1-D4}).
The RSch independence of the series (\ref{Dpt}) implies
specific transformation rules of the expansion coefficients
$d_j$ under the change of $\beta_j$'s ($j\geq 2$) \cite{PMS}
\ba
d_1 & = & {\overline d}_1 \ , \qquad 
d_2 = {\overline d}_2 - \frac{1}{\beta_0} (\beta_2 - {\overline \beta}_2 ) \ ,
\nonumber\\
d_3 & = & {\overline d}_3 - 
2 {\overline d}_1 \frac{1}{\beta_0} (\beta_2 - {\overline \beta}_2 )
- \frac{1}{2 \beta_0} (\beta_3 - {\overline \beta}_3 ) \ , \ldots
\label{RSchch}
\ea
where the bars denote the values with $\bMS$ RSch parameters
$\beta_k = {\overline b}_k = 
\sum {\overline b}_{kj} \beta_0^j$, and unchanged RScl.
This implies, in view of relations (\ref{dns}),
(\ref{c2ij}), (\ref{c311}),
specific transformation rules for $c^{(s)}_{nk}$
coefficients.
We will consider that the first term 
in skeleton-motivated expansion (\ref{D1-D4}) has a known
characteristic function, cf.~Eq.~(\ref{LS1}), and that
at most the first three nonleading coefficients 
of the perturbation expansion (\ref{Dpt}) of observable
${\cal D}$ are known: ${\overline d}_1$, ${\overline d_2}$, and
${\overline d_3}$ -- in $\bMS$ RSch and at RScl $\mu^2=Q^2$.
Since each term in expansion
(\ref{D1-D4}) is RScl-independent, we can re-expand
each ${\cal D}^{(j)}(Q^2)_{\rm pt}$ ($j \geq 2$) in powers of
$a(Q_j^2)$, i.e., at different chosen RScl's $Q_j$,
in a chosen common RSch ($\beta_2, \beta_3, \ldots$). 
The resulting subseries, however, will now be truncated since
$d_j$'s for $j \geq 4$ are not known. 
This leads to the following form of the
skeleton-motivated expansion (\ref{D1-D4}):
\ba
{\cal D}(Q^2)_{\rm pt} & = & {\cal D}(Q^2)_{\rm (TPS)}  
 + {\cal {O}}(\beta_0^3 a^5) \ ,
\label{Dpt2a}
\\
{\cal D}(Q^2)_{\rm (TPS)} & = & {\cal D}^{(1)}(Q^2)
+ t_2^{(2)} a^2(Q_2^2) + \sum_{j=2}^3 t_3^{(j)} a^3(Q_j^2) +
\sum_{j=2}^4 t_4^{(j)} a^4(Q_j^2) \ ,
\label{Dpt2b}
\ea
where the coefficients $t_i^{(j)}$ depend on the
scale ratios $Q_j^2/Q^2$ and the RSch parameters $\beta_k$
(\ref{betak}), and are written explicitly in Appendix \ref{app1} in
terms of the coefficients ${\overline c}^{(1)}_{ij}$,
the latter comprising
via Eq.~(\ref{dns}) the coefficients ${\overline d}_n$ of the original
perturbation series (\ref{Dpt}) in $\bMS$ RSch and at
the RScl $\mu^2 = Q^2$.

We now turn to the question of analytization of
the perturbation series (\ref{Dpt2b}), within a given
anQCD model with known analytic couplings
$\A_k$, Eqs.~(\ref{A1disp}), (\ref{tAn})-(\ref{A2A3A4}). 
For the first (LS) term, the natural analytization
procedure is to replace the perturbative coupling
$a(t e^{\cal C} Q^2)$ by its
anQCD counterpart $\A_1(t e^{\cal C} Q^2)$:\footnote{
A different approach to considering the perturbative
LS term (\ref{LS1}) was developed by the authors
of Ref.~\cite{Brooks:2006it}. They present a novel
version of the leading-$\beta_0$ renormalon calculus,
and consider that an OPE-term exists whose
$Q^2$-dependence is the same as that of the renormalon
ambiguity of the perturbative LS term and that the
ambiguity cancels in the sum (``PT+NP''). This
sum can be presented in the LS form (\ref{LS1})
with the perturbative coupling 
$a(t e^{\cal C} Q^2)$ there replaced by 
a modified (but nonanalytic) coupling with
one parameter. 
Since they work in the OPE framework, the latter parameter
is observable-dependent.} 
\be
{\cal D}^{(1)}(Q^2)_{\rm an} \equiv {\cal D}^{\rm (LS)}(Q^2) =
\int_0^\infty \frac{dt}{t}\: F_{\cal D}^{\cal {E}}(t) \: 
\A_1 (t e^{\cal C} Q^2) \ .
\label{LS2}
\ee
In contrast to expression (\ref{LS1}) which is an ill-defined integral
due to the Landau singularities of $a$, expression (\ref{LS2}) is
a well-defined integral in any given anQCD [unless $\A_1(Q^2)$
diverges too strongly when $Q^2 \to 0$]. We can adopt the
viewpoint that any anQCD model is defined:
(a) by a specific expression for $\A_1(Q^2)$, {\it and\/} 
(b) by prescription (\ref{LS2}) for calculation
of the LS-terms of any space-like observable.
The analytization of
the other terms in (\ref{Dpt2b}), after the choice of
an anQCD model, i.e., of $\A_1(Q^2)$, can be performed
in different ways. For example, the
replacements $a^k(Q_j^2) \mapsto \A_1^k(Q_j^2),
\A_1^{k-2}(Q_j^2) \A_2(Q_j^2), \ldots, \A_k(Q_j^2)$
all appear equally natural at first, since the
perturbative parts of these expressions are all the same
to the order considered -- cf.~relations (\ref{Antr})
and (\ref{tAntrmod}).
However, construction of the higher order
couplings $\A_k$ ($k \geq 2$) on the
basis of the anQCD coupling $\A_1$, as presented in Sec.~\ref{analytiz},
suggests that it is the replacement
\be
\left[ {\widetilde a}_k(Q_j^2) \mapsto \tA_k(Q_j^2) \ \Rightarrow \ 
\right] \
a^k(Q_j^2) \mapsto \A_k(Q_j^2) \qquad (k \geq 1)
\label{analyt}
\ee
that appears to be the most natural from the point of view
of the requirement of the RScl- and RSch-invariance
of the observables. Namely, $\A_k(\mu^2; \beta_2,\ldots)$'s 
fulfill, to the order considered,
the same evolution equations under the changes 
of the RScl and of RSch as $a^k(\mu^2; \beta_2, \ldots)$'s
when the replacements (\ref{analyt}) are performed everywhere.
Further, the LS-analytization (\ref{LS2}) of the 
first term ${\cal D}^{(1)}_{\rm pt}$
of (\ref{Dpt2b}) is also equivalent to
the term-by-term analytization (\ref{analyt}) of
the perturbation expansion of ${\cal D}^{(1)}_{\rm pt}$,
as is explicitly shown in Appendix \ref{app2}.
The analytization (\ref{analyt}) of the TPS (\ref{Dpt2b}),
which results in the ``truncated analytic series'' (TAS)
\ba
{\cal D}(Q^2) & = & 
{\cal D}(Q^2)_{\rm (TAS)} + {\cal {O}}(\beta_0^3 \A_5) 
\ ,
\label{TASa}
\\
{\cal D}(Q^2)_{\rm (TAS)} & = &
{\cal D}^{\rm (LS)}(Q^2)  
+ t_2^{(2)} \A_2(Q_2^2) + \sum_{j=2}^3 t_3^{(j)} \A_3(Q_j^2) +
\sum_{j=2}^4 t_4^{(j)} \A_4(Q_j^2) 
\ , 
\label{TASb}
\ea
has, as a consequence, the suppression of the
RScl- and RSch-dependence just as is
known for the corresponding TPS in pQCD, but with $a^k \mapsto \A_k$:
\ba
\frac{ \partial {\cal D}(Q^2)_{\rm (TAS)} }{\partial \ln Q_j^2 }
& = & {\cal {O}}(\beta_0^{5 - j} \A_5) 
\qquad (j=2,3,4) \ ,
\label{TASRScl}
\\
\frac{ \partial {\cal D}(Q^2)_{\rm (TAS)} }{\partial \beta_k }
& \leq & {\cal {O}}(\beta_0^{3-k} \A_5) 
\qquad (k=2,3) \ .
\label{TASRSch}
\ea
We are allowed, in principle, to vary
in the TAS series (\ref{TASb}) three different RScl's
$Q_j$ and $3+4$ RSch parameters $b_{2j}$ and $b_{3j}$
appearing in $\beta_2$ and $\beta_3$.
One may want to have, for given chosen RScl's $Q_j$,
such a RSch that effectively only the first coefficient $t_2^{(2)}$
in the beyond-the-LS contribution is nonzero. This
implies various conditions involving the other five $t_i^{(j)}$'s
[Eqs.~(\ref{t32})-(\ref{t44})]:
\ba
t^{(2)}_3 &=& t^{(3)}_3 = 0 \ ; \quad
\sum_{j=2}^4 t^{(j)}_4 = 0 \ ,
\label{vpcon}
\\
\Rightarrow \quad {\cal D}(Q^2) & = & 
{\cal D}^{\rm (LS)}(Q^2)  
+ t_2^{(2)} \A_2(Q_2^2) + {\cal {O}}(\beta_0^3 \A_5) \ .
\label{vp}
\ea
Specifically, if we choose for all three
${\cal D}^{(j)}(Q^2;\mu^2=Q_j^2)_{\rm (TAS)}$ ($j=2,3,4$) the same
RScl 
\be
Q_2^2=Q_3^2=Q_4^2 = Q^2 \exp({\cal C}) \ ,
\label{RScls}
\ee 
the corresponding
$\beta_k = b_{kj} \beta_0^j$ ($k=2,3$) have the following
$\delta b_{kj} \equiv b_{kj} - {\overline b}_{kj}$: 
\ba
\delta b_{22} & = & \ovc^{(1)}_{10} ( \ovc^{(2)}_{11} + 2 {\cal C} ) \ ,
\label{db22vp}
\\
\delta b_{21} & = & \ovc^{(1)}_{10} \ovc^{(2)}_{10} \ ,
\qquad \delta b_{20} = 0 \ ,
\label{db2120vp}
\\
\frac{1}{2} \delta b_{33} & = &
\ovc^{(1)}_{10} \ovc^{(2)}_{22} 
+ 3 \ovc^{(1)}_{10} {\cal C} (\ovc^{(2)}_{11} + {\cal C})
- \delta b_{22} 3 (\ovc^{(1)}_{11} + {\cal C}) \ ,
\label{db33vp}
\\
\frac{1}{2} \delta b_{32} & = &
\ovc^{(1)}_{10} \ovc^{(2)}_{21} + 
{\cal C} ( 3 \ovc^{(1)}_{10} \ovc^{(2)}_{10}
+ 2 b_{11} \ovc^{(1)}_{10}) 
- \delta b_{22} 2 \ovc^{(1)}_{10}
-\delta b_{21} 3 (\ovc^{(1)}_{11} + {\cal C}) \ ,
\label{db32vp}
\\
\frac{1}{2} \delta b_{31} & = &
\ovc^{(1)}_{10} \ovc^{(2)}_{20} 
+ {\cal C} 2 b_{10} \ovc^{(1)}_{10} 
- \delta b_{21} 2 \ovc^{(1)}_{10}
-\delta b_{20} 3 (\ovc^{(1)}_{11} + {\cal C}) \ ,
\label{db31vp}
\\
\frac{1}{2} \delta b_{30} & = &
- \delta b_{20} 2 \ovc^{(1)}_{10} (= 0) \ .
\label{db30vp}
\ea
Here, $\ovc^{(k)}_{ij} \equiv c^{(k)}_{ij}(\mu^2=Q^2; \bMS)$.
Results (\ref{db22vp})-(\ref{db30vp}) are obtained
by using explicit expressions (\ref{t32})-(\ref{t44})
obtained in Appendix \ref{app1}, applying to them conditions
(\ref{vpcon}) for the RScl choice (\ref{RScls}).
Specifically, result (\ref{db22vp}) is obtained by
the requirement $t^{(2)}_3=0$; results (\ref{db2120vp})
from the requirement $t^{(3)}_3=0$, being zero
both the coefficient at $\beta_0^0$ and 
at $1/\beta_0$, respectively; results (\ref{db33vp})-(\ref{db30vp})
are obtained from requirement $\sum t^{(j)}_4 = 0$,
being zero all the coefficients at the $\beta_0$-powers 
$\beta_0^2$, $\beta_0^1$, $\beta_0^0$, $1/\beta_0$,
respectively. 

Our evaluation method (\ref{TASb}), with the
choice of the scheme described above [Eqs.~(\ref{vp}), (\ref{RScls}),
(\ref{db22vp})-(\ref{db30vp})], emphasizes in the
beyond-the-LS parts the role of the analytic couplings
$\A_k(\mu^2)$ ($k \geq 2$) constructed in Sec.~\ref{analytiz}
from the couplings $\tA_n(\mu^2)$, Eq.~(\ref{tAn})
[see Eqs.~(\ref{A2A3A4})]. The couplings $\A_k(\mu^2)$ ($k \geq 2$)
were constructed in such a way as to have, at perturbative level,
their equivalence with $a^n(\mu^2)$. However, the
construction in Sec.~\ref{analytiz} strongly suggests that
the couplings $\tA_n(\mu^2)$ ($n \geq 2$) are more
basic since they are constructed as derivatives
of $\A_1(\mu^2)$ which is the basic quantity in any anQCD
model. Further, the skeleton-expansion arguments
presented in Appendix \ref{app2} show that
$\tA_n(\mu^2)$ are the basic elements for the expansion
of each term in the skeleton expansion. Therefore,
a more natural choice for RSch
($\beta_2, \beta_3$) in the evaluation method (\ref{TASb}), 
with RScl's (\ref{RScls}), would be such that
the resulting TAS expression is
\ba
{\cal D}(Q^2) & = & 
{\cal D}(Q^2)_{\rm (TAS)} + {\cal {O}}(\beta_0^3 \tA_5) 
\ ,
\label{TASav}
\\
{\cal D}(Q^2)_{\rm (TAS)} & = &
{\cal D}^{\rm (LS)}(Q^2)  
+ \tlt_2 \tA_2(Q^2 e^{\cal C}) 
\ . 
\label{TASbv}
\ea
To obtain the $\beta_k$'s ($k=2,3$) necessary for this result,
we first re-express all $\A_k$'s ($k \geq 2$) in
TAS (\ref{TASb}) in terms of $\tA_n$'s, Eqs.~(\ref{A2A3A4}).
Keeping the RScl's according to (\ref{RScls}), this
implies that, in a general RSch ($\beta_2, \beta_3$) 
expression (\ref{TASb}) can be re-expressed as
\ba
{\cal D}(Q^2)_{\rm (TAS)} & = &
{\cal D}^{\rm (LS)}(Q^2)  
+ \tlt_2 \tA_2(Q^2 e^{\cal C}) + 
\tlt_3 \tA_3(Q^2 e^{\cal C}) +
\tlt_4 \tA_4(Q^2 e^{\cal C}) 
\ , 
\label{TASbvgen}
\ea
where the coefficients $\tlt_i$ are certain combinations
of $t^{(k)}_s$, and are written explicitly in Appendix \ref{app1},
Eqs.~(\ref{tlt2})-(\ref{btlt4}). Requiring
the form (\ref{TASbv}), i.e.,
\be
\tlt_3=\tlt_4=0 \ ,
\label{tltszero}
\ee
implies, by Eqs.~(\ref{tlt2})-(\ref{btlt4}), that
the corresponding
$\beta_k = b_{kj} \beta_0^j$ ($k=2,3$) have the following
$\delta b_{kj} \equiv b_{kj} - {\overline b}_{kj}$: 
\ba
\delta b_{22} & = & \ovc^{(1)}_{10} ( \ovc^{(2)}_{11} + 2 {\cal C} ) \ ,
\label{db22v}
\\
\delta b_{21} & = & \ovc^{(1)}_{10} \ovc^{(2)}_{10} - b_{11} \ovc^{(1)}_{10}
\ ,
\qquad \delta b_{20} = - b_{10} \ovc^{(1)}_{10} \ ,
\label{db2120v}
\\
\frac{1}{2} \delta b_{33} & = &
\ovc^{(1)}_{10} \ovc^{(2)}_{22} 
+ 3 \ovc^{(1)}_{10} {\cal C} (\ovc^{(2)}_{11} + {\cal C})
- \delta b_{22} 3 (\ovc^{(1)}_{11} + {\cal C}) \ ,
\label{db33v}
\\
\frac{1}{2} \delta b_{32} & = &
\ovc^{(1)}_{10} \ovc^{(2)}_{21} 
- \frac{5}{2} b_{11} \ovc^{(1)}_{10} \ovc^{(2)}_{11}
-  \ovc^{(1)}_{10} {\overline b}_{22} 
+ 3 {\cal C} \ovc^{(1)}_{10} ( \ovc^{(2)}_{10}
- b_{11}) 
+ \delta b_{22} \left( - 3 \ovc^{(1)}_{10} + \frac{5}{2} b_{11} \right) 
- \delta b_{21} 3 (\ovc^{(1)}_{11} + {\cal C}) \ ,
\label{db32v}
\\
\frac{1}{2} \delta b_{31} & = &
\ovc^{(1)}_{10} \ovc^{(2)}_{20} 
- \frac{5}{2} b_{10}  \ovc^{(1)}_{10} \ovc^{(2)}_{11} 
- \frac{5}{2} b_{11} \ovc^{(1)}_{10} \ovc^{(2)}_{10}
+ \ovc^{(1)}_{10} \left( \frac{5}{2} b_{11}^2 - {\overline b}_{21} 
- 3 b_{10} {\cal C} \right)
\nonumber\\ &&
+ \delta b_{22} \frac{5}{2} b_{10} 
+ \delta b_{21} \left( - 3 \ovc^{(1)}_{10} + \frac{5}{2} b_{11} \right) 
- \delta b_{20} 3 (\ovc^{(1)}_{11} + {\cal C}) \ ,
\label{db31v}
\\
\frac{1}{2} \delta b_{30} & = &
- \frac{5}{2} b_{10} \ovc^{(1)}_{10} \ovc^{(2)}_{10} 
+ 5 b_{10} b_{11} \ovc^{(1)}_{10} -  {\overline b}_{20} \ovc^{(1)}_{10}
+ \frac{5}{2} b_{10} \delta b_{21} +
\delta b_{20} \left( - 3 \ovc^{(1)}_{10} + \frac{5}{2} b_{11} \right) \ . 
\label{db30v}
\ea
In these expressions, ${\overline b}_{2j}$ 
are the coefficients 
$b_{2j}$ in $\bMS$: ${\overline b}_{22} = 325/96$,
${\overline b}_{21} = 243/32$,
${\overline b}_{20} = - 37117/1536$ (and $b_{11}=19/4$,
$b_{10}=-107/16$).
We will apply, as a rule, our evaluation approach
in the RSch (\ref{db22v})-(\ref{db30v}), i.e,
where the resulting formula is (\ref{TASav})-(\ref{TASbv}),
and will use the RScl's (\ref{RScls}) with 
${\cal C} = {\overline {\cal C}} = -5/3$.
The RSch evidently depends on the observable.
Our starting point will be
this RSch for the massless Adler function
${\cal D}(Q^2) = d_v(Q^2)$, where the
STPS is known to a large degree of accuracy
up to $\sim a^4$ (up to $\sim a^3$ it's known
exactly) -- we will call this RSch A ('A' for 
Adler).\footnote{
The difference between this RSch A
and the RSch A' (\ref{db22vp})-(\ref{db30vp}) for the Adler function
is small. For example, for $n_f=3$, the values are
$\beta_2^{\rm (A)}=-18.92$, $\beta_2^{\rm (A')}=-18.59$;
$\beta_3^{\rm (A)}=-33.84$, $\beta_3^{\rm (A')}=-32.72$.
In Ref.~\cite{Cvetic:2006mk}, we used RSch A'
(\ref{db22vp})-(\ref{db30vp}) [with RScl's
(\ref{RScls}) with ${\cal C} = {\overline {\cal C}} = -5/3$],
and denoted there this approach as 'v2'.}
If an observable is known in STPS only up to $\sim a^3$,
only formulas (\ref{db22v})-(\ref{db2120v}) are 
to be applied, as ${\widetilde t}_4$ is not known;
in that case, in Eq.~(\ref{TASav}) the unknown
rest term is ${\cal O}(\beta_0^2 \tA_4)$.
For example, Bjorken polarized sum rule $d_b(Q^2)$
is such an observable.

In Appendix \ref{app2}, a different
method of evaluation is presented, which would be an evaluation
of the skeleton expansion itself if such an expansion existed in
the considered RSch. The
RSch-dependence of that method is numerically stronger,
which may be a reflection of the fact that this
expansion, if it exists, is valid only in a specific
('skeleton') RSch that is hitherto unknown \cite{Gardi:1999dq,Brodsky}.

\section{Numerical results}
\label{numres}

In this Section, we take the position that the 
anQCD models M1 and M2, introduced in Sec.~\ref{anQCDs}, 
the form of $\A_1(Q^2)$ there, 
Eqs.~(\ref{M1c}) and (\ref{M2}), is achieved 
in the aforementioned ``optimal'' RSch 
(\ref{db22v})-(\ref{db30v}) 
for the massless Adler function $d_v(Q^2)$ -- 
RSch A. We must keep in mind that models 
M1 and M2 change the form of $\A_1(Q^2)$ when the RSch 
($\beta_2, \beta_3,\ldots$) is changed.\footnote{
When $\beta_j$'s ($j \geq 2$) change, the change of
$\A_1(Q^2)$ in general cannot be described just by
running of the parameters of the model with $\beta_j$'s,
since new terms appear that depend on those parameters.} 

We will calculate numerically various low-energy QCD 
observables in the anQCD models MA, M1 and M2, 
with $n_f=3$, by using the skeleton-motivated 
evaluation method presented
in the previous Section, Eq.~(\ref{TASbvgen}).
One such quantity is the massless Adler function
$d_v(Q^2)$ whose pQCD expansion coefficients 
$d_1$ and $d_2$ (in $\bMS$ RSch and at RScl
$\mu^2=Q^2$) are known exactly \cite{d1,d2},
and $d_3$ has been estimated as a polynomial in $n_f$
to a high degree of accuracy \cite{Baikov:2002uw}
(see Appendix \ref{app4} for explicit expressions
of $d_1$, $d_2$, $d_3$).
The normalization of $d_v$ is taken according to
Eq.~(\ref{Dpt}) when $n_f=3$. 
The additional light-by-light contributions \cite{d2} 
do not contribute when $n_f=3$. Further,
the LS characteristic function $F_{v}^{\cal {E}}(t)$
for $d_v(Q^2)$ was obtained in \cite{Neubert}, and
is given in Appendix \ref{app3} in 
Eqs.~(\ref{FLSvIR})-(\ref{FLSvUV}). 
Evaluation method (\ref{TASbvgen}) can thus be applied
by including terms $\sim\tA_4$ in the case of the massless 
Adler function
(for a different approach to evaluating Adler function,
see Ref.~\cite{Nesterenko:2005wh}). 
The optimal RSch for the massless Adler
function $d_v(Q^2)$ is then obtained 
by requiring disappearance of 
$\sim \tA_3$ and $\sim \tA_4$ terms,
Eq.~(\ref{tltszero}), where we choose RScl according to
(\ref{RScls}) with ${\cal C} = {\overline {\cal C}} = -5/3$. 
We call this RSch Adler (A), 
and it can be obtained from $\bMS$ RSch by applying relations
(\ref{db22v})-(\ref{db30v}), resulting in
\ba
\beta_2^{\rm (A)} & = & -23.6074 - 16.0248 \beta_0 + 
8.04784 \beta_0^2 \ ,
\nonumber\\
\beta_3^{\rm (A)} & = & 127.38 - 35.8577 \beta_0 
- 12.8734 \beta_0^2  - 1.34926 \beta_0^3 \ .
\label{ARSch}
\ea
The values for $n_f=3$ are $\beta_2=-18.9211$ and
$\beta_3=-33.8404$ (in $\bMS$ RSch, at $n_f=3$, 
the values are $10.0599$ and $47.2281$, respectively).
In RSch A, the evaluated massless $d_v(Q^2)$ is thus
\be
d_v(Q^2)_{\rm TAS} =  
\int_0^\infty \frac{dt}{t}\: F_{v}^{\cal {E}}(t) \: 
\A_1 (t e^{\overline {\cal C}} Q^2; 
\beta_2^{\rm (A)}, \beta_3^{\rm (A)}) + 
\frac{1}{12} \tA_2( e^{\overline {\cal C}} Q^2 ) \ ,
\label{dvARSch}
\ee
and the difference between the (massless) true
$d_v(Q^2)$ and $d_v(Q^2)_{\rm TAS}$ is formally
${\cal O}(\beta_0^3 \tA_5)$.   

The ($V+A$-channel) semihadronic $\tau$ decay rate
ratio $r_{\tau}$ is one of the best measured
low-energy QCD quantities, its massless part
for non-strange hadron production has the value
$r_{\tau}(\triangle S=0,m_q=0) = 0.204 \pm 0.005$
\cite{ALEPH2,ALEPH3} 
(cf.~Appendix \ref{app5}, Eq.~(\ref{rtauexp})). 
The heavy quarks ($c$ and $b$) do not contribute,
since $r_{\tau}$ is a Minkowskian observable, 
and the $\tau$ particle cannot decay to charmed mesons
because their masses are larger than $m_{\tau}$.\footnote{
The contributions of heavy quarks in
Euclidean observables ${\cal D}(Q^2)$, such as the Adler function,
can be more important, even though $Q^2 < m_c^2$ --
see the discussion later in this Section.}
Our evaluation approach for $r_{\tau}(\triangle S=0,m_q=0)$
uses the aforementioned evaluation (\ref{dvARSch})
of the (massless) Adler function
$d_v(Q^2)$ which is then inserted in the contour integral
(\ref{rtaudv}). The LS-part can then be written
in the form (\ref{LSrt}) with the time-like
LS characteristic function (\ref{FLSrIR})-(\ref{FLSrUV}).
The beyond-the-LS (bLS) contribution is the contour integral 
\be
r_{\tau}(\triangle S=0, m_q=0)^{\rm (bLS)} =
\frac{1}{2 \pi}
\int_{-\pi}^{+\pi} d \phi \: (1 + e^{i \phi})^3 (1 -e^{i \phi})
\frac{1}{12} \tA_2( e^{\overline {\cal C}} m_{\tau}^2 e^{i \phi}) \ .
\label{rtbLS}
\ee

Yet another low-energy QCD observable that we will consider
is the Bjorken polarized sum rule (BjPSR) $d_b(Q^2)$. 
Its LS-characteristic
function is obtained in Appendix \ref{app3}, on the basis of the
known leading-$\beta_0$ coefficients \cite{Broadhurst:1993ru}
using the technique of \cite{Neubert}. The full perturbation
coefficients $d_1$ and $d_2$ for the massless $d_b(Q^2)$, in
$\bMS$ RSch and at RScl $\mu^2=Q^2$, were obtained in
Refs.~\cite{LV} (see Appendix \ref{app4} for explicit
expressions for $d_1$ and $d_2$).
For the coefficient $d_3$, only the leading-$n_f$ part
($\propto n_f^3$) is known exactly \cite{Broadhurst:1993ru};
based on this, estimates of $d_3$ as a polynomial in $\beta_0$ were
performed in Ref.~\cite{Broadhurst:2002bi} using naive nonabelianization
(NNA) $n_f \mapsto - 6 \beta_0$ \cite{Broadhurst:1994se}.
For the evaluation of (the massless part of) $d_b(Q^2)$ we will
not use estimates of the full $d_3$, i.e.,
we will use method (\ref{TASbvgen}) with terms up to 
${\widetilde t}_3 \tA_3$ included,
in any chosen RSch and with RScl's (\ref{RScls})
with ${\cal C} = {\overline {\cal C}} = -5/3$. 
The formal difference between the
evaluated and the true value is then ${\cal O}(\beta_0^2 \tA_4)$.
The experimental values of $d_b(Q^2)$ at low $Q^2$ are
much less precise than those of $r_{\tau}(\triangle S=0)$.
At $Q^2=2$ and $1 \ {\rm GeV}^2$ they are 
$d_b(2 {\rm GeV}^2) = 0.16 \pm 0.11$ and
$d_b(1 {\rm GeV}^2) = 0.17 \pm 0.07$  \cite{Deur:2004ti}
(for an application, cf.~Ref.~\cite{Campanario:2005np}).
The contributions of massive quarks ($m_c$, $m_b$)
are $|\delta d_b(Q^2; m_q \not=0)| < 10^{-3}$ for 
$Q^2 \leq 2 \ {\rm GeV}^2$ \cite{Buza:1996xr}, thus negligible.
We recall that both $d_v$ and $d_b$ are massless observables which
are normalized here according to the convention (\ref{Dpt})
for $n_f=3$. Although the uncertainty of the measured values of
$d_b(Q^2)$ is significantly lower at $Q^2= 1 \ {\rm GeV}^2$
than at $Q^2= 2 \ {\rm GeV}^2$, we will use both central values.
We expect the theoretical predictions of our evaluations
in general to be more reliable at higher momenta
$Q^2 > 1 \ {\rm GeV}^2$.

Now we will fix the parameters of models M1 and M2.
Model M1 (\ref{M1tl})-(\ref{M1MA}) has three independent
parameters $c_f, c_r, c_0$ (and $\bL = 0.4$ GeV as in MA). 
Requiring the reproduction of the aforementioned experimental 
central values $r_{\tau}(\triangle S=0,m_q=0) = 0.204$,
$d_b(2 {\rm GeV}^2) = 0.16$ and  $d_b(1 {\rm GeV}^2) = 0.17$,
we obtain a solution for the three parameters,
with the following values: $c_f=1.08$, $c_r=0.45$, $c_0=2.94$.
We will use these parameter values in M1 (in RSch A).
In general, the predicted values of observables do not
change a lot when $c_0$ is varied in the regime $\sim 1$;
they change more when $c_r$ and/or $c_f$ are varied. 
The experimental values of various higher-energy QCD observables
${\cal D}(Q^2), {\cal R}(s)$
($Q^2, s \agt 10 \ {\rm GeV}^2$) should be well
reproduced in M1, because condition (\ref{M1MA})
ensures that M1 and MA merge at higher energies
$Q^2, s \gg \bL^2$, and it has been demonstrated
that MA with $\bL_{(n_f =3)}=0.4$ GeV ($\Rightarrow
\bL_{(n_f =5)} = 0.26$ GeV) reproduces well those values \cite{Sh}.
We note that model  MA 
(with $\bL = 0.4$ GeV) predicts
$r_{\tau}(\triangle S=0,m_q=0) \approx 0.14$,
which is significantly too low.

Model M2 (\ref{M2tl})-(\ref{M2}) has two free parameters
$c_v$ and $c_p$, both assumed to be $\sim 1$. 
Requiring reproduction of the central value of 
$r_{\tau}(\triangle S=0,m_q=0) = 0.204$,
and requiring $|c_p|, |c_v| \geq 0.1$,
it turns out that the model then always predicts values
$d_b(2 {\rm GeV}^2) > 0.19$.
Requiring the minimal possible value 
$d_b(2 {\rm GeV}^2)\approx 0.19$
gives us the parameter values $c_v=0.1$ and $c_p=3.4$.
We will use these parameter values in M2 (in RSch A).

\begin{table}
\caption{\label{table1} Results of evaluation of
the semihadronic tau decay ratio
$r_{\tau}(\triangle S = 0,m_q=0)$ and of 
BjPSR $d_b(Q^2=2 \ {\rm GeV}^2)$,
in various anQCD models,
using evaluation method (\ref{TASbvgen}) in
RSch A (\ref{ARSch}). The basis for calculation
of $\rho_1^{\rm (pt)}(\sigma)$ is expansion
(\ref{apt}) at loop-level=4 (i.e., when $\beta_3^{\rm (A)}$ 
included) and with $k_{\rm max}=6$. In parentheses
are the results at loop-level=3 and $k_{\rm max}=5$
(in that case, the $d_3$-term of the Adler function
is not included). Presented are the results of
the full evaluation (leading-skeleton and beyond: LS+bLS), 
Eq.~(\ref{TASbvgen}), and for $r_{\tau}(\triangle S = 0,m_q=0)$
also the results of LS.
The experimental values are $r_{\tau}(\triangle S = 0,m_q=0)
= 0.204 \pm 0.005$, $d_b(Q^2=2 \ {\rm GeV}^2) = 0.16 \pm 0.11$
and $d_b(Q^2=1 \ {\rm GeV}^2) = 0.17 \pm 0.07$.
See the text for further details.}
\begin{ruledtabular}
\begin{tabular}{lllll}
 & $r_{\tau}(\triangle S = 0,m_q=0)$ &
$r_{\tau}(\triangle S = 0,m_q=0)$ [LS] &
$d_b(Q^2=2 \ {\rm GeV}^2)$ &
$d_b(Q^2=1 \ {\rm GeV}^2)$ 
\\
\hline
MA & 0.141 (0.142) & 0.139 (0.141) & 0.137 (0.138) & 0.155 (0.155)
\\
M1 & 
0.204 (0.205) & 0.197 (0.198) & 0.160 (0.161) & 0.170 (0.171)
\\
M2 & 
0.204 (0.206) & 0.203 (0.204) & 0.189 (0.190) & 0.219 (0.220)
\end{tabular}
\end{ruledtabular}
\end{table}
In Table 1 we present results of calculations
of $r_{\tau}(\triangle S=0,m_q=0)$ and
$d_b(Q^2=2 {\rm GeV}^2)$ with our evaluation method
(\ref{TASbvgen}), in the aforementioned RSch A  
(\ref{ARSch})-(\ref{dvARSch}) and at loop-level=4
and 3, in various anQCD models: M1,
M2, and MA. When loop-level=4 (and $k_{\rm max}=6$),
we used in the calculation of $r_{\tau}(\triangle S=0,m_q=0)$ 
the estimated ${\rm N}^3{\rm LO}$ perturbation coefficient
$d_3$ of Ref.~\cite{Baikov:2002uw} for the Adler function
(cf.~Appendix \ref{app4}),
as mentioned earlier. In the case of
$d_b(Q^2=2 {\rm GeV}^2)$, when loop-level=3 or 4,
evaluation formula (\ref{TASbvgen}) was used in RSch A
by inclusion of terms up to $\tA_3$ only, as
the ${\rm N}^3{\rm LO}$ coefficient $d_3$ is not known there.
We note that MA (with $\bL_{(n_f =3)}=0.4$ GeV),
with light quark masses $m_u, m_d, m_s \ll \bL$
($m_u, m_d, m_s \approx 0$),
does not reproduce the well-measured experimental value
$r_{\tau}(\triangle S=0,m_q=0) = 0.204 \pm 0.005$,
as already mentioned in the Introduction. This fact
led us to suggest alternative versions of anQCD
(e.g., M1, M2).

\begin{figure}[htb] 
\begin{minipage}[b]{.49\linewidth}
 \centering\epsfig{file=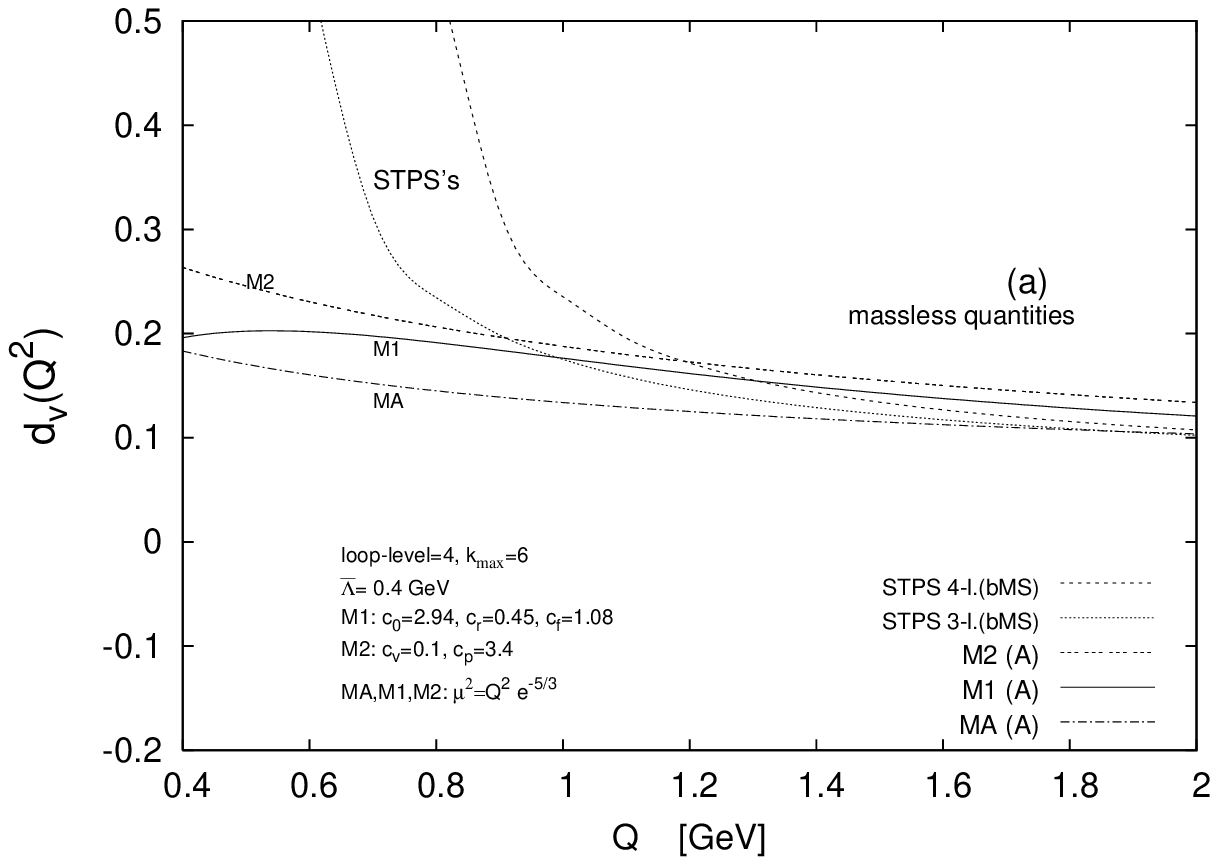,width=\linewidth}
\end{minipage}
\begin{minipage}[b]{.49\linewidth}
 \centering\epsfig{file=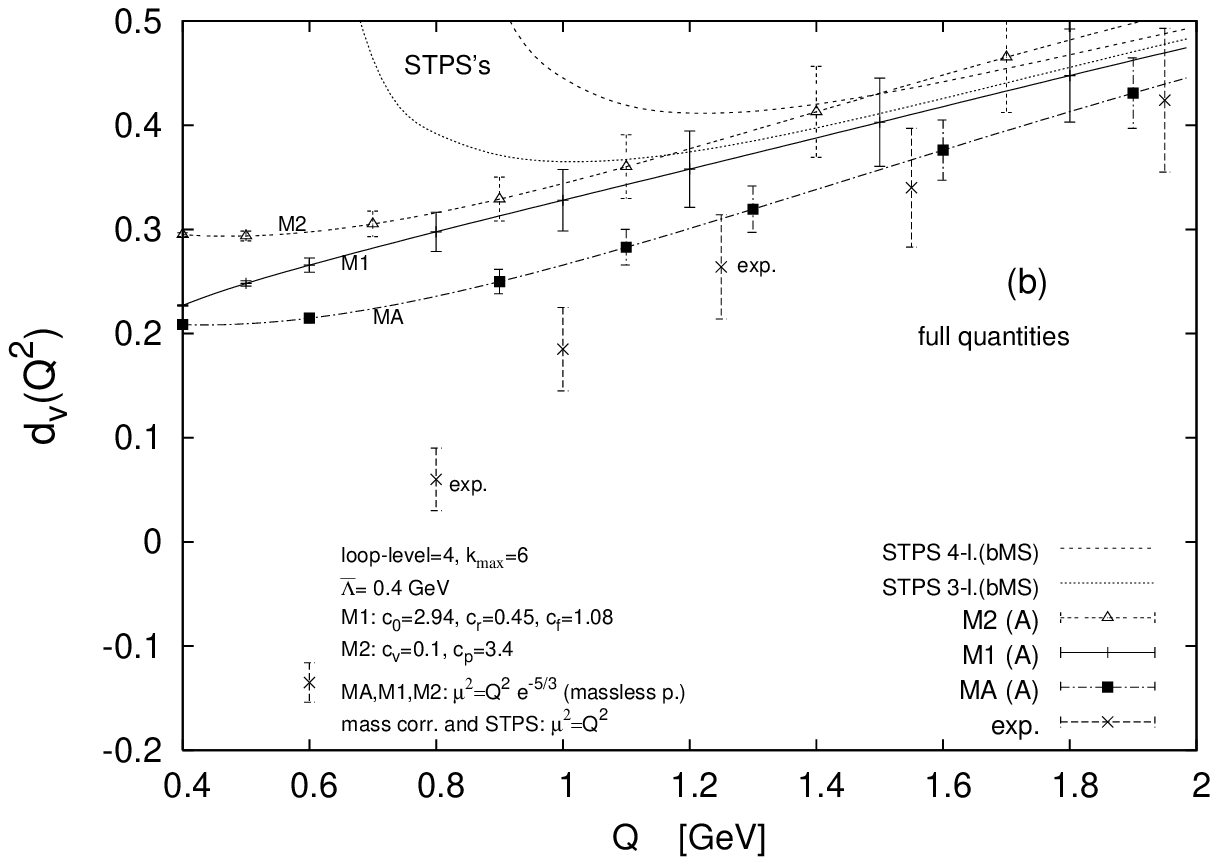,width=\linewidth}
\end{minipage}
\vspace{-0.4cm}
 \caption{\footnotesize  Adler function as predicted by pQCD,
and by our approach in several analytic QCD models (see the text):
(a) the massless part ($n_f=3$); 
(b) the full quantity, with the contribution of massive
quarks included.}
\label{Figdv}
 \end{figure}
Now that the parameters of the presented anQCD models
have been fixed, we can present various results of
these models, evaluated with the method (\ref{TASbvgen}).
In Fig.~\ref{Figdv}(a) we present curves for the
massless Adler function $d_v(Q^2)$ (with $n_f=3$) as functions
of energy $Q$, in models M1, M2, and MA.
The RSch used is RSch A (\ref{ARSch})-(\ref{dvARSch}).
Loop-level is four, i.e., we include the value
$\beta_3^{\rm (A)}$ in our calculation for
$\rho_1^{\rm (pt)}$, with $k_{\rm max}=6$
[cf.~Eq.~(\ref{apt})], and use the estimated
${\rm N}^3{\rm LO}$ perturbation coefficient
$d_3$ of Ref.~\cite{Baikov:2002uw} (cf.~Appendix \ref{app4}).
The light-by-light contributions, which have a different
topology of diagrams and should probably be resummed separately
(cf.~Ref.~\cite{KS}),
appear for the first time at $\sim a^3$ and are proportional
to the square of the sum of the quark charges $(\sum Q_f)^2$ \cite{d2}. 
This sum is zero in the case $n_f=3$ considered here.
Fig.~\ref{Figdv}(b) represents the results for
the full Adler function,
i.e., the $V$-channel heavy quark corrections 
$\delta d_v(Q^2; m_c,m_b)$ have been added there.
For the calculation of the latter, we follow 
the procedure of \cite{Eidelman}, including
the $a^2$-contributions
(note that $d_v(Q^2) \equiv (1/2) D(Q^2)\!-\!1$, 
where $D$ is defined in \cite{Eidelman}).
The first seven coefficients of the low-momentum Taylor expansion 
for the heavy quark $a^2$-contributions are calculated in \cite{ChHKS1}. 
Through a conformal mapping together with Pad\'e 
improvement, as proposed in \cite{FT},  an approximant is obtained. 
The approximant reproduces the low-momentum behavior and
fits very well the large-momentum 
expansion \cite{ChHKS2} for this quantity 
up to energies $Q^2 \approx 16 m_q^2$ 
(see also Fig.~4 of Ref.~\cite{Eidelman}). 
Thus, this method can be safely 
used for the $q=c,b$ quarks in the 
energy range we are interested in.\footnote{
Some contributions from heavy quarks are not considered here as 
we base our analysis on the expressions of Ref.~\cite{ChHKS1}.
The relevant diagrams are shown in Fig.~2 of Ref.~\cite{ChHKS1};
the contributions with internal heavy and external light quarks are
not included. These type of ($a^2$-)contributions have been
obtained for the $R_{e^+e^-}(s)$ function in Refs.~\cite{CHKST,HJKT,Kniehl}.
We checked that these contributions, when translated 
into the corresponding contributions for $d_v(Q^2)$ via
the usual integral transformation relating $R$ and $d_v$, 
result in $a^2$-contributions which are an order of
magnitude smaller than the heavy quark $a^2$-contributions included in our curves.} 
In the heavy quark  contributions, we simply replaced $a(Q^2)$
and $a^2(Q^2)$ by $\A_1(Q^2)$ and $\A_2(Q^2)$
(using $\Lambda = \bL = 0.4$ GeV). The indicated
$\pm$ uncertainties in the full Adler function curves
are those $c$ quark contributions which are proportional
to $\A_2$. 
In Figs.~\ref{Figdv}(a),(b) we included the
STPS's [truncated forms of Eq.~(\ref{Dpt})]
in $\bMS$ RSch and with RScl $\mu^2=Q^2$.
In Fig.~\ref{Figdv}(b) we included experimental values, for comparison.
The experimental values of $d_v(Q^2)$ are taken from 
Ref.~\cite{Eidelman} where the integral expression for 
$d_v(Q^2)$ in terms of the  $e^+ e^-$ QCD ratio 
$R_{e^+ e^- }(s)$ is evaluated.
All the values of $R_{e^+ e^- }(s)$ are needed -- 
from the two-pion threshold to infinity.
The evaluation is based on the data compilation of Ref.~\cite{EJ}. 
The pQCD result for $R_{e^+ e^- }(s)$ is used in the integral 
where it can be trusted, and data in the rest of the energy interval. 
Resonances are included separately. 
In Fig.~\ref{Figdv}(b) we can see that various anQCD
models predict at low energies ($Q < 1.2$ GeV)
values which are significantly closer to the
experimental values than STPS's. Further,
STPS's lose any predictability at $Q < 1.2$ GeV,
mainly because of the vicinity of the unphysical
Landau pole in the pQCD coupling $a(Q^2)$.

\begin{figure}[htb]
\begin{minipage}[b]{.49\linewidth}
 \centering\epsfig{file=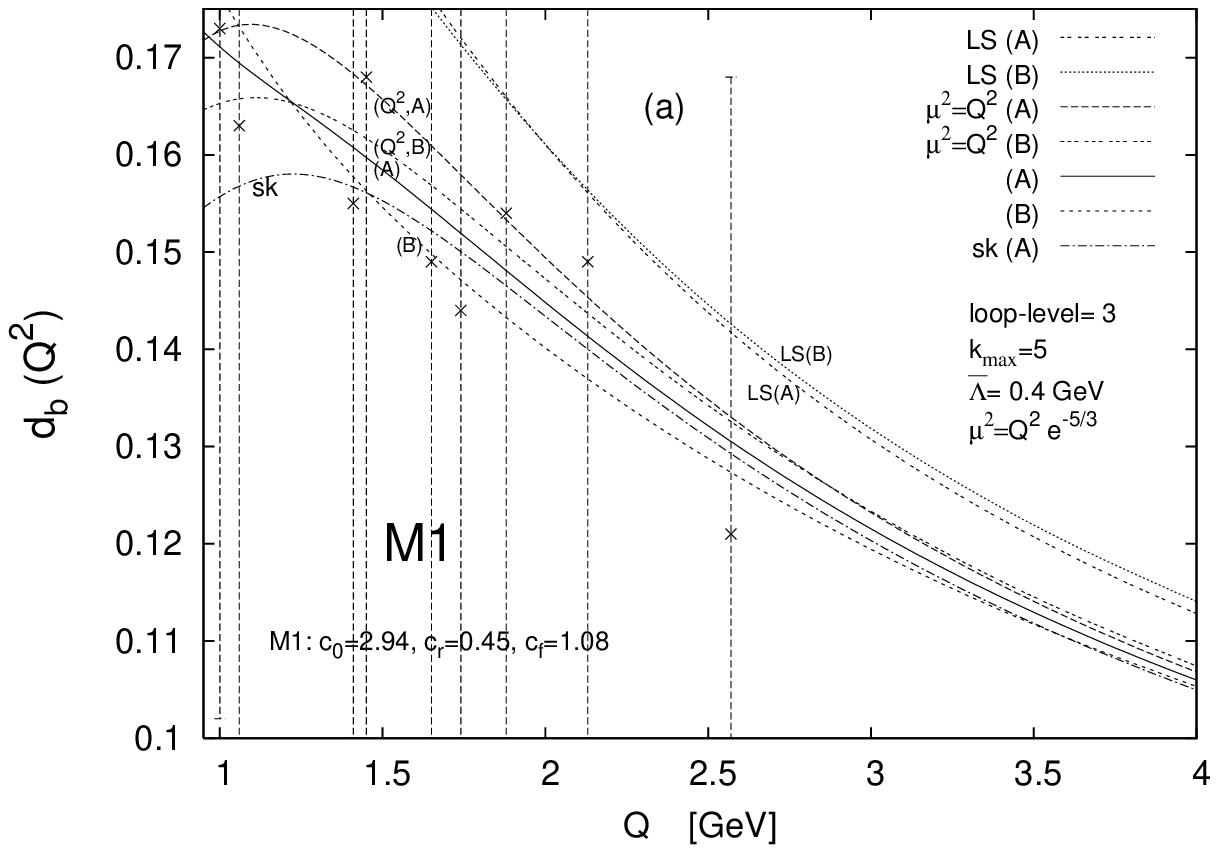,width=\linewidth}
\end{minipage}
\begin{minipage}[b]{.49\linewidth}
 \centering\epsfig{file=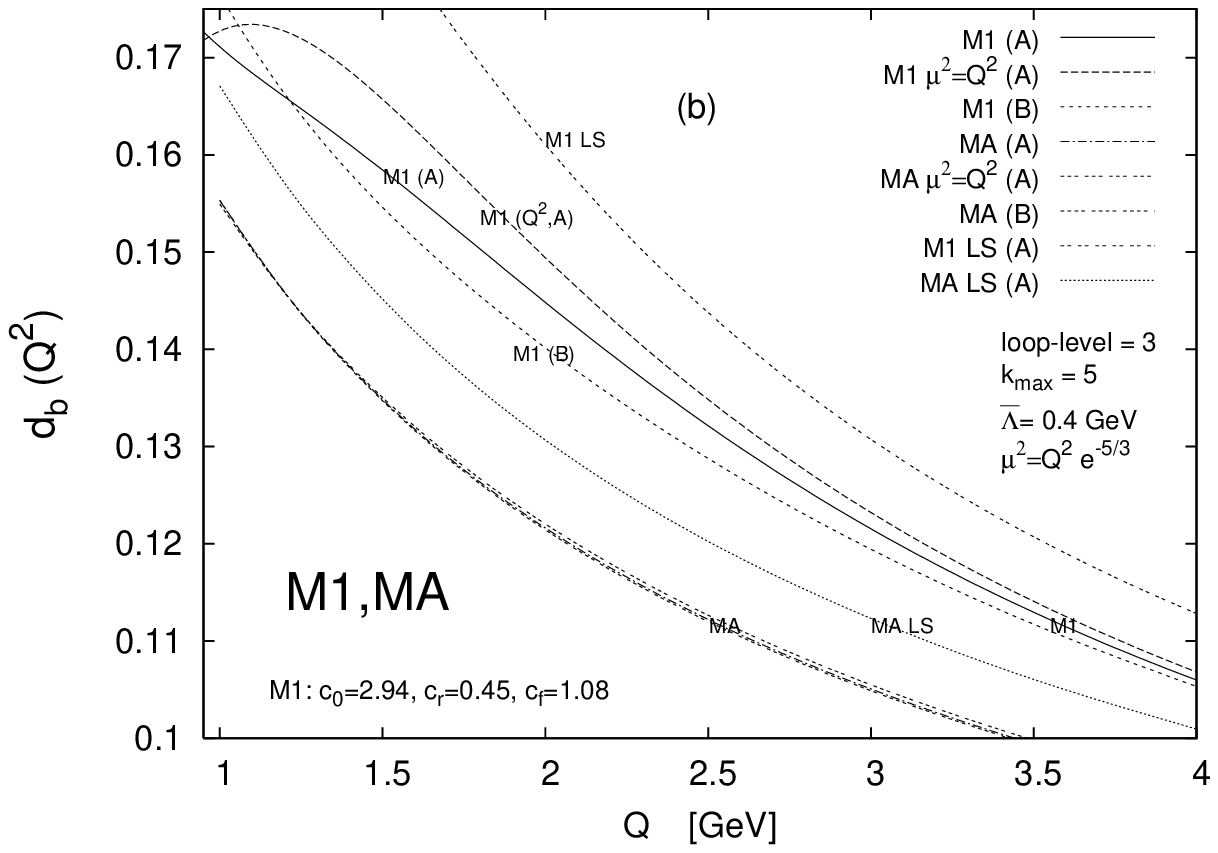,width=\linewidth}
\end{minipage}
\vspace{-0.4cm}
\caption{Bjorken polarized sum rule (BjPSR)
$d_b(Q^2)$ in (a) model M1, and (b) comparison of M1 and MA; 
in various RSch's and at various RScl's. The vertical lines in (a) represent
experimental data, with errorbars in general covering the entire
depicted range of values.}
\label{dbM1MAFig}
\end{figure}
In Fig.~\ref{dbM1MAFig}(a), we present results of 
calculation of BjPSR $d_b(Q^2)$ at low energies in model
M1, at loop-level=3 (and $k_{\rm max}=5$), in two different RSch's:
RSch A (\ref{ARSch}), and RSch B which
is the ''optimal'' RSch for $d_b(Q^2)$, i.e.,
$\beta_2^{\rm (B)}$ is obtained from the
requirement ${\widetilde t}_3 = 0$ for $d_b$,
Eqs.~(\ref{db22v})-(\ref{db2120v})
\be
\beta_2^{\rm (B)} = -30.2949 - 10.4415 \beta_0 + 
7.44582 \beta_0^2 \ .
\label{b2B}
\ee
At $n_f=3$ we have $\beta_2^{\rm (B)}(n_f=3)=-16.0938$.
The analytic couplings in RSch B are obtained
from those in RSch A by applying the 
looplevel=3 RSch-evolution equations (\ref{dA1dbll3}).
In addition, we present in Fig.~\ref{dbM1MAFig}(a) results
when the RScl in the beyond-the-LS terms
($Q_2^2, Q_3^2$) is increased from
$Q^2 \exp(-5/3)$ to $Q^2$ (note that coefficients
${\widetilde t}_2$ and ${\widetilde t}_3$ then
change accordingly). We see that at low
energies $Q < 2$ GeV the results in M1
change moderately but not insignificantly under 
the variation of RSch and RScl.
For comparison, we included the curve
obtained from the skeleton evaluation
(\ref{vsk}) in RSch A [with $Q_2^2=Q_3^2=Q^2 \exp(-5/3)$],
assuming that the skeleton expansion exists
in RSch A (which is probably not true).
We include the present experimental data,
with the crosses representing the central values; the errorbars
extend in general over the entire depicted range of values,
most of the experimental uncertainties are of the order of $\pm 0.1$.
The experimental data were deduced from Fig.~2 of 
Ref.~\cite{Deur:2004ti}, with the neutron decay parameter
value $|g_A| = 0.21158 \pm 0.00048$ taken from \cite{Yao:2006px}.
The present experimental errors are
too high to discriminate between various
evaluation methods.
In Fig.~\ref{dbM1MAFig}(b) we compare the
results for of MA and M1. 
The RSch- and RScl-dependence of
MA-results remains very weak in all the shown region. 

\begin{figure}[htb]
\begin{minipage}[b]{.49\linewidth}
 \centering\epsfig{file=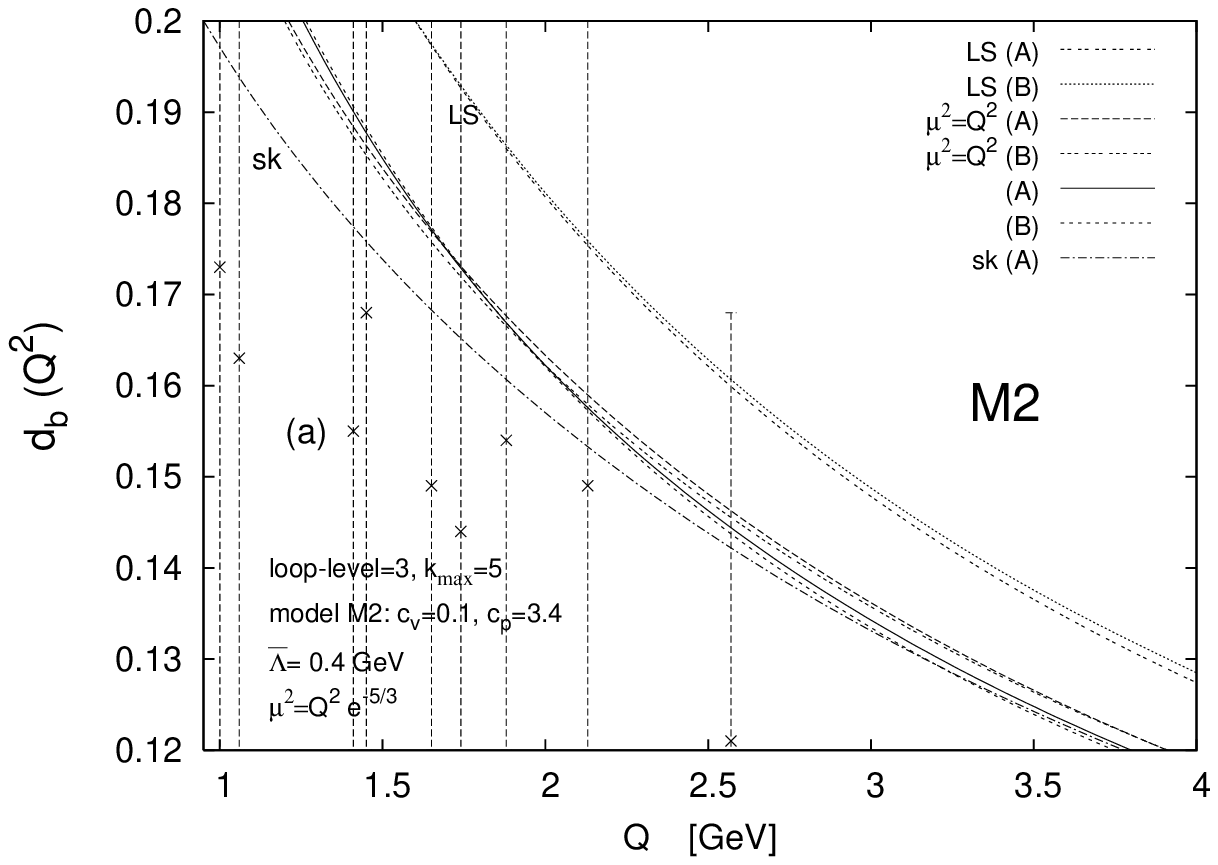,width=\linewidth}
\end{minipage}
\begin{minipage}[b]{.49\linewidth}
 \centering\epsfig{file=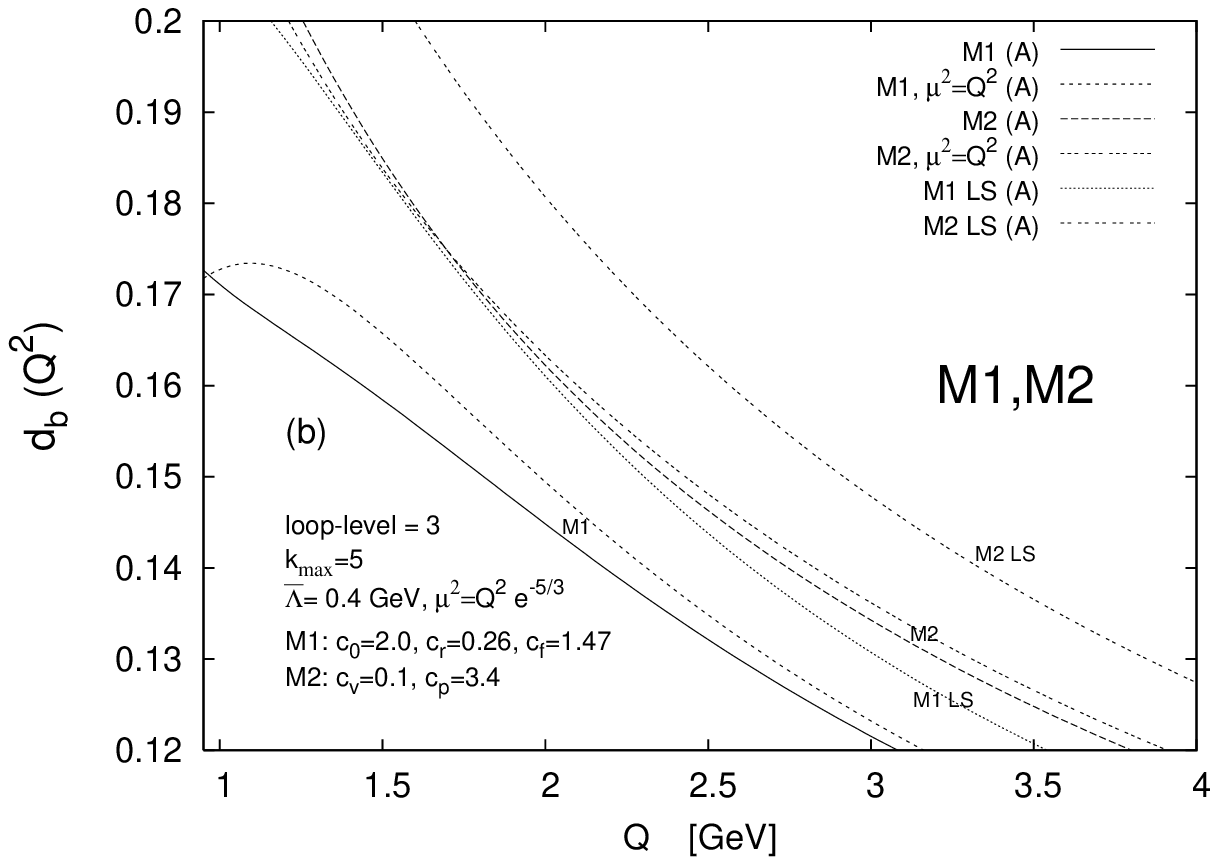,width=\linewidth}
\end{minipage}
\vspace{-0.4cm}
\caption{\footnotesize BjPSR
$d_b(Q^2)$ in (a) model M2, and (b) comparison of M2 and M1; 
at various RScl's (a,b) and in various RSch's (a). The vertical
lines in (a) represent the experimental data.}
\label{dbM2M1Fig}
\end{figure}
In Fig.~\ref{dbM2M1Fig}(a) we present the same type of
curves for M2 model. We see that the
RSch- and RScl-dependence in M2 remains
quite weak down to low energies. 
In Fig.~\ref{dbM2M1Fig}(b) we compare the results
of M2 and M1 models.
Only the curves in RSch A are presented in Fig.~\ref{dbM2M1Fig}(b).

Up until now, we applied the (skeleton-motivated) method
(\ref{TASbvgen}) for the evaluation of QCD
observables, in various anQCD models for $\A_1(\mu^2)$,
with the higher-order couplings $\tA_k$
($k \geq 2$) constructed by Eqs.~(\ref{tAn})
in a certain RSch (usually RSch A)
and equivalently the higher-order couplings 
$\A_k$ by Eqs.~(\ref{A2A3}) [Eqs.~(\ref{A2A3A4})
if loop-level=4].
There remains a question of how this method of
evaluation compares with the APT evaluation
approach of  Milton {\em et al.\/} and Shirkov 
\cite{Milton:1997mi,Sh}. We recall that the
APT approach was defined for the MA anQCD model,
and it consists of using the available
(NLO or ${\rm N}^2{\rm LO}$) STPS of an observable (\ref{Dpt})
and replacing there $a^k(Q^2) \mapsto \A_k(Q^2)^{\rm (MA)}$
($k \geq 1$), where the higher-order MA couplings
$\A_k(Q^2)^{\rm (MA)}$ were constructed according to
formula (\ref{MAAkdisp}). In the ${\rm N}^2{\rm LO}$ STPS
case [e.g., for $d_b(Q^2)$], this reads
\be
{\cal D}_{\text{APT}}(Q^2) =
\A_1(Q^2)^{\rm (MA)} + d_1 \A_2(Q^2)^{\rm (MA)} + 
d_2 \A_3(Q^2)^{\rm (MA)} \ .
\label{MSSSh}
\ee
The RSch is usually taken to be $\bMS$, but could in
principle be any RSch. One of the differences between
our and APT evaluation method here is 
the construction of the higher-order 
couplings $\A_k(Q^2)^{\rm (MA)}$
of the model MA, where comparison with our construction
has been made in Figs.~\ref{FigMSSShbMS} and
\ref{FigMSSShARSch} in Sec.~\ref{analytiz}. 
\begin{figure}[htb]
\begin{minipage}[b]{.49\linewidth}
 \centering\epsfig{file=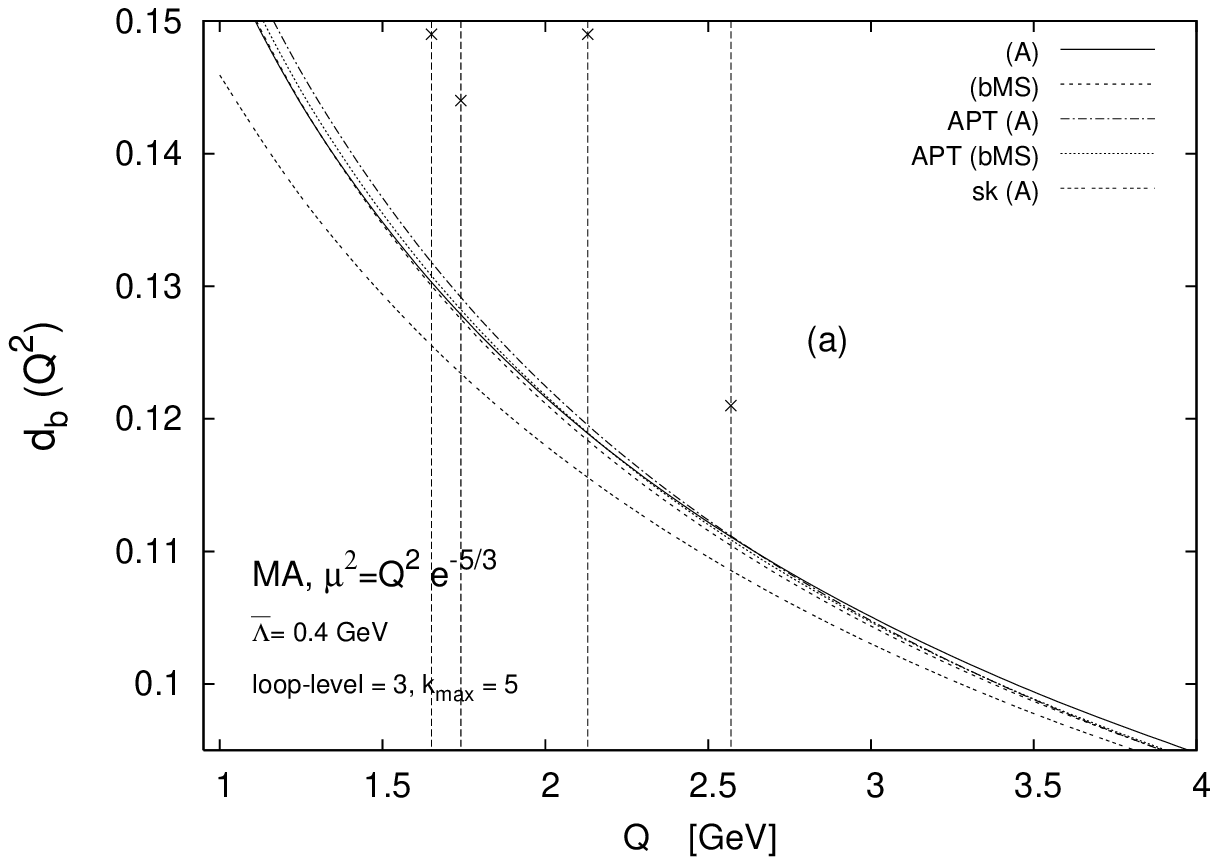,width=\linewidth}
\end{minipage}
\begin{minipage}[b]{.49\linewidth}
 \centering\epsfig{file=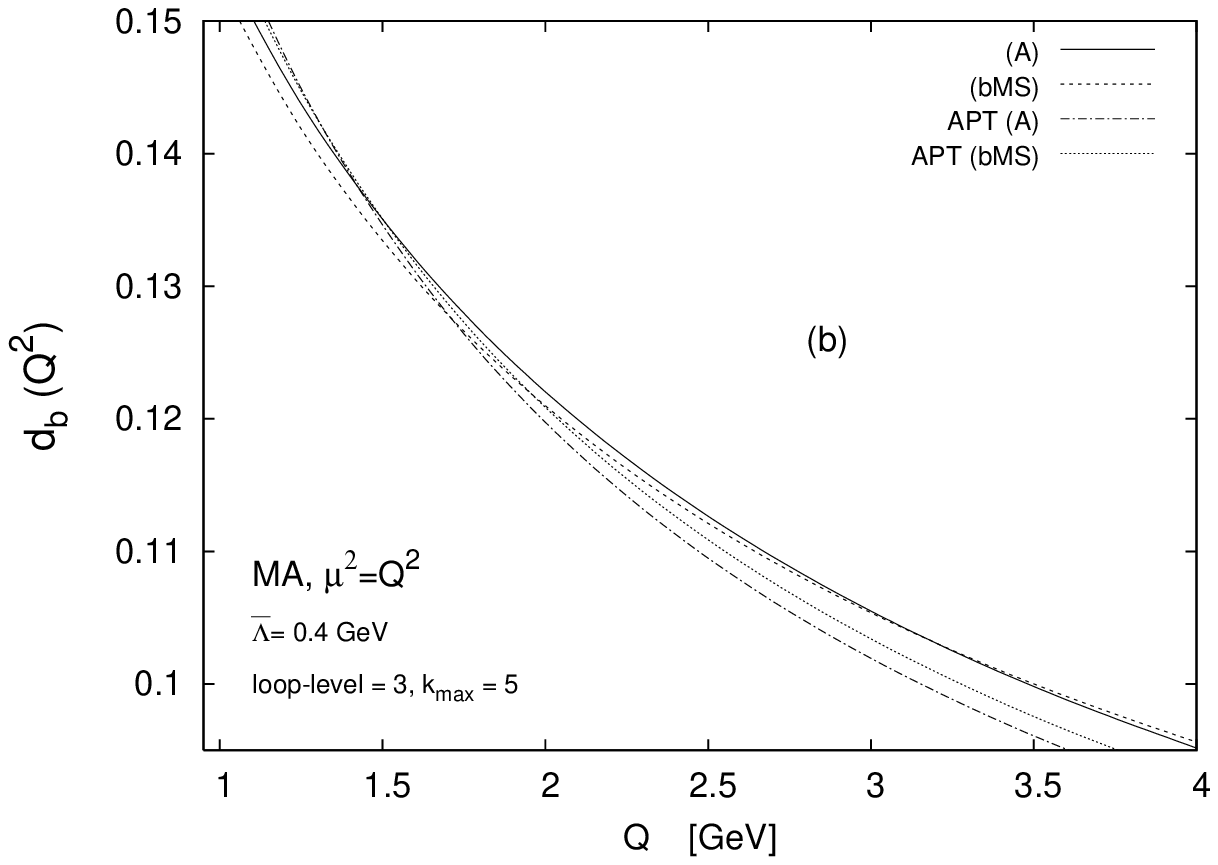,width=\linewidth}
\end{minipage}
\vspace{-0.4cm}
\caption{\footnotesize BjPSR
$d_b(Q^2)$ in the MA anQCD, with our evaluation method
(\ref{TASbvgen}) and that of  
Milton {\em et al.\/} and Shirkov 
\cite{Milton:1997mi,Sh} (APT), in RSch A (\ref{ARSch}),
and in $\bMS$; the RScl is chosen to be 
(a) $\mu^2 = Q^2 \exp(-5/3)$,
(b) $\mu^2 = Q^2$. The vertical
lines in (a) represent the experimental data.}
\label{dbMSSShFig}
\end{figure}
Another difference is that our evaluation method
(\ref{TASbvgen}) includes, in addition, the
leading-$\beta_0$ contributions to all orders.
We compare in
Figs.~\ref{dbMSSShFig}(a), (b) the results
of our method (\ref{TASbvgen}) and APT-method (\ref{MSSSh})
for BjPSR $d_b(Q^2)$,
in MA model, in two different RSch's: A
(\ref{ARSch}), and $\bMS$ (relevant for $\beta_2$ 
coefficient only). In Figs.~\ref{dbMSSShFig}(a) and (b)
the renormalization scale was taken to be
$\mu^2 = Q^2 \exp(-5/3)$ and $\mu^2=Q^2$,
respectively, in both 
beyond-the-LS terms of our approach
($\propto \tA_2, \tA_3$), and in
APT approach. The RSch-change from
RSch A to $\bMS$ was performed in our approach
according to the loop-level=3 evolution equations
(\ref{dA1dbll3}), while in APT approach
the corresponding values of $\beta_2$ were inserted
directly in (\ref{apt}) and thus in
all $\rho_k^{\rm (pt)}$'s. For additional
comparison, we included the skeleton evaluation
(in RSch A), Eq.~(\ref{vsk}).
We see in Figs.~\ref{dbMSSShFig} that 
in both our and APT evaluation approaches,
in MA anQCD model, the RSch- and RScl-dependence
of $d_b(Q^2)$ is very weak down to quite low energies.
More detailed inspection reveals that our evaluation
approach (\ref{TASbvgen}) gives for $d_b(Q^2)$
even somewhat less RScl-and RSch-dependent results
than APT approach.

\section{Summary and conclusions}
\label{summary}

In this work we suggested various models of
analytic QCD (anQCD), i.e., models for construction
of the anQCD coupling $\A_1(Q^2)$ which is an analytic
analog of the perturbative QCD coupling $a(Q^2) \equiv \alpha_s(Q^2)/\pi$.
The main reason why we suggest alternatives to 
the minimal analytic (MA) model, i.e., to the coupling
$\A^{\rm (MA)}_1(Q^2)$ of Shirkov and Solovtsov \cite{ShS},
is that it cannot correctly reproduce simultaneously various
higher-energy QCD observables on the one hand 
and the low-energy observable
$r_{\tau}$ (semihadronic $\tau$ decay rate ratio) on the other hand,
unless large masses of $u$, $d$ and $s$ quarks are introduced
\cite{Milton:2000fi}. The described alternative models (M1 and M2)
have $\A_1(Q^2)$ with additional dimensionless parameters in it,
which can be adjusted in order to modify the behavior at low $Q^2$.
Furthermore, we presented, for any anQCD model, an algorithm
which allows construction of higher-order analytic couplings
$\A_k(Q^2)$ ($k \geq 2$) which are the analytic analogs of 
$a^k(Q^2)$. In addition, we presented a method of evaluation
of Euclidean QCD observables in anQCD models, a method which is
(partly) motivated by the so-called skeleton expansion structure
but does not depend on the existence of such a skeleton expansion.
The evaluation method sums up all the leading-$\beta_0$
contributions (LS: leading-skeleton) 
and adds those contributions beyond the LS which are
known by the knowledge of a first few perturbation expansion
coefficients of the considered observable.
We tested this evaluation method, for three anQCD models,
in the case of the Adler function, semihadronic $\tau$
decay ratio, and the Bjorken polarized sum rule (BjPSR) at low
energies. The results show in general good stability under
variation of the renormalization scale and scheme
down to low energies $Q \sim 1$ GeV. We further carried
out comparison of our evaluation method with that of
Milton {\it et al.} (APT) \cite{Milton:1997mi,Sh,Milton:2000fi},
for the BjPSR, in the MA model where the latter method can be applied.
The two methods give results which at low energies 
differ in general by only a few per cent for this observable.

\begin{acknowledgments}
The authors acknowledge helpful communication with
F.~Jegerlehner.
This work was supported by Fondecyt grant 1050512 (G.C.),
Mecesup grant USA0108 and Conicyt Fellowship 3060106 (C.V.).
\end{acknowledgments}

\appendix

\section{Relevant coefficients of the skeleton-motivated expansion}
\label{app1}

In this Appendix, we present explicit formulas for the
coefficients $t_i^{(j)}$ appearing in the 
skeleton-motivated expansion (\ref{Dpt2b}), which is
a slightly reorganized form of expansion (\ref{D1-D4}).
We consider the case when, in $\bMS$ RSch
($\beta_k = {\overline b}_k = 
\sum {\overline b}_{kj} \beta_0^j$, $k \geq 2$)
and at RScl $\mu^2=Q^2$, the first three
coefficients in expansion (\ref{Dpt}) are
explicitly known (${\overline d}_j$, $j=1,2,3$),
and all the leading-$\beta_0$ parts
${\overline c}^{(1)}_{nn} {\beta_0}^n$ 
of coefficients ${\overline d}_n$ ($n \geq 1$) in 
expansion (\ref{dns}) are known
(we note that ${\overline c}^{(1)}_{n,-1} = 0$ in $\bMS$).
The RSch ($\beta_2, \beta_3, \ldots$) is chosen
and common to all terms ${\cal D}^{(j)}$ ($j \geq 1$),
and belongs to the class of RSch's of Eq.~(\ref{betak}).
The RScl's used in the resulting truncated versions
of ${\cal D}^{(j)}(Q^2)$ ($j \geq 2$) are $Q_j^2$,
they may be mutually different as each ${\cal D}^{(j)}(Q^2)$
(and $k_j$) is RScl-independent. For the RSch and the
RScl's we will use notations
\begin{eqnarray}
\delta b_{kj} & \equiv & b_{kj} - {\overline b}_{kj} \ ,
\label{dbs}
\\
Q_j^2  & \equiv & Q^2 \exp ( {\cal C}_j ) \ .
\label{Cjs}
\end{eqnarray} 
We then obtain for the coefficients $t_i^{(j)}$ of
expansion (\ref{Dpt2b}) the following expressions,
on the basis of relations (\ref{D1})-(\ref{k4}),
as well as (\ref{betak}), (\ref{dns}) and (\ref{RSchch}):
\begin{eqnarray}
t_2^{(2)} & = & {\overline t}_2^{(2)} = {\overline c}^{(1)}_{10} \ ,
\label{t22}
\\
t_3^{(2)} & = & {\overline t}_3^{(2)} - {\beta_0} \delta b_{22} +
{\beta_0} 2 {\overline c}^{(1)}_{10} {\cal C}_2 \ ,
\label{t32}
\\
t_3^{(3)} & = & {\overline t}_3^{(3)} - \delta b_{21} - 
\frac{1}{\beta_0} \delta b_{20} \ ,
\label{t33}
\end{eqnarray}
\begin{eqnarray}
t_4^{(2)} & = & {\overline t}_4^{(2)} + 
(b_{11} \beta_0 + b_{10}) \left( - \delta b_{22} + 
2 {\overline c}^{(1)}_{10} {\cal C}_2 \right) 
\nonumber\\
&& + \beta_0^2 \left( - \delta b_{22} 3 {\overline c}^{(1)}_{11}
- \frac{1}{2} \delta b_{33} + 3 \ovc^{(1)}_{10} \ovc^{(2)}_{11} {\cal C}_2
- \delta b_{22} 3 {\cal C}_2 + 3 \ovc^{(1)}_{10} {\cal C}_2^2 \right) \ ,
\label{t42}
\\
t_4^{(3)} & = & {\overline t}_4^{(3)} +
\beta_0 \left( - \delta b_{22} 2 \ovc^{(1)}_{10}
- \delta b_{21} 3 \ovc^{(1)}_{11} - \frac{1}{2} \delta b_{32} +
b_{11} \delta b_{22} \right) 
\nonumber\\
&& - \delta b_{20} (\ovc^{(1)}_{10} \ovc^{(2)}_{10}-\delta b_{21} )^{-1}
 \left( \ovc^{(1)}_{10} \ovc^{(2)}_{21} - \delta b_{22} 2 \ovc^{(1)}_{10} -
\delta b_{21} 3 \ovc^{(1)}_{11} - \frac{1}{2} \delta b_{32} -
b_{11} \ovc^{(1)}_{10} \ovc^{(2)}_{11} + b_{11} \delta b_{22} \right) 
\nonumber\\
&& + \beta_0 \left( \ovc^{(1)}_{10} \ovc^{(2)}_{10} - \delta b_{21}
- \frac{1}{\beta_0} \delta b_{20} \right) 3 {\cal C}_3 \ ,
\label{t43}
\end{eqnarray}
\begin{eqnarray}
t_4^{(4)} & = & {\overline t}_4^{(4)} +
\left( - \delta b_{21} 2 \ovc^{(1)}_{10}
- \delta b_{20} 3 \ovc^{(1)}_{11} - \frac{1}{2} \delta b_{31} +
b_{10} \delta b_{22} \right) 
\nonumber\\
&& + \delta b_{20} (\ovc^{(1)}_{10} \ovc^{(2)}_{10}-\delta b_{21} )^{-1}
 \left( \ovc^{(1)}_{10} \ovc^{(2)}_{21} - \delta b_{22} 2 \ovc^{(1)}_{10} -
\delta b_{21} 3 \ovc^{(1)}_{11} - \frac{1}{2} \delta b_{32} -
b_{11} \ovc^{(1)}_{10} \ovc^{(2)}_{11} + b_{11} \delta b_{22} \right) 
\nonumber\\
&& + \frac{1}{\beta_0} \left( - \delta b_{20} 2 \ovc^{(1)}_{10}
- \frac{1}{2} \delta b_{30} \right) \ .
\label{t44}
\end{eqnarray}
Here, ${\overline t}_i^{(j)}$ are the values of
the $t_i^{(j)}$ in $\bMS$ RSch and with RScl $\mu^2=Q^2$:
\ba
{\overline t}_2^{(2)} & = & \ovc^{(1)}_{10} \ ,
\label{bt22}
\\
{\overline t}_3^{(2)} & = & \beta_0 \ovc^{(1)}_{10} \ovc^{(2)}_{11} \ ,
\label{bt32}
\\
{\overline t}_3^{(3)} & = & \ovc^{(1)}_{10} \ovc^{(2)}_{10} \ ,
\label{bt33}
\\
{\overline t}_4^{(2)} & = & 
(b_{11} \beta_0 + b_{10}) \ovc^{(1)}_{10} \ovc^{(2)}_{11} +
\beta_0^2 \ovc^{(1)}_{10} \ovc^{(2)}_{22} \ ,
\label{bt42}
\\
{\overline t}_4^{(3)} & = & \beta_0 \ovc^{(1)}_{10}
( \ovc^{(2)}_{21} - b_{11} \ovc^{(2)}_{11} ) \ , 
\label{bt43}
\\
{\overline t}_4^{(4)} & = & \ovc^{(1)}_{10}
( \ovc^{(2)}_{20} - b_{10} \ovc^{(2)}_{11} ) \ . 
\label{bt44}
\ea
Coefficients $\ovc^{(2)}_{ij}$ appearing in the
above formulas can be obtained directly from  
coefficients $\ovc^{(1)}_{k \ell}$ by using
relations (\ref{c2ij}) in $\bMS$ scheme
(with RScl $\mu^2=Q^2$):
\ba
\ovc^{(1)}_{10} \ovc^{(2)}_{1j} & = & \ovc^{(1)}_{2j} - b_{1j} \ovc^{(1)}_{11} 
\quad (j=1,0,-1) \ ,
\nonumber\\
\ovc^{(1)}_{10} \ovc^{(2)}_{2j} & = & \ovc^{(1)}_{3j} - 
\frac{5}{2} b_{1,j-1} \ovc^{(1)}_{22} - b_{2j} \ovc^{(1)}_{11}
\quad (j=2,1,0,-1) \ .
\label{bc2ij}
\ea
Formulas (\ref{bc2ij}), (\ref{bt22})-(\ref{bt44}),
(\ref{t22})-(\ref{t44}), with notations (\ref{dbs}) and
(\ref{Cjs}), allow us to obtain all the coefficients $t_i^{(j)}$
of the skeleton-motivated expansion (\ref{Dpt2b})
in any chosen RSch and with chosen RScl's $Q_j^2$,
if we know in $\bMS$ RSch at RScl $\mu^2=Q^2$
all the leading-$\beta_0$ parts ${\ovc}^{(1)}_{nn} \beta_0^n$
of the expansion coefficients ${\overline d}_n =
\sum_0^n \ovc^{(1)}_{nk} \beta_0^k$ of observable
${\cal D}(Q^2)$ Eq.~(\ref{Dpt}), and we know
exactly the full coefficients ${\overline d}_j$ for $j=1,2,3$,
i.e., we know $\ovc^{(1)}_{jk}$ for $j=1,2,3$ and $k=0,\ldots,j$.
If, on the other hand, we do not know ${\overline d}_3$,
the above formulas can be applied for $t_i^{(j)}$
for $i=2,3$ only.

When the beyond-the-LS contributions in our approach (\ref{TASb})
are expressed in terms of $\tA_k$'s, Eq.~(\ref{TASbvgen}),
with the RScl choice (\ref{RScls}) [${\cal C}_j={\cal C}$], 
coefficients $\tlt_i$ can be expressed in terms of the
above coefficients $t^{(k)}_s$ via relations (\ref{A2A3A4})
between $\A_k$'s and $\tA_n$'s. After some straightforward
algebra, we obtain
\begin{eqnarray}
\tlt_2 & = & {\overline \tlt}_2 = {\overline c}^{(1)}_{10} \ ,
\label{tlt2}
\\
\tlt_3 & = & {\overline \tlt}_3 
 - {\beta_0} \delta b_{22} +
{\beta_0} 2 {\overline c}^{(1)}_{10} {\cal C} 
- \delta b_{21} - \frac{1}{\beta_0} \delta b_{20}
\ ,
\label{tlt3}
\end{eqnarray}
\begin{eqnarray}
\tlt_4 & = & {\overline \tlt}_4 + 
\beta_0^2 \left[  - \frac{1}{2} \delta b_{33}
- \delta b_{22} 3 \left( \ovc^{(1)}_{11} + {\cal C} \right) 
+ 3 {\cal C} \ovc^{(1)}_{10} \left( \ovc^{(2)}_{11} 
+ {\cal C} \right) \right]
\nonumber\\
&& 
+ \beta_0 \left[
- \frac{1}{2} \delta b_{32} 
+ \delta b_{22} \left( - 3  \ovc^{(1)}_{10} + \frac{5}{2}  b_{11} \right)
- \delta b_{21} 3 \ovc^{(1)}_{11} 
+ 3 {\cal C} \ovc^{(1)}_{10} \left(
\ovc^{(2)}_{10} - b_{11} \right) \right]
\nonumber\\
&& 
+ \left[
- \frac{1}{2} \delta b_{31} 
+ \frac{5}{2} b_{10} \delta b_{22}
+ \delta b_{21} \left(
- 3 \ovc^{(1)}_{10} + \frac{5}{2} b_{11} \right) 
- 3 \delta b_{20} \left( \ovc^{(1)}_{11} + {\cal C} \right)
- 3 b_{10} \ovc^{(1)}_{10} {\cal C} \right]
\nonumber\\
&&
+ \frac{1}{\beta_0} \left[
- \frac{1}{2} \delta b_{30} + \frac{5}{2} b_{10} \delta b_{21}
+ \delta b_{20} \left( 
- 3 \ovc^{(1)}_{10} + \frac{5}{2} b_{11} \right) \right]
+ \frac{1}{\beta_0^2} \frac{5}{2} b_{10} \delta b_{20} \ ,
\label{tlt4}
\end{eqnarray}
where ${\overline \tlt}_{i}$ are the values of $\tlt_i$
in $\bMS$ and with RScl $\mu^2 = Q^2$:
\ba
{\overline \tlt}_2 &=& {\overline c}^{(1)}_{10} \ ,
\label{btlt2}
\\
{\overline \tlt}_3 &=& 
\beta_0 \ovc^{(1)}_{10} \ovc^{(2)}_{11} 
+ \ovc^{(1)}_{10} ( \ovc^{(2)}_{10} - b_{11} )
- \frac{1}{\beta_0} \ovc^{(1)}_{10} b_{10} \ ,
\label{btlt3}
\\
{\overline \tlt}_4 &=& 
\beta_0^2  \ovc^{(1)}_{10} \ovc^{(2)}_{22} +
\beta_0 \ovc^{(1)}_{10} \left(
 \ovc^{(2)}_{21} - \frac{5}{2} b_{11} \ovc^{(2)}_{11} - 
{\overline b}_{22} \right) 
+ \ovc^{(1)}_{10} \left(
\ovc^{(2)}_{20} - \frac{5}{2} b_{10} \ovc^{(2)}_{11} 
- \frac{5}{2} b_{11} \ovc^{(2)}_{10}
+ \frac{5}{2} b_{11}^2 - {\overline b}_{21} \right) 
\nonumber\\
&& + \frac{1}{\beta_0} \ovc^{(1)}_{10} \left(
5 b_{11} b_{10} - \frac{5}{2} b_{10} \ovc^{(2)}_{10}
- {\overline b}_{20} \right) 
+ \frac{1}{\beta_0^2} \frac{5}{2} \ovc^{(1)}_{10} b_{10}^2 \ .
\label{btlt4}
\ea 

\section{Skeleton expansion}
\label{app2}

In this Appendix we will construct an expression for evaluation of QCD
space-like observables ${\cal D}(Q^2)$ (for any anQCD model)
which will be derived directly from the QCD skeleton expansion.
Here we will take the position that such an expansion exists
in the class of schemes with the QCD scale 
$\Lambda_{\cal C}^2 = \Lambda_{0}^2 \exp({\cal C})$ 
where $\Lambda_{0}$ is the so-called $V$-scheme scale
and ${\cal C}$ is an arbitrary $n_f$-independent constant,
and with $\beta_k$ of Eq.~(\ref{betak}) where $b_{kj}$ are
arbitrary constants. In this context, choosing the
$\bMS$ scale parameter ${\cal C} = {\overline {\cal C}} \equiv -5/3$
($\Lambda = \bL$) for scaling the RScl $\mu^2$
represents no additional restriction.
This expansion involves in the integrands the
characteristic functions $F_{\cal D}^{\cal {E}}(t_1,\!\ldots\!,t_n)$,
which are considered $n_f$-independent, and the
(singular) pQCD coupling $a(\mu^2)$. We replace $a(\mu^2)$ by
an anQCD coupling $\A_1(\mu^2)$ in the skeleton integrals
which makes the integrals unambiguous
\ba
{\cal D}(Q^2)_{\rm skel.} & = &
\int_0^\infty \frac{dt}{t}\: F_{\cal D}^{\cal E}(t) \: 
\A_1(t Q^2 e^{\overline {\cal C}})   
+ \sum_{n=2}^{\infty} s_{n-1}^{\cal D} 
\left[ \prod_{j=1}^{n} \!\int_0^{\infty}\!\frac{d t_j}{t_j} 
\A_1(t_j Q^2 e^{\overline {\cal C}}) \right]
F_{\cal D}^{\cal E}(t_1,\!\ldots\!,t_n) 
\label{sk1}		
\\
& = & {\cal D}^{\rm (LS)}(Q^2) + 
s_1^{\cal D} {\cal D}^{\rm (NLS)}(Q^2) +
s_2^{\cal D} {\cal D}^{( {\rm N}^2{\rm LS})}(Q^2) +
s_3^{\cal D} {\cal D}^{( {\rm N}^3{\rm LS})}(Q^2) + \cdots \ .
\label{sk2}
\ea
Here, $F_{\cal D}^{\cal {E}}(t_1,\!\ldots\!,t_n)$ are the
characteristic functions and have the normalizations
\begin{equation}
\int_0^\infty \frac{dt}{t} \ F_{\cal D}^{\cal E}(t)=1,
\qquad
\int \frac{dt_1}{t_1}\frac{dt_2}{t_2} \ 
F_{\cal D}^{\cal E}(t_1,t_2)=1, \  \dots,
\label{normcf}
\end{equation}
implying for the perturbative parts 
\ba
{\cal D}^{\rm (\kappa)}(Q^2)_{\rm pt} & = &
a^{n_{\kappa}} \left[ 1  + {\cal O}(a) \right] \ ,
\label{Dsnorm}
\ea
where $n_{\kappa}=1$ for $\kappa=$LS,
$n_{\kappa}=2$ for $\kappa=$NLS, etc.  
The perturbative part of $\A_1(\mu^2)$ is
$a(\mu^2)$ [$\A_1(\mu^2) = a(\mu^2) + {\rm NP}$, 
where NP involves non-analytic in $a=0$ functions of $a$,
cf.~Eq.~(\ref{tAntr})].
We will use RGE evolution series (\ref{avsast})
for expansion of $a(t e^{\overline {\cal C}} Q^2)$
around $a(\mu^2)\equiv a$ 
\ba
a(t e^{\overline {\cal C}} Q^2) &=& 
a + 
\sum_{n=1}^{\infty} {\widetilde a}_{n+1} \beta_0^n (- \ln {\cal T})^n
\nonumber\\
& = & a + a^2 \beta_0 (- \ln {\cal T}) + 
a^3 \left[ \beta_0^2 \ln^2 {\cal T} - \beta_1 \ln {\cal T} \right] +
a^4 \left[ - \beta_0^3 \ln^3 {\cal T} + 
\frac{5}{2} \beta_0 \beta_1 \ln^2 {\cal T} - 
\beta_2 \ln {\cal T} \right]
+ \cdots \ ,
\label{atvsa}
\ea
where ${\cal T} \equiv t Q^2 e^{\cal C}/\mu^2$,
and ${\widetilde a}_n$ are defined in Eq.~(\ref{tan}).
Using expansion (\ref{atvsa}) in
the leading-skeleton (LS) term in (\ref{sk1}), this term can be shown
to have the following expansion for its perturbative part:
\ba
{\cal D}^{\rm (LS)}(Q^2)_{\rm pt} & = &
a + a^2 \beta_0 \langle - \ln {\cal T} \rangle_{(1)} +
a^3 \left[ \beta_0^2 \langle (- \ln {\cal T})^2 \rangle_{(1)}
+ \beta_1 \langle - \ln {\cal T} \rangle_{(1)} \right] 
\nonumber\\ &&
+ a^4 \left[ \beta_0^3 \langle (- \ln {\cal T})^3 \rangle_{(1)}
+ \frac{5}{2} \beta_1 \beta_0 \langle (- \ln {\cal T})^2 \rangle_{(1)}
+ \beta_2 \langle - \ln {\cal T} \rangle_{(1)} \right]
+ {\cal O}(\beta_0^4 a^5) \ ,
\label{LSpt}
\ea
where we adhere to notations summarized in the following:
\ba
{\cal T} & = & \frac{t Q^2 e^{\overline {\cal C}}}{\mu^2} \ ,
\qquad a \equiv a(\mu^2) \ ,
\label{calt}
\\
\langle f(t_1,\!\ldots\!,t_n) \rangle_{(n)} &\equiv &
\prod_{j=1}^{n} \!\int_0^{\infty}\!\frac{d t_j}{t_j} 
F_{\cal D}^{\cal {E}}(t_1,\!\ldots\!,t_n)
f(t_1,\!\ldots\!,t_n) \ .
\label{avnot}
\ea
Requiring that the perturbative part of the LS-term absorb
all the leading-$\beta_0$ parts of ${\cal D}(Q^2)_{\rm pt}$
[see Eqs.~(\ref{Dpt})-(\ref{dns})] implies that
\be
\langle (- \ln {\cal T})^n \rangle_{(1)} = c^{(1)}_{nn}
\qquad (n=0,1,2,\ldots) \ .
\label{av1}
\ee
This, in conjunction with expansion (\ref{LSpt}),
implies that ${\cal D}^{\rm (LS)}(Q^2)_{\rm pt}$
is precisely ${\cal D}^{(1)}(Q^2)_{\rm pt}$
of construction in Sec.~\ref{skmotexp}, Eq.~(\ref{D1}),
i.e., we really have for ${\cal D}^{(1)}(Q^2)_{\rm pt}$
the resummed form (\ref{LS1}). 
Taylor expansion of $\A_1(t e^{\overline {\cal C}} Q^2)$ around
$Q^2$ is completely analogous to expansion (\ref{atvsa})
\ba
\lefteqn{
\A_1(t e^{\overline {\cal C}} Q^2) = 
\A_1 + \sum_{n=1}^{\infty} \tA_{n+1} \beta_0^n (- \ln {\cal T})^n
}
\nonumber\\
& = &
\A_1 + \A_2 \beta_0 (- \ln {\cal T}) + 
\A_3 \left[ \beta_0^2 \ln^2 {\cal T} - \beta_1 \ln {\cal T} \right] +
\A_4 \left[ - \beta_0^3 \ln^3 {\cal T} + 
\frac{5}{2} \beta_0 \beta_1 \ln^2 {\cal T} - 
\beta_2 \ln {\cal T} \right]
+ \cdots \ ,
\label{AtvsA}
\ea
where $\tA_k \equiv \tA_k(\mu^2)$ and
$\A_k \equiv \A_k(\mu^2)$. 
In the last identity we used the fact that $\tA_n$'s
appear on the left-hand side of RGE's (\ref{AkRGEtr}), 
which are analogous
to pQCD RGE's with ${\widetilde a}_n$'s on the
left-hand side [analogy valid up to terms
${\cal O}(\tA_{n_{\rm m}})$ where $n_{\rm m}=$loop-level]
when the correspondence $a^k \leftrightarrow \A_k$ is made.
Eq.~(\ref{AtvsA}) implies
for the (full analytic) LS-term of the skeleton
expansion (\ref{sk1}) a nonpower analytic expansion
\ba
{\cal D}^{\rm (LS)}(Q^2) & \equiv &
\int_0^\infty \frac{dt}{t}\: F_{\cal D}^{\cal {E}}(t) \: 
\A(t Q^2 e^{\overline {\cal C}}) 
\nonumber\\ 
& = &
\A_1 + \sum_{n=1}^{\infty} \tA_{n+1} \beta_0^n 
\langle (- \ln {\cal T})^n \rangle_{(1)}
\nonumber\\ 
& = & \A_1 + \A_2 \beta_0 \langle - \ln {\cal T} \rangle_{(1)} +
\A_3 \left[ \beta_0^2 \langle (- \ln {\cal T})^2 \rangle_{(1)}
+ \beta_1 \langle - \ln {\cal T} \rangle_{(1)} \right] 
\nonumber\\ &&
+ \A_4 \left[ \beta_0^3 \langle (- \ln {\cal T})^3 \rangle_{(1)}
+ \frac{5}{2} \beta_1 \beta_0 \langle (- \ln {\cal T})^2 \rangle_{(1)}
+ \beta_2 \langle - \ln {\cal T} \rangle_{(1)} \right]
+ {\cal O}(\A_5) \ ,
\label{LSanexp}
\ea
which is just the analyticized analog [according
to the rule (\ref{analyt})] of perturbation expansion
(\ref{LSpt}) and (\ref{D1}).

Now we will investigate the beyond-the-LS contributions
of the skeleton expansion (\ref{sk1}). 
In view of normalization conditions (\ref{Dsnorm}),
it follows immediately that 
\be
s_1^{\cal D} = c^{(1)}_{10} \ ,
\label{s1}
\ee
which is just the coefficient $k_2$ (\ref{DD1})
in the approach of Sec.~\ref{skmotexp}.
In analogy with the LS-part, we now require that 
${\cal D}^{\rm (NLS)}(Q^2)_{\rm pt}$ be such as
to absorb all the leading-$\beta_0$ parts of the difference 
$(1/s_1^{\cal D}) 
[ {\cal D}(Q^2) - {\cal D}^{\rm (LS)}(Q^2)]_{\rm pt}$. 
In completely analogous way as before, we can show that
this is equivalent to
\ba
(-1)^n {\big \langle} \ln^n {\cal T}_1  +
\ln^{n-1} {\cal T}_1 \ln {\cal T}_2 + \cdots
+ \ln^n {\cal T}_2 {\big \rangle}_{(2)} & = & c^{(2)}_{nn} 
\quad (n=1,2 \ldots) \ ,
\label{av2}
\ea
where ${\cal T}_j = t_j Q^2 e^{\overline {\cal C}}/\mu^2$,
and coefficients $c^{(2)}_{nn}$ are defined in
Eqs.~(\ref{DD1})-(\ref{c2ij}). These coefficients are
known if the perturbative coefficients $d_j$ (\ref{Dpt}) are
known. The (nonpower) expansion in $\A_k \equiv \A_k(\mu^2)$
of the NLS-term is then
\ba
s_1^{\cal D} {\cal D}^{\rm (NLS)}(Q^2) & = &
c^{(1)}_{10} {\Big \{} \A_1^2 + \A_1 \A_2 \beta_0 c^{(2)}_{11} +
\A_1 \A_3 \left[ \beta_0^2 c^{(2)}_{22} + \beta_1 c^{(2)}_{11} \right]
\label{skNLS1}
\\ &&
+ \left[ \A_2^2 - \A_1 \A_3 \right] \beta_0^2 
\langle \ln {\cal T}_1 \ln {\cal T}_2 \rangle_{(2)} +
{\cal {O}}(\A_1 \A_4, \A_2 \A_3, \ldots) 
{\Big \}} \ .
\label{skNLS2}
\ea
The last term in brackets has a coefficient $\propto
\langle \ln {\cal T}_1 \ln {\cal T}_2 \rangle_{(2)}$
which cannot be obtained on the basis of the perturbative
coefficients $d_j$ (\ref{Dpt}). 
The perturbative part of this last term is zero.
We know $c^{(1)}_{10}$ if we know
the NLO coefficient $d_1$ 
of the perturbation expansion (\ref{Dpt}) of observable
${\cal D}(Q^2)$; for the knowledge of $c^{(2)}_{11}$
we need, in addition, the knowledge of $d_2$, and for
$c^{(2)}_{22}$ the knowledge of $d_3$.

We now continue analogously one step further. 
In view of the normalization conditions (\ref{Dsnorm}),
it follows immediately  
\be
s_2^{\cal D} = c^{(1)}_{10} \left( c^{(2)}_{10} + 
\frac{1}{\beta_0} c^{(2)}_{1,-1} \right) \ ,
\label{s2}
\ee
which is identical to the coefficient $k_3$ (\ref{k3})
in the approach of Sec.~\ref{skmotexp}.
We require that the third (${\rm N}^2 {\rm LS}$) term 
$s^{\cal D}_2 {\cal D}^{({\rm N}^2{\rm LS})}(Q^2)$
in skeleton expansion (\ref{sk2}) satisfy the condition:
${\cal D}^{({\rm N}^2{\rm LS})}(Q^2)_{\rm pt}$ be such as to absorb
all the leading-$\beta_0$ parts of the difference
$(1/s_2^{\cal D}) 
[ {\cal D}(Q^2) - {\cal D}^{\rm (LS)}(Q^2) -
s^{\cal D}_1 {\cal D}^{\rm (NLS)}(Q^2)]_{\rm pt}$.
This then implies 
\be
\langle - \ln {\cal T}_1 - \ln {\cal T}_2 - \ln {\cal T}_3
\rangle_{(3)} = c^{(3)}_{11} \ ,
\label{c311sk}
\ee
where $c^{(3)}_{11}$ is given in Eq.~(\ref{c311}); and
similarly for higher terms ($c^{(3)}_{22}$, etc.). The
(nonpower) expansion in $\A_k \equiv \A_k(\mu^2)$
of the ${\rm N}^2 {\rm LS}$-term is then
\ba
s_2^{\cal D} {\cal D}^{({\rm N}^2 {\rm LS})}(Q^2) & = &
s_2^{\cal D} {\Big \{} \A_1^3 + \A_1^2 \A_2 \beta_0 c^{(3)}_{11}
+ {\cal {O}}(\A_1^2 \A_3, \A_1 \A_2^2, \ldots) 
{\Big \}} \ .
\label{skNNLS}
\ea
We know the quantity $s_2^{\cal D}$ if we know
the coefficients $d_1$ and $d_2$ in
the perturbation expansion (\ref{Dpt}) of observable
${\cal D}(Q^2)$; for the knowledge of $c^{(3)}_{11}$
we need, in addition, the knowledge of $d_3$.

Normalization conditions (\ref{Dsnorm}) now imply that
the coefficient $s_3^{\cal D}$ of the ${\rm N}^3 {\rm LS}$-term
in the skeleton expansion (\ref{sk1})-(\ref{sk2}) is
\ba
s_3^{\cal D} & = &
c^{(1)}_{10} \left[
c^{(2)}_{20} - b_{10} c^{(2)}_{11} - 
\frac{c^{(2)}_{1,-1}}{c^{(2)}_{10}} 
( c^{(2)}_{21} - b_{11} c^{(2)}_{11} )
+ \frac{1}{\beta_0} c^{(2)}_{2,-1} \right] \ ,
\label{s3}
\ea 
which is identical to the coefficient $k_4$ (\ref{k4})
in the approach of Sec.~\ref{skmotexp}. The 
(nonpower) expansion in $\A_k \equiv \A_k(\mu^2)$ of the
${\rm N}^3 {\rm LS}$-term is then
\ba
s_3^{\cal D} {\cal D}^{({\rm N}^3 {\rm LS})}(Q^2) & = &
s_3^{\cal D} \A_1^4 +
{\cal {O}}(\A_1^3 \A_2, \A_1^2 \A_3, \ldots) \ .
\label{skNNNLS}
\ea
We know the quantity $s_3^{\cal D}$ if we know
the coefficients $d_1$, $d_2$ and $d_3$ in
the perturbation expansion (\ref{Dpt}) of observable
${\cal D}(Q^2)$. 

Finally, we can combine the LS-term (\ref{LS2}),
whose characteristic function
is usually known, with all the beyond-the-LS
terms written hitherto (\ref{skNLS1})-(\ref{skNNNLS})
which are known if $d_1$, $d_2$ and $d_3$ are known;
since each of these terms is RScl-independent, we can use
in the most general case various (space-like) RScl's
$Q_j^2 = Q^2 \exp( {\cal C}_j )$ as Eq.~(\ref{Cjs})
($j=2,3,4$ for the NLS, ${\rm N}^3 {\rm LS}$
and ${\rm N}^3 {\rm LS}$ terms, respectively). This
then results in
\ba
{\cal D} & = &
{\cal D}^{\rm (LS)}(Q^2) + t^{(2)}_2 \left[ \A_1(Q_2^2) \right]^2 +
\left\{ t^{(2)}_3 \A_1(Q_2^2) \A_2(Q_2^2) +
t^{(3)}_3 \left[ \A_1(Q_3^2) \right]^3 \right\} 
\nonumber\\
&& + 
\left\{ t^{(2)}_4 \A_1(Q_2^2) \A_3(Q_2^2) +
t^{(3)}_4 \left[ \A_1(Q_3^2) \right]^2 \A_2(Q_3^2) +
t^{(4)}_4 \left[ \A_1(Q_4^2) \right]^4 \right\} 
\nonumber\\
&&+
\left\{ \left[ \A_2(Q_2^2) \right]^2 - \A_1(Q_2^2) \A_3(Q_2^2) \right\} 
c^{(1)}_{10} \beta_0^2 
\langle \ln {\cal T}_1 \ln {\cal T}_2 \rangle_{(2)} +
{\cal {O}}(\A_1^5, \A_1^3 \A_2, \ldots) \ ,
\label{vsk}
\ea  
where the coefficients $t^{(i)}_{(j)}$ are
precisely those given in Appendix \ref{app1},
Eqs.~(\ref{t22})-(\ref{bt44}). Therefore, the evaluation
method presented in the present Appendix, which is a
representation of an assumed skeleton expansion 
(\ref{sk1})-(\ref{sk2}), reduces to the evaluation method
presented in Sec.~\ref{skmotexp} when the following
replacements are made:
\be
\left[ \A_1(\mu^2) \right]^{k_1} 
\left[ \A_2(\mu^2) \right]^{k_2}
\cdots 
\left[ \A_s(\mu^2) \right]^{k_s} \mapsto
\A_{k_1 + 2 k_2 \cdots + s k_s}(\mu^2) \ .
\label{vsktov}
\ee
In the present method, the coefficient at the 
last term in brackets in 
expression (\ref{vsk}) can be evaluated only if certain
assumptions about the NLS characteristic function
$F_{\cal D}^{\cal {E}}(t_1,t_2)$ are made.
For simplicity, we will make the factorization assumption
\ba
F_{\cal D}^{\cal {E}}(t_1,t_2) &=&
w_{\cal D}^{\cal {E}}(t_1) w_{\cal D}^{\cal {E}}(t_2)
\Rightarrow
\langle \ln {\cal T}_1 \ln {\cal T}_2 \rangle_{(2)} = 
\frac{1}{4} \left( c^{(2)}_{11} \right)^2 \ ,
\label{factF}
\ea
where the last identity is obtained on the basis of
identity (\ref{av2}) for $n=1$.

Similarly as the skeleton-motivated 
evaluation method (\ref{TASa})-(\ref{TASb}),
the skeleton evaluation method (\ref{vsk}) can
be performed in principle at any chosen RScl's $Q_j$ and
in any RSch of the class (\ref{betak}).
This method was denoted as 'v1' in Ref.~\cite{Cvetic:2006mk}.
However, skeleton method (\ref{vsk}) makes sense
only if the skeleton expansion (\ref{sk2}) really exists.
If the latter exists, it probably does so only in a specific
('skeleton') scheme \cite{Gardi:1999dq,Brodsky}. In contrast, the
skeleton-motivated evaluation method (\ref{TASa})-(\ref{TASb})
does not rely on the existence of the skeleton expansion.

\section{Leading-skeleton characteristic functions in the space-like and time-like form}
\label{app3}

In this Appendix we summarize the knowledge of the 
LS characteristic functions for the space-like observables 
${\cal D}(Q^2)$. In the space-like formulation (\ref{LS2}),
which involves the space-like coupling $\A_1$, the characteristic
function can be obtained from the knowledge of the leading-$\beta_0$
coefficients $c^{(1)}_{nn}$ -- cf.~Eqs.~(\ref{Dpt}) and (\ref{dns}),
following the formalism of Neubert \cite{Neubert}. 

For example, in the case of the Bjorken polarized sum rule (BjPSR)
$d_b(Q^2)$, the leading-$\beta_0$ coefficients
were obtained in Ref.~\cite{Broadhurst:1993ru}. In $\bMS$ RSch and
at RScl $\mu^2 = Q^2 \exp({\overline {\cal C}})$
(we use $\Lambda = \bL$ throughout, i.e.,
${\cal C} = {\overline {\cal C}} \equiv -5/3$) , they are
\be 
c^{(1)}_{nn} = n! \left[ 
\frac{8}{9} + \frac{4}{9} (-1)^n 
- \frac{5}{18} \frac{1}{2^n} 
- \frac{1}{18} \frac{1}{2^n} (-1)^n \right] \quad
(n=0,1, \ldots) \ .
\label{c1nnBj}
\ee
This implies that the (leading-$\beta_0$)
Borel transform is
\ba
{\hat S}_b(u; Q^2; \mu^2= Q^2 e^{\overline {\cal C}}) 
& \equiv &
\sum_{n=0}^{\infty} \frac{1}{n!} c^{(1)}_{nn} u^n
= \frac{1}{3} \frac{(3 + u)}{(1 - u^2) (1 - u^2/4)} \ .
\label{BTBj}
\ea 
The renormalon poles are at $u=\pm 1, \pm 2$.
The LS characteristic function appearing in
(\ref{LS2}) is then obtained by the general formula
\be
F_{\cal D}^{\cal {E}}(\tau) = \frac{1}{2 \pi i}
\int_{u_0 - i \infty}^{u_0 + i \infty} du \: 
{\hat S}_{\cal D}(u) \tau^u \ ,
\label{FLSBj1}
\ee
where $u_0$ is any real number closer to the origin than the
leading renormalon ($-1 < u_0 < 1$). We can choose
$u_0=0$ and introduce a new integration variable 
$r = - i u$. The integral, with ${\hat S}(u)$ of
Eq.~(\ref{BTBj}), then reduces to
\be
F_{b}^{\cal {E}}(\tau) = \frac{2}{3 \pi}
\int_{-\infty}^{+\infty} dr \: e^{i r \ln \tau}
\frac{ (3 + i r)}{(r+i)(r-i)(r + 2 i)(r - 2 i)} \ ,
\label{FLSBj2}
\ee
which can be calculated by the use of the Cauchy theorem 
in the complex $r$-plane:
when $\tau > 1$, we close the path with a large
semicircle in the upper half plane; when $\tau < 1$,
in the lower half plane. This gives us the result
\ba
F_{b}^{\cal {E}}(\tau) = 
\left\{
\begin{array}{ll}
\frac{8}{9} \tau \left( 1 - \frac{5}{8} \tau \right) 
& \tau \leq 1 \\ 
\frac{4}{9 \tau} \left( 1 - \frac{1}{4 \tau} \right)
& \tau \geq 1
\end{array}
\right\} \ ,
\label{FLSBj3}
\ea
which we already used in \cite{Cvetic:2006mk}.

The LS characteristic function for the
Adler function $d_v(Q^2)$ was obtained in Ref.~\cite{Neubert},
on the basis of the large-$\beta_0$ expansion
of the Borel transform of $d_v$ obtained in
Refs.~\cite{Broadhurst:1992si,Beneke:1993ee}

\ba
F_{v}^{\cal {E}}(t) &=& 
2 C_F t \left[ \left( \frac{7}{4} - \ln t \right) t +
		  (1 + t) \left( {\rm PolyLog}_2(-t) + 
\ln t \ln (1+t) \right) \right] \quad (t \leq 1)
\label{FLSvIR}
\\
& = &
2 C_F \left[ t ( 1 + \ln t) + 
\left( \frac{3}{4} + \frac{1}{2} \ln t \right)  +
t ( 1 + t) \left( {\rm PolyLog}_2(-1/t) - \ln t \ln(1 + 1/t) 
\right) \right] \quad (t \geq 1) \ ,
\label{FLSvUV}
\ea
where $C_F = (N_c^2 - 1)/(2 N_c) = 4/3$.

The semihadronic $\tau$ decay ratio $r_{\tau}$ is a
time-like observable. The LS term of $r_{\tau}$
can be obtained from the LS-term of the Adler function
on the basis of the relation
\ba
r_{\tau}(\Delta S=0,m_q=0) & = & \frac{1}{2 \pi}
\int_{-\pi}^{+\pi} d \phi \: (1 + e^{i \phi})^3 (1 -e^{i \phi})
d_v(Q^2=m_{\tau}^2 e^{i \phi}) \ .
\label{rtaudv}
\ea
This implies for the LS term of $r_{\tau}(\Delta S=0,m_q=0)$
\be
r_{\tau}(\Delta S=0,m_q=0)^{\rm (LS)} = 
\int_0^\infty \frac{dt}{t}\: F_{r}^{\cal {M}}(t) \: 
\tlA_1 (t e^{\cal C} m_{\tau}^2) \ ,
\label{LSrt}
\ee
where $\tlA_1$ is the time-like coupling appearing
in Eqs.~(\ref{tlA1A1})-(\ref{tlA1rho1}), and superscript
${\cal {M}}$ in the characteristic function means that it is
Minkowskian (time-like). The latter was obtained in 
Ref.~\cite{Neubert2}\footnote{
We use a different normalization,
so an additional factor of $t/4$ appears, in comparison to \cite{Neubert2}.}
\ba
F_{r}^{\cal {M}}(t) & = &
4 C_F \; t {\big [} 
4 - \frac{73}{12} t - \frac{23}{24} t^2 - \frac{259}{432} t^3  -
2 \; {\rm PolyLog}_3(-t) - 3 \zeta(3) 
\nonumber\\
&&+ \left(  \frac{17}{6} t + \frac{1}{3} t^2 + {\rm PolyLog}_2(-t) \right) 
\ln t +
\left( \frac{3}{4} t^2 + \frac{1}{6} t^3 \right) \ln^2 t
\nonumber\\
&&- \left( \frac{11}{6} + 3 t + \frac{3}{2} t^2 + 
\frac{1}{3} t^3 \right) \left( \ln t \ln(1+t) +
{\rm PolyLog}_2(-t) \right) {\big ]} \quad (t \leq 1) \ ,
\label{FLSrIR}
\\
F_{r}^{\cal {M}}(t)
& = &
4 C_F  {\big [} - \frac{575}{216}t + \frac{37}{48}  
- \frac{17}{12} t^2
- \frac{1}{3} t^3 + 2 t \; {\rm PolyLog}_3(-1/t) 
\nonumber\\
&&-
\left( \frac{85}{36} t - \frac{1}{4} + \frac{4}{3} t^2
+ \frac{1}{3} t^3 - t \; {\rm PolyLog}_2(-1/t) \right) \ln t
\nonumber\\
&& +  
\left( \frac{11}{6} t + 3 + \frac{3}{2} t^3 +
\frac{1}{3} t^4 \right) \left(
\ln t \ln (1 + 1/t) - {\rm PolyLog}_2(-1/t) \right)
{\big ]} \quad (t \geq 1) \ .
\label{FLSrUV}
\ea
Here, ${\rm PolyLog}_n$ is the polylogarithm function of $n$'th
order (using notation of \cite{math}).

\begin{figure}[htb] 
\centering
\epsfig{file=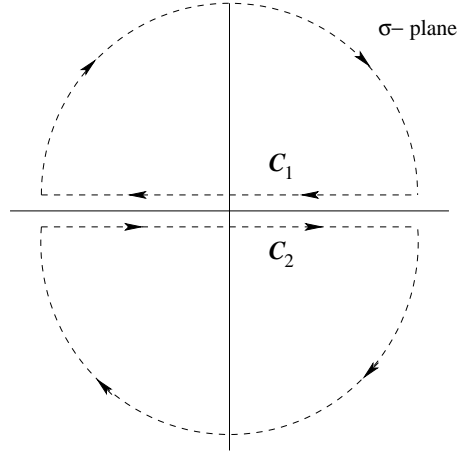,width=6.cm,height=6.cm}
\vspace{-0.4cm}
 \caption{\footnotesize  The path of integration of integral (\ref{Cauchy}) in the complex $\sigma$-plane: $\sigma > 0$
semiaxis is the cut of $\A_1(-\sigma e^{\cal C})$ factor,
and $\sigma < 0$ is the cut of the 
$[{\cal F}_{\cal D}(\sigma/Q^2) - {\cal F}_{\cal D}(0)]$
factor in the integral.}
\label{C1C2}
 \end{figure}

The LS part of any space-like observable ${\cal D}(Q^2)$ can be
written in two equivalent forms -- the form involving the space-like
coupling $\A_1$, Eq.~(\ref{LS2}), and the form involving the time-like
$\tlA_1$ of Eqs.~(\ref{tlA1A1})-(\ref{tlA1rho1})
\be
{\cal D}^{\rm (LS)}(Q^2) =
\int_0^\infty \frac{dt}{t}\: F_{\cal D}^{\cal {E}}(t) \: 
\A_1 (t e^{\cal C} Q^2) =
\int_0^\infty \frac{dt}{t}\: F_{\cal D}^{\cal M}(t) \: 
\tlA_1 (t e^{\cal C} Q^2) \ ,
\label{LS3}
\ee
where the superscript ${\cal M}$ 
stands for the ``Minkowskian'' (time-like) formulation, and the
two characteristic functions are related via relations
\ba
F_{\cal D}^{\cal M}(t) & = & 
- \pi \frac{d}{d \ln t} {\cal F}_{\cal D}(t) =
t \: \int_0^\infty \frac{dt'}{(t'+t)^2} \: F_{\cal D}^{\cal {E}}(t')  
\label{LSME}
\\
F_{\cal D}^{\cal {E}}(t) & = & 
{\rm Im} {\cal F}_{\cal D}(-t - i \varepsilon) \quad {\rm where:}
\quad {\cal F}_{\cal D}(t) \equiv \frac{1}{\pi}  \: \int_0^\infty
\frac{d \tau}{(\tau +t)} \: F_{\cal D}^{\cal {E}}(\tau) \ .
\label{LSEM}
\ea
Identity (\ref{LSEM}) is a direct consequence of the
definition of ${\cal F}_{\cal D}$ there. On the other hand, 
relation (\ref{LS3}) is a direct consequence of 
identity (\ref{LSME}) and of the following identity in the complex
$\sigma$-plane (where $\sigma = k^2$ is square of a four-vector):
\be
\int_{{\cal C}_1 + {\cal C}_2} 
\frac{d \sigma}{\sigma} \: \A_1(K^2=-\sigma e^{\cal C}) \left[
{\cal F}_{\cal D}(\sigma/Q^2) - {\cal F}_{\cal D}(0) \right] = 0 \ ,
\label{Cauchy}
\ee
where function ${\cal F}_{\cal D}$ is defined by identity
(\ref{LSEM}), and the path ${\cal C}_1 + {\cal C}_2$ is depicted 
in Fig.~(\ref{C1C2}). 
In the $\sigma$-plane, the only singularities
of the integrand in Eq.~(\ref{Cauchy}) are the cut of
$\A_1(-\sigma)$ along the positive semiaxis, and the cut of 
$[{\cal F}_{\cal D}(\sigma/Q^2) - {\cal F}_{\cal D}(0)]$ along
the negative semiaxis. Identity (\ref{Cauchy}) thus follows
from the Cauchy theorem.

When applying relation (\ref{LSME}) to the characteristic
function (\ref{FLSBj3}) of BjPSR,
we obtain for the time-like characteristic function of that
observable
\ba
F_b^{\cal M}(t) & = & t \left[ - \frac{10}{9}
- \frac{1}{3 t}  - \frac{2}{9 t^2} 
- \frac{2}{9} (5 t + 4) \ln t + 
\frac{2}{9} \left(
5t + 4 + \frac{2}{t^2} + \frac{1}{t^3} \right) \ln (1 + t)
\right] \ ,
\label{FLSBj4}
\ea
which agrees with the corresponding expression in Ref.~\cite{DMW} after identifying in their Eq.~(4.57): 
${\dot {\cal F}}_3( \epsilon, N=1) \equiv (-3/2)
F_b^{\cal M}(t)$, and $\epsilon \equiv t
= \mu^2/Q^2$ (Ref.~\cite{DMW} uses apparently ${\cal C}=0$).
 
\section{Explicit expressions for various coefficients}
\label{app4}

Expansion (\ref{apt}) is solution of the perturbative
RGE equation (\ref{pRGE}).
If the conventional (``$\bMS$'') 
reference scale $\bL$ \cite{Buras:1977qg,Bardeen:1978yd}
is adopted, 
and RGE (\ref{pRGE}) is iteratively solved for large $Q^2/\bL^2$
[$\ln(Q^2/\bL^2) \gg 1$] in an arbitrary RSch 
($\beta_2, \beta_3, \ldots$), this results in expansion (\ref{apt})
with coefficients $K_{k \ell}$ given in Eqs.~(\ref{kijbL})
for $k \leq 3$, and for $k=4,5,6$ given below
(notations: $c_j \equiv \beta_j/\beta_0$):

\noindent
For $k=4$:
\ba
K_{40} &=& - \frac{1}{2} \left( \frac{c_1^3}{\beta_0^4} \right)
\left( 1 - \frac{c_3}{c_1^3} \right), \qquad
K_{41} = - \left( \frac{c_1^3}{\beta_0^4} \right)
\left( -2 + 3 \frac{c_2}{c_1^2} \right),
\nonumber\\
K_{42} &=& \frac{5}{2} \left( \frac{c_1^3}{\beta_0^4} \right),
\qquad K_{43} = - \left( \frac{c_1^3}{\beta_0^4} \right) \ .
\label{K4ell}
\ea
For $k=5$:
\ba
K_{5 0} &=& \frac{1}{6 \beta_0^5 } \left(
7 c_1^4 - 18 c_1^2 c_2 + 10 c_2^2 - c_1 c_3 + 2 c_4 \right),
\nonumber\\
K_{5 1} &=& \frac{c_1}{\beta_0^5}
\left( 4 c_1^3 - 3 c_1 c_2 - 2 c_3 \right),
\qquad
K_{5 2} = - \frac{3}{2 \beta_0^5} 
\left( c_1^4 - 4 c_1^2 c_2 \right),
\nonumber\\
K_{5 3} &=& - \frac{13 c_1^4}{3 \beta_0^5}, 
\qquad
K_{5 4}= \frac{c_1^4}{\beta_0^5} \ .
\label{K5ell}
\ea
For $k=6$:
\ba
K_{6 0} &=& \frac{1}{12 \beta_0^6 } 
\left( 17 c_1^5 - 18 c_1^3 c_2 - c_1 c_2^2 - 23 c_1^2 c_3 + 
24 c_2 c_3 - 2 c_1 c_4 + 3 c_5 \right),
\nonumber\\
K_{6 1} &=& \frac{1}{6 \beta_0^6}
\left( -11 c_1^5 + 72 c_1^3 c_2 - 50 c_1 c_2^2 - 7 c_1^2 c_3 
- 10 c_1 c_4 \right),
\nonumber\\
K_{6 2} &=& \frac{1}{2 \beta_0^6}
\left( -23 c_1^5 + 27 c_1^3 c_2 + 10 c_1^2 c_3 \right),
\nonumber\\
K_{6 3} &=& \frac{1}{6 \beta_0^6}
\left( -11 c_1^5 - 60 c_1^3 c_2 \right),
\nonumber\\
K_{6 4} &=& \frac{77 c_1^5}{12 \beta_0^6},
\qquad K_{6 5} = -\frac{c_1^5}{\beta_0^6} \ .
\label{K6ell}
\ea
In practical calculations, we use: (a) at loop-level=3:
$c_3=c_4=c_5=0$ and we include in expansion (\ref{apt})
terms $K_{k \ell}$ up to $k_{\rm max}=5$; (b) at loop-level=4:
$c_4=c_5=0$ and we include terms up to $k_{\rm max}=6$.

The perturbation coefficients $d_j$ ($j=1,2$)
of the perturbation expansion for the massless Adler function
$d_v(Q^2)$, cf.~Eq.~(\ref{Dpt}), 
in $\bMS$ RSch and at RScl $\mu^2=Q^2$, are known
exactly, Refs.~\cite{d1,d2}, respectively
\ba
d_1^{\rm (Adl.)} & = & \frac{1}{12} + 0.691772 \beta_0,
\qquad
d_2^{\rm (Adl.)} = -27.849 + 8.22612 \beta_0 + 3.10345 \beta_0^2 \ .
\label{d1d2Adl}
\ea
The ${\rm N}^3{\rm LO}$ coefficient $d_3$, in the
aforementioned RSch and RScl, was obtained in
an approximate form in Ref.~\cite{Baikov:2002uw}
[(Eqs.~(20) and (12) in \cite{Baikov:2002uw})]:
\ba
d_3^{\rm (Adl.)} & = & 46.1992- 131.04 \beta_0 + 49.5237 \beta_0^2  
+ 2.18004 \beta_0^3 \ ,
\label{d3Adl}
\ea
where the coefficients at $\beta_0^3$ and at $\beta_0^2$
are known exactly 
(\cite{Broadhurst:1992si,Beneke:1993ee}, \cite{Bardeen:1978yd}), 
and the
other two coefficients were estimated in Ref.~\cite{Baikov:2002uw}
by using the methods of the principle of minimal
sensitivity (PMS) \cite{PMS}, and of the
effective charge (ECH) \cite{ECH,KKP}.

The light-by-light contributions are not included in
the coefficients (\ref{d1d2Adl} and (\ref{d3Adl}).
They have a different topology of diagrams and 
should probably be resummed separately (cf.~Ref.~\cite{KS}),
and they appear for the first time at $\sim a^3$ \cite{d2}.
They are proportional to the sum of the charges $\sum Q_f$. 
This sum is zero in the case $n_f=3$ considered here.

Coefficients $d_1$ and $d_2$ for BjPSR $d_b(Q^2)$,
in the aforementioned RSch and RScl, were obtained
in Ref.~\cite{LV} and are
\ba
d_1^{\rm (Bj.)} & = & - \frac{11}{12} + 2 \beta_0, \qquad
d_2^{\rm (Bj.)}   = 
-35.7644 + 10.5048 \beta_0 + 6.38889 \beta_0^2 \ .
\label{d1d2Bj} 
\ea
In the coefficient $d_3^{\rm (Bj.)}$, only the leading-$n_f$ part
($\propto n_f^3$) is known exactly \cite{Broadhurst:1993ru}
($\Leftrightarrow$ the leading-$\beta_0$ part, $\propto \beta_0^3$).
On this basis, the authors of Ref.~\cite{Broadhurst:2002bi} obtained
estimates of $d_3^{\rm (Bj.)}$ as a polynomial in $\beta_0$ 
by using naive nonabelianization (NNA) \cite{Broadhurst:1994se}: 
$n_f \mapsto - 6 \beta_0$. 
Several relations between BjPSR, Bjorken unpolarized sum rule, 
and Gross-Llewellyn Smith sum rule were found out and
investigated in Refs.~\cite{Kataev:2005ci}.

\section{Massless part of the strangeless tau decay ratio}
\label{app5}

In this Appendix we extract the measured value of the 
massless part of the QCD-canonical strangeless ratio
$r_{\tau}(\triangle S=0, m_q=0)$ for the semihadronic
decay, on the basis of the results of the final
ALEPH data analysis \cite{ALEPH2,ALEPH3}.\footnote{
For an extraction of $r_{\tau}(\triangle S=0, m_q=0)$ based
on the older set of measured results \cite{ALEPH1},
see for example Ref.~\cite{rtgctl}.}
This quantity is related to the ALEPH-measured \cite{ALEPH2,ALEPH3}
(V+A)-decay ratio 
\begin{eqnarray}
R_{\tau}(\triangle S\!=\!0)  &\equiv& 
\frac{ \Gamma (\tau^- \to \nu_{\tau} {\rm hadrons} (\gamma) )}
{ \Gamma (\tau^- \to \nu_{\tau} e^- {\overline {\nu}_e} (\gamma))}
- R_{\tau}(\triangle S\!\not=\!0)
\label{Rtaudef}
\\
& = & \frac{(1 - B_e - B_{\mu} )}{B_e} - R_{\tau}(\triangle S\!\not=\!0)
= 3.482  \pm 0.014 \ .  
\label{Rtauexp}
\end{eqnarray}
These values were obtained in Ref.~\cite{ALEPH3} from the
measured leptonic branching ratio
$B_e \equiv B(\tau^- \to \nu_{\tau} e^- {\overline \nu}_e)
= (17.810 \pm 0.039) \%$ (ALEPH, \cite{ALEPH2}),
from $B_{\mu} \equiv 
B(\tau^- \to \nu_{\tau} \mu^- {\overline \nu}_{\mu})
= (17.332 \pm 0.049) \%$ (world average, \cite{ALEPH3}),
and from the strangeness-changing branching
ratio $B_S = (2.85 \pm 0.11) \%$ (ALEPH,\cite{ALEPH2}).
The relation between the canonic massless quantity 
$r_{\tau}(\triangle S=0, m_q=0)$ and
quantity (\ref{Rtaudef})-(\ref{Rtauexp}) is
\ba
r_{\tau}(\triangle S=0, m_q=0) &\equiv&
r_{\tau}(\triangle S=0, m_q) - 
\delta r_{\tau}(\triangle S=0, m_{u,d}\not=0)
\label{rtgen1} 
\\
&=& \frac{ R_{\tau}(\triangle S=0) }
{ 3 |V_{ud}|^2 (1 + {\delta}_{{\rm EW}}) } -
(1 + \delta_{\rm EW}^{\prime} ) - 
\delta r_{\tau}(\triangle S=0, m_{u,d}\not=0) \ .
\label{rtgen2}
\ea
Here, $r_{\tau}(\triangle S=0, m_q=0)$ is QCD-canonical,
i.e., its pQCD expansion is 
$r_{\tau}(\triangle S=0, m_q=0)_{\rm pt} = a + {\cal O}(a^2)$;
the Cabibbo-Kobayashi-Maskawa (CKM) matrix element $|V_{ud}|$
has the value largely dominated by $0^+ \to 0^+$ nuclear beta decays
\cite{Blucher:2005dc}
\be
|V_{ud}| = 0.9738 \pm 0.0003 \ ,
\label{Vud}
\ee
the electroweak (EW) correction parameter is
$1+\delta_{\rm EW} = 1.0198 \pm 0.0006$
\cite{ALEPH2,ALEPH3}; the residual EW correction
parameter is $\delta_{\rm EW}^{\prime} = 0.0010$
\cite{Braaten:1990ef}; the (V+A)-channel corrections
$\delta r_{\tau}(\triangle S=0, m_{u,d}\not=0)$
due to the nonzero quark masses are
\cite{Braaten:1992qm,ALEPH3}
the sum of the $D=2,4,6,$ and 8-dimensional
corrections $(\delta_{ud,V}^{(D)} + \delta_{ud,A}^{(D)})/2$   
and their value is \cite{ALEPH3} either 
$\delta r_{\tau}(\triangle S=0, m_{u,d}\not=0) = (-5.2 \pm 1.7)
\times 10^{-3}$ if the gluon condensate contribution 
is included, and $(- 5.0 \pm 1.7) \times 10^{-3}$ if the gluon condensate
contribution is not included (using for the gluon condensate
the ALEPH-value $\langle a G G \rangle = (-0.5 \pm 0.3) \times
10^{-2}$ \cite{ALEPH2,ALEPH3}).

Inserting all the aforementioned values in relation
(\ref{rtgen2}) and taking into account the value
(\ref{Rtauexp}), we extract the experimental prediction
for $r_{\tau}(\triangle S=0, m_q=0)$
based on the most recent ALEPH data 
\be
r_{\tau}(\triangle S=0, m_q=0)_{\rm exp.} =
0.204 \pm 0.005 \ ,
\label{rtauexp}
\ee
where the uncertainties have been added in quadrature.
The uncertainty in Eq.~(\ref{rtauexp}) is dominated by the experimental
uncertainty $\delta R_{\tau} = \pm 0.014$ (\ref{Rtauexp}).
The value (\ref{rtauexp}) remains unaffected up to the displayed
digits when we either include or exclude from the
above quantity the gluon condensate contribution.


\begin{thebibliography}{9}

\bibitem{ShS}
  D.~V.~Shirkov and I.~L.~Solovtsov,
  hep-ph/9604363;
  Phys.~Rev.~Lett.~{\bf 79}, 1209 (1997)
[hep-ph/9704333].

\bibitem{Milton:1997mi}
  K.~A.~Milton, I.~L.~Solovtsov and O.~P.~Solovtsova,
  Phys.\ Lett.\ B {\bf 415}, 104 (1997)
[hep-ph/9706409].

\bibitem{Sh}
  D.~V.~Shirkov,
  Theor.\ Math.\ Phys.\  {\bf 127}, 409 (2001)
[hep-ph/0012283];
  Eur.\ Phys.\ J.\ C {\bf 22}, 331 (2001)
  [hep-ph/0107282].

\bibitem{Milton:2000fi}
  K.~A.~Milton, I.~L.~Solovtsov, O.~P.~Solovtsova and V.~I.~Yasnov,
  Eur.\ Phys.\ J.\ C {\bf 14}, 495 (2000)
  [hep-ph/0003030].

\bibitem{Bakulev}
  A.~P.~Bakulev, S.~V.~Mikhailov and N.~G.~Stefanis,
  Phys.\ Rev.\ D {\bf 72}, 074014 (2005)
  [Erratum-ibid.\ D {\bf 72}, 119908 (2005)]
[hep-ph/0506311];
  hep-ph/0607040;
  A.~P.~Bakulev, A.~I.~Karanikas and N.~G.~Stefanis,
  Phys.\ Rev.\ D {\bf 72}, 074015 (2005)
[hep-ph/0504275].

\bibitem{Broadhurst:2000yc}
  D.~J.~Broadhurst, A.~L.~Kataev and C.~J.~Maxwell,
  Nucl.\ Phys.\ B {\bf 592}, 247 (2001)
  [hep-ph/0007152].

\bibitem{ALEPH1}
  R.~Barate {\it et al.}  [ALEPH Collaboration],
  Eur.\ Phys.\ J.\ C {\bf 4}, 409 (1998);
  K.~Ackerstaff {\it et al.}  [OPAL Collaboration],
  Eur.\ Phys.\ J.\ C {\bf 7}, 571 (1999)
[hep-ex/9808019].

\bibitem{ALEPH2}
  S.~Schael {\it et al.}  [ALEPH Collaboration],
  Phys.\ Rept.\  {\bf 421}, 191 (2005)
  [hep-ex/0506072];

\bibitem{ALEPH3}
  M.~Davier, A.~H\"ocker and Z.~Zhang,
  hep-ph/0507078.

\bibitem{Milton:2001mq}
K.~A.~Milton, I.~L.~Solovtsov, O.~P.~Solovtsova,
  Phys.\ Rev.\ D {\bf 64}, 016005 (2001)
[hep-ph/0102254];
Mod.\ Phys.\ Lett.\ A {\bf 21}, 1355 (2006)
  [hep-ph/0512209].

\bibitem{KS}
  A.~L.~Kataev and V.~V.~Starshenko,
  Mod.\ Phys.\ Lett.\ A {\bf 10}, 235 (1995)
  [hep-ph/9502348].

\bibitem{Cvetic:2006mk}
  G.~Cveti\v c and C.~Valenzuela,
  J.\ Phys.\ G {\bf 32}, L27 (2006)
  [hep-ph/0601050].

\bibitem{math}
{\it Mathematica\/} 5.2, Wolfram Research, Inc. 

\bibitem{Buras:1977qg}
  A.~J.~Buras, E.~G.~Floratos, D.~A.~Ross and C.~T.~Sachrajda,
  Nucl.\ Phys.\ B {\bf 131}, 308 (1977).

\bibitem{Bardeen:1978yd}
  W.~A.~Bardeen, A.~J.~Buras, D.~W.~Duke and T.~Muta,
  Phys.\ Rev.\ D {\bf 18}, 3998 (1978).


\bibitem{Magradze}
  D.~S.~Kurashev and B.~A.~Magradze,
  Theor.\ Math.\ Phys.\  {\bf 135}, 531 (2003);
  hep-ph/0104142.

\bibitem{Magradze2}
  B.~A.~Magradze,
  hep-ph/0305020.

\bibitem{Alekseev:2002zn}
  A.~I.~Alekseev,
  Few Body Syst.\  {\bf 32}, 193 (2003)
  [hep-ph/0211339];
  J.\ Phys.\ G {\bf 27}, L117 (2001)
[hep-ph/0105338].


\bibitem{Gardi:1998qr}
  E.~Gardi, G.~Grunberg and M.~Karliner,
  JHEP {\bf 9807}, 007 (1998)
  [hep-ph/9806462].


\bibitem{Nesterenko}
  A.~V.~Nesterenko,
  Phys.\ Rev.\ D {\bf 62}, 094028 (2000)
[hep-ph/9912351];
  Phys.\ Rev.\ D {\bf 64}, 116009 (2001)
[hep-ph/0102124];
  Int.\ J.\ Mod.\ Phys.\ A {\bf 18}, 5475 (2003)
[hep-ph/0308288];
  A.~V.~Nesterenko and J.~Papavassiliou,
  Phys.\ Rev.\ D {\bf 71}, 016009 (2005)
[hep-ph/0410406].


\bibitem{Raczka}
P.~A.~R\c{a}czka, 
hep-ph/0512339, presented at QCD05, Montpellier, July 2005;
hep-ph/0602085;
hep-ph/0608196.

\bibitem{Alekseev}
  A.~I.~Alekseev,
  Theor.\ Math.\ Phys.\  {\bf 145}, 1559 (2005)
  [Teor.\ Mat.\ Fiz.\  {\bf 145}, 221 (2005)];
  hep-ph/0503242.

\bibitem{Prosperi:2006hx}
  G.~M.~Prosperi, M.~Raciti and C.~Simolo,
  hep-ph/0607209.

\bibitem{time-like1}
A.~V.~Radyushkin, Dubna JINR preprint E2-82-159 (1982);
  JINR Rapid Commun.\  {\bf 78}, 96 (1996)
  [hep-ph/9907228].

\bibitem{time-like2}
  N.~V.~Krasnikov and A.~A.~Pivovarov,
  Phys.\ Lett.\ B {\bf 116} (1982) 168;
  A.~A.~Pivovarov,
  Sov.\ J.\ Nucl.\ Phys.\  {\bf 54}, 676 (1991)	
  [hep-ph/0302003].


\bibitem{time-like3}
  K.~A.~Milton and I.~L.~Solovtsov,
  Phys.\ Rev.\ D {\bf 55}, 5295 (1997)
[hep-ph/9611438].

\bibitem{PPR}
  S.~Peris, M.~Perrottet and E.~de Rafael,
  JHEP {\bf 9805}, 011 (1998)
[hep-ph/9805442].

\bibitem{Cvetic:2005my}
  G.~Cveti\v c, C.~Valenzuela and I.~Schmidt,
  hep-ph/0508101, presented at QCD05, Montpellier, France, July 2005.

\bibitem{PMS}
P.~M.~Stevenson, Phys. Rev. D {\bf 23}, 2916 (1981);
Phys. Lett. {\bf 100B}, 61 (1981);
Nucl. Phys. {\bf B203}, 472 (1982).

\bibitem{Cvetic:2000mz}
  G.~Cveti\v c and R.~K\"ogerler,
  Phys.\ Rev.\ D {\bf 63}, 056013 (2001)
  [hep-ph/0006098].


\bibitem{Neubert}
  M.~Neubert,
  Phys.\ Rev.\ D {\bf 51}, 5924 (1995)
  [hep-ph/9412265].

\bibitem{Neubert2}
M.~Neubert, hep-ph/9502264.

\bibitem{Brooks:2006it}
  P.~M.~Brooks and C.~J.~Maxwell,
hep-ph/0604267.


\bibitem{Gardi:1999dq}
  E.~Gardi and G.~Grunberg,
  JHEP {\bf 9911}, 016 (1999)
[hep-ph/9908458].

\bibitem{Brodsky}
  S.~J.~Brodsky, E.~Gardi, G.~Grunberg and J.~Rathsman,
  Phys.\ Rev.\ D {\bf 63}, 094017 (2001)
[hep-ph/0002065].

\bibitem{d1}
  K.~G.~Chetyrkin, A.~L.~Kataev and F.~V.~Tkachov,
  Phys.\ Lett.\ B {\bf 85}, 277 (1979);
  M.~Dine and J.~R.~Sapirstein,
  Phys.\ Rev.\ Lett.\  {\bf 43}, 668 (1979);
  W.~Celmaster and R.~J.~Gonsalves,
  Phys.\ Rev.\ Lett.\  {\bf 44}, 560 (1980).

\bibitem{d2}
  S.~G.~Gorishnii, A.~L.~Kataev and S.~A.~Larin,
  Phys.\ Lett.\ B {\bf 259}, 144 (1991);
  L.~R.~Surguladze and M.~A.~Samuel,
  Phys.\ Rev.\ Lett.\  {\bf 66}, 560 (1991)
  [Erratum-ibid.\  {\bf 66}, 2416 (1991)].


\bibitem{Baikov:2002uw}
  P.~A.~Baikov, K.~G.~Chetyrkin and J.~H.~K\"uhn,
  Phys.\ Rev.\ D {\bf 67}, 074026 (2003)
[hep-ph/0212299].

\bibitem{Nesterenko:2005wh}
  A.~V.~Nesterenko and J.~Papavassiliou,
  J.\ Phys.\ G {\bf 32}, 1025 (2006)
  [hep-ph/0511215].

\bibitem{Broadhurst:1993ru}
  D.~J.~Broadhurst and A.~L.~Kataev,
  Phys.\ Lett.\ B {\bf 315}, 179 (1993)
[hep-ph/9308274].

\bibitem{LV}
S.~G.~Gorishny and S.~A.~Larin, Phys. Lett. B {\bf 172}, 109 (1986);
E.~B.~Zijlstra and W.~Van Neerven, Phys. Lett. B {\bf 297}, 377 (1992);
S.~A.~Larin and J.~A.~M. Vermaseren, Phys. Lett. B {\bf 259}, 
345 (1991).

\bibitem{Broadhurst:2002bi}
  D.~J.~Broadhurst and A.~L.~Kataev,
  Phys.\ Lett.\ B {\bf 544}, 154 (2002)
  [hep-ph/0207261].

\bibitem{Broadhurst:1994se}
  D.~J.~Broadhurst and A.~G.~Grozin,
  Phys.\ Rev.\ D {\bf 52}, 4082 (1995)
  [hep-ph/9410240].

\bibitem{Deur:2004ti}
  A.~Deur {\it et al.},
  Phys.\ Rev.\ Lett.\ {\bf 93}, 212001 (2004)
[hep-ex/0407007].

\bibitem{Campanario:2005np}
  F.~Campanario and A.~Pineda,
  Phys.\ Rev.\ D {\bf 72}, 056008 (2005)
[hep-ph/0508217].

\bibitem{Buza:1996xr}
  M.~Buza, Y.~Matiounine, J.~Smith and W.~L.~van Neerven,
  Nucl.\ Phys.\ B {\bf 485}, 420 (1997)
  [hep-ph/9608342];
  J.~Bl\"umlein and W.~L.~van Neerven,
  Phys.\ Lett.\ B {\bf 450}, 417 (1999)
  [hep-ph/9811351].


\bibitem{Eidelman}
  S.~Eidelman, F.~Jegerlehner, A.~L.~Kataev and O.~Veretin,
  Phys.\ Lett.\ B {\bf 454}, 369 (1999)
[hep-ph/9812521].


\bibitem{ChHKS1}
  K.~G.~Chetyrkin, J.~H.~K\"uhn and M.~Steinhauser,
  Nucl.\ Phys.\ B {\bf 482}, 213 (1996)
  [hep-ph/9606230].


\bibitem{FT}
J.~Fleischer and O.~V.~Tarasov,
Z.\ Phys.\ C {\bf 64}, 413 (1994)
[hep-ph/9403230].

\bibitem{ChHKS2}
K.~G.~Chetyrkin, R.~Harlander, J.~H.~K\"uhn and M.~Steinhauser,
Nucl.\ Phys.\ B {\bf 503}, 339 (1997)
[hep-ph/9704222].


\bibitem{CHKST}
  K.~G.~Chetyrkin, A.~H.~Hoang, J.~H.~K\"uhn, M.~Steinhauser and T.~Teubner,
  Eur.\ Phys.\ J.\ C {\bf 2}, 137 (1998)
  [hep-ph/9711327].

\bibitem{HJKT}
  A.~H.~Hoang, M.~Je\v zabek, J.~H.~K\"uhn and T.~Teubner,
  Phys.\ Lett.\ B {\bf 338}, 330 (1994)
  [hep-ph/9407338].

\bibitem{Kniehl}
  B.~A.~Kniehl,
  Phys.\ Lett.\ B {\bf 237}, 127 (1990).

\bibitem{EJ}
 S.~Eidelman and F.~Jegerlehner,
 Z.\ Phys.\ C {\bf 67}, 585 (1995)
 [hep-ph/9502298].

\bibitem{Yao:2006px}
  W.~M.~Yao {\it et al.}  [Particle Data Group],
  J.\ Phys.\ G {\bf 33}, 1 (2006).

\bibitem{Broadhurst:1992si}
  D.~J.~Broadhurst,
  Z.\ Phys.\ C {\bf 58}, 339 (1993).

\bibitem{Beneke:1993ee}
  M.~Beneke,
  Phys.\ Lett.\ B {\bf 307}, 154 (1993);
  M.~Beneke,
  Nucl.\ Phys.\ B {\bf 405}, 424 (1993).

\bibitem{DMW}
  Y.~L.~Dokshitzer, G.~Marchesini and B.~R.~Webber,
  Nucl.\ Phys.\ B {\bf 469}, 93 (1996)
  [hep-ph/9512336].

\bibitem{ECH}
G.~Grunberg, Phys. Lett. {\bf 95B}, 70 (1980),
{\bf 110B}, 501(E) (1982);
{\bf 114B}, 271 (1982);
Phys. Rev. D {\bf 29}, 2315 (1984).

\bibitem{KKP}
  N.~V.~Krasnikov,
  Nucl.\ Phys.\ B {\bf 192} (1981) 497
  [Yad.\ Fiz.\  {\bf 35} (1982) 1594];
A.~L.~Kataev, N.~V. Krasnikov, and A.~A. Pivovarov,
Nucl. Phys. {\bf B198}, 508 (1982).

\bibitem{Kataev:2005ci}
  A.~L.~Kataev,
  JETP Lett.\  {\bf 81}, 608 (2005)
  [Pisma Zh.\ Eksp.\ Teor.\ Fiz.\  {\bf 81}, 744 (2005)]
  [hep-ph/0505108];
  Mod.\ Phys.\ Lett.\ A {\bf 20}, 2007 (2005)
  [hep-ph/0505230].

\bibitem{rtgctl}
  G.~Cveti\v c and T.~Lee,
  Phys.\ Rev.\ D {\bf 64}, 014030 (2001)
  [hep-ph/0101297];
  G.~Cveti\v c, C.~Dib, T.~Lee and I.~Schmidt,
  Phys.\ Rev.\ D {\bf 64}, 093016 (2001)
  [hep-ph/0106024].

\bibitem{Blucher:2005dc}
  E.~Blucher {\it et al.},
  hep-ph/0512039,
to appear in the Proceedings of 3rd Workshop on the Unitarity 
Triangle: CKM 2005, San Diego, California, March 15-18, 2005.


\bibitem{Braaten:1990ef}
E.~Braaten and C.~Li,
Phys.\ Rev.\ D {\bf 42}, 3888 (1990).

\bibitem{Braaten:1992qm}
E.~Braaten, S.~Narison, and A.~Pich,
Nucl.\ Phys.\ B {\bf 373}, 581 (1992).



\end{thebibliography}
\end{document}